\documentclass[prc,aps,floats,floatfix,twocolumn,showpacs,nofootinbib,%
superscriptaddress]{revtex4}

\usepackage{amsmath}
\usepackage{amssymb}
\usepackage{amsfonts}
\usepackage{graphicx}
\usepackage{tabularx}
\usepackage{multirow}
\usepackage{lscape}
\usepackage{rotating}
\usepackage{bbm}

\usepackage{epsfig}

\newcommand{\rmd}{\text{d}}

\allowdisplaybreaks

\newcommand{\cev}[1]{\reflectbox{\ensuremath{\vec{\reflectbox{\ensuremath{#1}}}}}}

\begin{document}

\title{Skyrme functional from a three-body pseudo-potential of second-order in gradients. \\
       Formalism for central terms.}


\author{J. Sadoudi}
\affiliation{CEA-Saclay DSM/Irfu/SPhN, F-91191 Gif sur Yvette Cedex, France}
\affiliation{Universit\'e Bordeaux, Centre d'Etudes Nucl\'eaires de
            Bordeaux Gradignan, UMR5797, F-33175 Gradignan, France}
\affiliation{CNRS/IN2P3, Centre d'Etudes Nucl\'eaires de Bordeaux
            Gradignan, UMR5797, F-33175 Gradignan, France}
         
\author{T. Duguet}
\affiliation{CEA-Saclay DSM/Irfu/SPhN, F-91191 Gif sur Yvette Cedex, France}
\affiliation{National Superconducting Cyclotron Laboratory and Department of Physics and Astronomy, Michigan State University, East Lansing, MI 48824, USA}
          
\author{J. Meyer}
\affiliation{Universit\'e de Lyon, F-69003 Lyon, France; Universit\'e Lyon 1,
             43 Bd. du 11 Novembre 1918, F-69622 Villeurbanne cedex, France   \\
             CNRS-IN2P3, UMR 5822, Institut de Physique Nucl{\'e}aire de Lyon}
  
\author{M. Bender}
\affiliation{Universit\'e Bordeaux, Centre d'Etudes Nucl\'eaires de
            Bordeaux Gradignan, UMR5797, F-33175 Gradignan, France}
\affiliation{CNRS/IN2P3, Centre d'Etudes Nucl\'eaires de Bordeaux
            Gradignan, UMR5797, F-33175 Gradignan, France}


\begin{abstract}
\begin{description}

\item[Background]
In one way or the other, all modern parametrizations of the nuclear
energy density functional (EDF) do not respect the exchange symmetry
associated with Pauli's principle. It has been recently shown that
this practice jeopardizes multi-reference (MR) EDF calculations by
contaminating the energy with spurious self-interactions that, for
example, lead to finite steps or even divergences when plotting 
it as a function of collective coordinates
[J.\ Dobaczewski \textit{et al}., Phys.\ Rev.\ C \textbf{76}, 054315 (2007);
D.\ Lacroix \textit{et al}., Phys.\ Rev.\ C \textbf{79}, 044318 (2009)].
As of today, the only viable option to bypass these pathologies is to 
rely on EDF kernels that enforce Pauli's principle from the outset by 
strictly and exactly deriving from a \textit{genuine}, i.e.\ 
density-independent, Hamilton operator.
\item[Purpose]
The objective is to build cutting-edge parametrizations of the EDF kernel 
deriving from a pseudo potential that can be safely employed in 
symmetry restoration and configuration mixing calculations. 
\item[Methods]
We wish to develop the most general Skyrme-like EDF parametrization 
containing linear, bilinear and trilinear terms in the density matrices 
with up to two gradients, under the key constraint that it derives
strictly from an effective Hamilton operator. While linear and bilinear 
terms are obtained from a standard one-body kinetic energy operator and a
(density-independent) two-body Skyrme pseudo-potential, the most general 
three-body Skyrme-like pseudo-potential containing up to two gradient 
operators is constructed to generate the trilinear part. The present 
study is limited to central terms. Spin-orbit and tensor will be addressed
in a forthcoming paper.
\item[Results]
The most general central Skyrme-type zero-range three-body interaction 
is built up to second order in derivatives. The complete trilinear energy 
density functional, including time-odd and \mbox{$T=1$} pairing parts, is 
derived along with the corresponding normal and anomalous fields entering the 
Hartree-Fock-Bogoliubov equations of motion. Its building blocks are the 
same local densities that the standard Skyrme functional is constructed from. 
The central three-body pseudo-potential is defined out of six independent 
parameters. Expressions for bulk properties of symmetric, isospin-asymmetric 
and spin-polarized homogeneous nuclear matter, as well as associated Landau 
parameters, are given.
\item[Conclusions]
This study establishes a first step towards a new generation of nuclear 
energy density functionals that respect Pauli's principle and that can 
be safely used in predictive and spuriousity-free SR and MR EDF calculations.
\end{description}
\end{abstract}


\pacs{21.60.Jz, 
      21.30.Fe  
}

\date{2 October 2013}


\maketitle

%
\section{Introduction}
\label{sect:intro}
%

Methods based on an energy density functional (EDF) are at present the only 
available microscopic tools to address all medium- and heavy-mass nuclei 
within one consistent framework. They allow for a unified description of 
many phenomena in nuclear structure and dynamics over the entire chart of 
nuclei~\cite{quentin78a,bennaceur03a,bender03a,erler11a}. 
The EDF method coexists on two levels: One level is usually characterized 
as "mean-field" and frequently identified with Hartree-Fock (HF), HF+BCS 
and Hartree-Fock-Bogoliubov (HFB) methods. The second
level is often qualified as "beyond mean-field", a notion used for both the 
random phase approximation (RPA) and its extensions on the one hand, and
symmetry restorations and the generator coordinate method (GCM) on the other 
hand. Throughout this article we refer to mean-field methods as 
\textit{single-reference} (SR) EDF methods, as all densities entering the 
kernel are constructed from a single product state. Symmetry restoration 
and GCM will be denoted as \textit{multi-reference} (MR) EDF 
methods, as the densities entering the kernel are constructed from 
pairs of product states belonging to a large set of reference states. 
An overview over the SR- and MR-EDF formalism can be found in 
Ref.~\cite{duguet13a}.

In the literature, one finds nuclear EDF kernels of many different forms, 
being  either local or non local, and being either relativistic or 
non-relativistic.  For the purpose of the present study there is a 
third categorization of energy functionals that has to be made that 
concerns the handling of Pauli's exclusion principle. Its most obvious 
consequence, namely that all single-particle levels are occupied by at 
most one nucleon, is always satisfied at the level of individual densities 
used to build the EDF given that one-body density matrices are explicitly 
computed from antisymmetric product states of either Slater or Bogoliubov 
type. However, a violation of Pauli's principle may arise when multiplying 
two or even more such densities to build bilinear, trilinear, etc.\ terms 
in the EDF kernel. Terms of a given power in the density matrices have to
combine in a very specific way to cancel out the unphysical interaction 
of a particle with itself or that of a pair of particles with 
itself~\cite{stringari78a,perdew81a,lacroix09a,bender09a,bender99n,chamel10a}, 
and to provide an antisymmetric residual interaction~\cite{stringari78a}.

As of today, the only practical way to enforce all aspects of exchange 
symmetry is to set up the off-diagonal EDF kernel as the matrix elements 
of a \textit{genuine}, i.e.\ density-independent, operator between two 
product states of Bogoliubov type, taking all exchange and pairing terms 
into account without any approximation or simplification. 
Such an effective Hamilton operator is typically meant to be the sum of the 
kinetic energy operator and a pseudo-potential. By virtue of the generalized 
Wick theorem~\cite{balian69a,ring80a,blaizotripka}, the resulting kernel 
takes the form of a specific functional of one-body transition density 
matrices built from the two product states. In what follows, such EDF 
kernels are said to be \textit{pseudo-potential based}. 

None of the modern, e.g.\ Skyrme, Gogny or relativistic, parametrizations 
belong to this category of pseudo-potential-based kernels. The most remote 
are kernels directly built on the level of combinations of one-body densities 
or density matrices without making reference to any underlying operator. 
In what follows, such EDF kernels are said to belong to the category of  
\textit{general functionals}. In that case, the form of the EDF kernel 
is typically constrained by all symmetries of the nuclear 
Hamiltonian~\cite{doba96b} but the exchange symmetry. Frequently used 
examples are the EDF kernels constructed by Fayans and 
collaborators~\cite{fayans00}, and the Barcelona-Catania-Paris 
parametrization~\cite{baldo08a}.

The large majority of existing parametrizations, however, falls in between 
the pseudo-potential-based kernels and the general functionals. We denote 
those as \textit{hybrid} parametrizations. The paradigm here consists of 
relating the interaction part of the EDF kernel to the expectation value 
of a \textit{density-dependent}, and therefore state-dependent, effective 
interaction, keeping all exchange and pairing terms. Prominent examples 
are the Gogny family of 
parametrizations~\cite{gogny75,decharge80a,chappert08a,goriely09a} and those 
derived from the density-dependent M3Y interaction by Nakada~\cite{nakada03a}.
Also a very few Skyrme parametrizations were constructed along this line; 
examples are SkP~\cite{dobaczewski84a}, SkS1-SkS4~\cite{gomez92a}, SkE2 and 
SkE4~\cite{waroquier79a}. For most of the other Skyrme parametrizations, the 
link to the density-dependent effective interactions is only kept for some 
terms, but not for all.\footnote{As a relic of the historical origin of 
the Skyrme effective interaction as a pseudo-potential, general or hybrid 
Skyrme parametrizations are frequently defined in terms of the parameters 
of a density-dependent effective interaction to which is added a list 
modifications of the EDF generated by an actual density-dependent  two-body 
effective interaction~\cite{bender03a,erler11a}. This ambivalence of the 
Skyrme EDF and the representation of its parameters is a source of confusion 
and frequently provokes its inconsistent use in RPA and in the calculation 
of infinite nuclear matter properties~\cite{bender02a,bender03a,chamel10a}.
} 
First of all, the particle-hole (i.e.\ normal) and particle-particle 
(i.e.\ pairing or anomalous) parts of almost  all Skyrme parametrizations are 
entirely unrelated. As a matter of fact, only the particle-hole part of the 
kernel is usually referred to as the \textit{Skyrme EDF}, which is then 
combined with the pairing EDF of the respective author's preference. Second, 
specific exchange terms in the energy functional are often modified or simply 
set to zero for reasons of phenomenology. The latter practice mainly 
concerns spin-orbit and spin-tensor 
terms~\cite{bender03a,reinhard95a,lesinski07a,zalewski08a,bender09b,erler11a} and the 
so-called time-odd terms entering the particle-hole part of the EDF~\cite{bender02a,duguet02a,bender03a,erler11a,pototzky10a,Schunck10a,hellemans12a}. 
Such modifications bring the functionals close to spirit of the general 
functionals. 

Among the existing functionals, Gogny or M3Y-based parametrizations are the 
closest to the concept of a pseudo-potential-based EDF kernel. Still, from 
the point of view of Pauli's exclusion principle, they are fundamentally 
different given that the exchange symmetry is not respected by the 
density-dependent term. In that respect, general and hybrid functionals all 
belong to the same category. Indeed, it makes no difference that the exchange 
symmetry is broken by just one term in the functional or by many of them.

In spite of the many successes of general and  hybrid functionals, there are 
good reasons to revisit pseudo-potential-based EDF kernels. Indeed, it was 
recently demonstrated~\cite{lacroix09a,bender09a} that any breaking of 
Pauli's principle contaminates the EDF kernel with spurious contributions 
that can jeopardize MR-EDF 
calculations~\cite{tajima92a,donau98a,almehed01a,anguiano01a,dobaczewski07a,zdunczuk09a}. The problem manifests itself through finite steps and/or even 
divergences when plotting the symmetry-restored energy as a function of 
a collective coordinate~\cite{tajima92a,dobaczewski07a,bender09a,duguet09a,zdunczuk09a}. 
Even more striking, contaminated kernels can lead to \textit{non-zero} 
(non-normalized) energies when restoring good \textit{negative} particle 
number \cite{bender09a}, which is impossible for projected operator
matrix elements.
Generally speaking, the results of MR-EDF calculations based on hybrid or 
general functionals depend on how the sums and/or integrals over the 
collective coordinates are discretized. Decreasing the step sizes often 
amplifies the contamination with spurious contributions as they become better 
resolved \cite{bender09a}. In addition,  using non-analytical functions for 
the density dependence, such as the popular  $\rho^\alpha(\vec{r})$ 
dependence, introduces a further problem into the MR EDF frame by making 
the EDF kernel a multivalued function of the collective 
coordinates~\cite{duguet09a}. 

In the use of the Gogny functional, whose only breaking of Pauli's 
principle relates to its density-dependent term, a special treatment of 
the latter has been used in some MR calculations to bypass the problem 
invoked above~\cite{anguiano01a,rodriguez05a}. Besides not being consistent
with the definition of the rest of the EDF kernel, this recipe cannot be 
expected to work for all configuration mixings of interest~\cite{robledo07a}.

Isospin and angular-momentum projected MR EDF calculations of 
Refs.~\cite{satula10a,satula12a} employ a kernel  that derives strictly 
from a simple two-body Skyrme pseudo-potential without density dependence. 
In doing so, the pathologies alluded to above are fully avoided by
construction. As these calculations neglect pairing correlations altogether, 
it is of no concern that the two-body Skyrme pseudo-potentials employed fail 
to provide reasonable pairing correlations. To the best of our knowledge, 
there has been only one recent attempt by us to construct a Skyrme 
pseudo-potential \textit{without any  density-dependence} that is meant to 
be used in both the particle-hole and the particle-particle parts of the 
EDF~\cite{sadoudi13a}. This was achieved by adding gradient-less three- 
and four-body contact terms to a standard velocity-dependent two-body 
Skyrme pseudo-potential. However, the best parameter fit generated in this 
way did, by far, not match the performance of modern hybrids or general 
Skyrme parametrizations. As a matter of fact, it turned out to be impossible 
to have simultaneously appropriate empirical nuclear matter properties, 
attractive pairing and stability against infinite- and finite-size 
instabilities~\cite{sadoudi13a}. This limitation points to the necessity 
of introducing additional higher-order terms.

Aiming at a strict pseudo-potential-based approach, there are two possible 
directions of enriching a Skyrme-like parametrization. One is to include 
terms containing an higher number of gradient operators in the two-body  
pseudo-potential. Terms of this kind have already been suggested in the 
seminal papers by Skyrme~\cite{skyrme56a,bell56a,skyrme59a}. Their most
general form has recently been worked out systematically up to sixth 
order~\cite{raimondi11a}. The associated EDF kernel remains strictly 
bilinear, but is expressed in terms of a larger set of local densities 
invoking more derivatives than is the case for the standard Skyrme functional.
An alternative is to stick to the second order in gradients, but to consider 
many-body operators, e.g.\ velocity-dependent three-body interactions. The 
associated EDF kernel can still be expressed in terms of the same set of 
local densities as the bilinear one deriving from the standard Skyrme 
two-body pseudo potential, but contains higher-order polynomials. 

The number of contributions to infinite nuclear matter properties that 
originate from fourth-order gradient operators in the two-body 
pseudo-potential is much smaller than the number of those originating from 
second-order terms~\cite{k4k6:INM}. In the end, one mainly obtains a multitude
of contributions that influence surface properties of finite nuclei. 
Consequently, many-body terms might offer an easier access to a decoupling of 
nuclear matter properties from  pairing and instabilities. This is thus the 
route we wish to pursue in the present work. Ultimately, one may of course 
combine both types of extensions. However, and although those two 
extensions are systematic, one should note that no strong formal argument 
exists at this point to declare one to be superior to the 
other. Even though "naturalness" can be invoked~\cite{Kortelainen:2010dt}
in the context of Skyrme EDF parametrizations, this concept has not yet been 
proven to provide a truly meaningful and systematic power counting at finite 
density.\footnote{We note that Weinberg's power counting based on naive 
dimensional analysis does already not provide an appropriate power 
counting for in-vacuum nuclear interactions based on chiral effective 
field theory~\cite{Long:2013lla}.} A formal framework that establishes
a hierarchy of terms in the EDF is currently missing and clearly deserves 
attention in the future.

Skyrme-type contact three-body pseudo potentials containing up to two 
gradient operators have been already used in the 
past~\cite{blaizot75a,liu75a,onishi78a,arima86a,zheng90a,waroquier76a,waroquier79a,waroquier83a,waroquier83b,liu91a,liu91b}. 
None of these developments has been systematic and aiming at the complete
set of possible terms. Also, not all of these studies have combined their 
three-body pseudo-potential  with a Skyrme-type two-body interaction. 
Additionally, all these studies limited themselves to central interactions, 
and none of them aimed at the most general structure. Only time-even 
contributions to the normal part of the resulting EDF were discussed, if at 
all, and spherical symmetry was assumed and exploited in all cases to simplify
the resulting energy functional and one-body fields.

The aim of the present study is to supersede the existing body of work in 
several respects by
\begin{enumerate}
\item
Constructing the most general contact three-body pseudo potential containing 
up to two gradient operators. In the present study, we focus on its 
\textit{central} part, i.e.\ on terms that do not couple the orientation of 
spins and momenta. Central terms are the most important ones for our goal 
of replacing the traditional density-dependence of the standard Skyrme EDF. 
At the SR-EDF level, only central terms contribute to properties of 
non-polarized infinite nuclear matter and therefore to bulk properties of 
even-even nuclei. 
By contrast, three-body spin-orbit and tensor interactions produce 
terms that allow for the fine-tuning of the nucleon-number dependence of 
shell structure. These will be discussed elsewhere \cite{sadoudi:LS+tensor}.

\item 
Deriving the complete trilinear EDF kernel from the three-body 
pseudo-potential, i.e.\ providing time-even and time-odd contributions to 
the normal part of the EDF along with the complete pairing part, in a 
form that is suited for symmetry-unrestricted SR and MR calculations. 
We limit ourselves to the case where single-particle states, and consequently
the one-body density matrices, retain a good neutron or proton character.

\item
Deriving the expressions of the corresponding one-body fields entering 
the HFB Hamiltonian matrix.

\item 
Computing infinite nuclear matter properties and associated Landau parameters.
\end{enumerate}
As will become clear below, there exists a large number of possible central  
three-body contact operators that respect symmetries of the exact nuclear 
Hamiltonian. Only a small subset of these, however, provide linearly 
independent contributions to the EDF kernel. To find 
a complete irreducible set of such operators, we proceed in the  following way
\begin{enumerate}
\item
Write down all possible operator structures consistent with symmetries of the 
underlying nuclear Hamiltonian.
\item
Derive the corresponding EDF kernel.
\item
Perform the singular-value decomposition (SVD) of the matrix expressing the 
coupling constants multiplying each contribution to the EDF kernel in terms 
of the parameters entering the underlying pseudo-potential and determine 
the number of independent parameters defining the latter.
\end{enumerate}
Owing to the large number of possible three-body terms, the above tasks 
cannot be accomplished safely on the basis of pen and paper. Consequently, a 
formal algebra code has been developed to carry them out~\cite{sadoudithese}.
The code also derives contributions to normal and anomalous one-body 
potentials, bulk properties of infinite nuclear matter and associated Landau 
parameters.

The paper is organized as follows. Section~\ref{sect:ingredients:pp} 
introduces the building blocks needed to construct the pseudo-potential in 
a pedestrian, though necessary for the following discussion, fashion. 
Section~\ref{sect:Skyrme_EDF} outlines the use of pseudo potentials with
gradients within the context of the nuclear EDF method. The central 
three-body Skyrme-like pseudo potential is then build in 
Sec.~\ref{sect:Skyrme_int}. The construction of the standard two-body 
Skyrme pseudo potential is provided as a reference throughout.  
Section~\ref{sec:general:EDF} details bilinear and trilinear contributions 
to the EDF kernel derived from two- and three-body  central pseudo 
potentials. Expressions are given there in the so-called isoscalar-isovector 
representation. Section~\ref{sect:concl} concludes the discussion and 
gives perspectives for future work, some already under way. Appendices 
provide further details on
(i) the derivation of infinite nuclear matter properties at zero 
and non-zero spin and isospin asymmetry as well as associated Landau 
parameters, (ii) the formulation of the EDF kernel in the so-called 
neutron-proton representation, (iii) the explicit expressions of 
normal and anomalous one-body fields entering the HFB Hamilton matrix, 
(iv) the algebraic steps needed to derive the EDF kernel from the pseudo 
potential and (v) a verification of the local gauge invariance of the 
functional.


\section{Basic ingredients}
\label{sect:ingredients:pp}


This section introduces the necessary ingredients to  set-up the 
three-body pseudo-potential and to compute the EDF kernel that derives 
from it.

\subsection{Introductory remarks}
\label{introductoryremarks}

In addition to invariances of the pseudo-potential under time-reversal, 
parity, rotational, translational, and Galilean transformations, we also 
assume it to be isospin invariant. 
Translational and Galilean transformations are special cases of local 
gauge transformations~\cite{raimondi11a}. The invariance of the EDF under 
the latter has been invoked as a possible guiding principle for the 
construction of pseudo-potentials and general EDF 
kernels~\cite{dobaczewski95a,carlsson08a,raimondi11b,raimondi11a}. We 
check the local gauge invariance for the pseudo-potential-based EDF 
constructed here in Appendix~\ref{sec:Skyrme_int:Gauge}. 

The EDF is derived assuming pure proton and neutron one-body density matrices,
which excludes at this stage the possibility of having \mbox{$T=0$} or 
\mbox{$T=1$} proton-neutron pairing~\cite{perlinska04a,rohozinski10a}. Such
correlations, however, have never been addressed so far in a systematic
and complete fashion in nuclear EDF calculations anyway. When needed, 
extensions of the present work to this case are straightforward.

The formulation of the three-body contact potentials, however, is 
not a straightforward generalization of the formalism usually used to 
set-up the standard Skyrme two-body interaction. To illustrate these 
differences and to validate our procedure, we describe two- and three-body 
terms side-by-side.

The Coulomb energy is omitted from the present discussion as its evaluation  
is standard. In a strict pseudo-potential-based framework, its exchange 
and pairing contributions have to be calculated 
exactly~\cite{anguiano01a,rodriguez05a}. By contrast, the kinetic energy 
is kept at various places throughout the paper as it contributes to 
nuclear matter properties discussed in Appendix~\ref{sect:INM}.


\subsection{Coordinate basis}

The coordinate representation
$\{| \vec{r} \sigma q\rangle\} 
\equiv \{ | \vec{r} \rangle \, \otimes \, | \sigma\rangle \, \otimes \, |q\rangle\}$
labels nucleon states with the position vector $\vec{r} \in \mathbb{R}^3$, 
the spin projection $\sigma=\pm 1/2$ and the isospin component\footnote{The 
quantum number $q$ will sometimes be labelled by a letter, i.e.\ $n$ for 
neutrons and $p$ for protons, instead of $+1/2$ for neutrons and $-1/2$ 
for protons.} \mbox{$q = \pm 1/2$} such that
\begin{subequations}
\label{eq:Skyrme_int:Coord_representation}
\begin{align}
\hat{\vec{r}} \, | \vec{r} \sigma q\rangle 
& = \vec{r} \, | \vec{r} \sigma q\rangle \, , \\
\hat{\vec{s}}^{\,2}\, | \vec{r} \sigma q\rangle 
& = \frac{3\hbar^2}{4} \,  | \vec{r} \sigma q\rangle \, ,  \\
\hat{s}_z| \vec{r} \sigma q\rangle  
& = \hbar \sigma \, | \vec{r} \sigma q\rangle  \, , \\
\hat{\tau}^{2} \, | \vec{r} \sigma q \rangle 
&= \frac{3\hbar^2}{4} \,  | \vec{r} \sigma q\rangle \, ,  \\
\hat{\tau}_z | \vec{r} \sigma q\rangle  
&= \hbar q \, | \vec{r} \sigma q\rangle \, .
\end{align}
\end{subequations}
This constitutes a continuous orthonormal direct-product basis of the 
one-body Hilbert space $\mathcal{H}_1 = \mathcal{H}_{1,\vec{r}} \otimes 
\mathcal{H}_{1,\sigma} \otimes \mathcal{H}_{1,q}$. Associated 
orthogonality and completeness relations are given by
\begin{subequations}
\label{eq:Skyrme_int:Coord_representation2}
\begin{align}
\langle  \vec{r} \sigma q |  \vec{r}\,' \sigma' q' \rangle 
& =  \delta(\vec{r} -\vec{r}\,') \, \delta_{\sigma\sigma'} \, \delta_{qq'}
      \,  ,  \\
\label{completeness1body}
\int \! d^3r \sum_{\sigma} \sum_{q} |  \vec{r} \sigma q\rangle
\langle  \vec{r} \sigma q | 
& = \hat{\ensuremath{\mathbbm{1}}}_1  \, , 
\end{align}
\end{subequations}
where $\ensuremath{\mathbbm{1}}_1=\ensuremath{\mathbbm{1}}_{1,\vec{r}}
\otimes \ensuremath{\mathbbm{1}}_{1 \sigma} \otimes 
\ensuremath{\mathbbm{1}}_{1,q}$ is the unity operator on $\mathcal{H}_1$. 
Introducing a complete set of orthogonal single-particle wave functions
\begin{equation}
\langle \, \vec{r} \sigma q \, | \, i \rangle 
\equiv \varphi^{\,}_i (\vec{r} \sigma q) \, ,
\end{equation}
creation and annihilation operators of a nucleon at coordinates 
$\{\vec{r} \sigma q\}$ are given by
\begin{subequations}
\label{eq:Skyrme_int:crtoci}
\begin{align}
a_{\vec{r} \sigma q} 
& \equiv \sum_i \varphi^{\,}_i (\vec{r} \sigma q) \; a^{\,}_{i}  \, , \\
a^{\dagger}_{\vec{r} \sigma q} 
& \equiv \sum_i \varphi_i ^\ast(\vec{r} \sigma q) \; a^{\dagger}_{i} \, .
\end{align}
\end{subequations}
The pseudo-potentials constructed below act on two- ($\mathcal{H}_2$) and 
three-body ($\mathcal{H}_3$) Hilbert spaces. We thus introduce bases of 
$\mathcal{H}_2$ and $\mathcal{H}_3$ through tensor products of the 
one-body basis $\{| \vec{r} \sigma q\rangle\}\equiv \{|\xi \rangle\}$. 
This provides non-antisymmetrized basis states
\begin{subequations}
\label{productbases}
\begin{eqnarray}
| \xi_3 \xi_4 \rangle 
& \equiv & | 1: \vec{r}_3 \sigma_3 q_3 \, , 2: \vec{r}_4 \sigma_4 q_4 \rangle  \label{productbases1} \\
& \equiv & | \vec{r}_3 \sigma_3 q_3 \, , \vec{r}_4 \sigma_4 q_4 \rangle \, , \label{productbases2}  \\
|\xi_4 \xi_5 \xi_6 \rangle 
& \equiv & | 1: \vec{r}_4 \sigma_4 q_4 \,, 2: \vec{r}_5 \sigma_5 q_5 \, , 3: \vec{r}_6 \sigma_6 q_6 \rangle \label{productbases3}  \\
& \equiv & | \vec{r}_4 \sigma_4 q_4 \,, \vec{r}_5 \sigma_5 q_5 \, , \vec{r}_6 \sigma_6 q_6 \rangle  \label{productbases4}
\, ,
\end{eqnarray}
\end{subequations}
where the shorthand notation will be used whenever possible. In such 
non-antisymmetrized states, each individual nucleon occupies a well-defined 
single-particle state. This is made very explicit in 
Eqs.~(\ref{productbases1}) and~(\ref{productbases3}), but only implicit in 
Eqs.~(\ref{productbases2}) and~(\ref{productbases4}) for brevity. It is clear 
from the former equations that the particle index (i.e.\ being the 
first, second or third particle in a two- or three-body state) should not 
be confused with the indices labeling different states in the
single-particle basis. For example, in Eq.~(\ref{productbases1}) nucleon~1
occupies single-particle state $|\vec{r}_3 \sigma_3 q_3 \rangle$, whereas
nucleon~2 occupies the state $| \vec{r}_4 \sigma_4 q_4 \rangle$. Associated 
orthogonality
\begin{subequations}
\label{eq:Skyrme_int:Coord_representation:2_3_body}
\begin{eqnarray}
\langle \xi_1 \xi_2 | \xi_3 \xi_4  \rangle 
& = & \delta_{\xi_1\xi_3}\delta_{\xi_2\xi_4} \, , 
   \\
\langle \xi_1 \xi_2 \xi_3 | \xi_4 \xi_5 \xi_6 \rangle 
& = & \delta_{\xi_1\xi_4}\delta_{\xi_2\xi_5}\delta_{\xi_3\xi_6} \, , 
\end{eqnarray}
\end{subequations}
and completeness relations
\begin{subequations}
\begin{eqnarray}
\label{completeness2body}
\iint \! \rmd \xi_1 \, \rmd \xi_2 \;
| \xi_1 \xi_2 \rangle \langle \xi_1 \xi_2 | 
& = & \hat{\ensuremath{\mathbbm{1}}}_2 \,\,, 
   \\
\label{completeness3body}
\iiint \! \rmd \xi_1 \, \rmd \xi_2 \, \rmd \xi_3 \;
| \xi_1 \xi_2 \xi_3 \rangle \langle \xi_1 \xi_2 \xi_3 | 
& = & \hat{\ensuremath{\mathbbm{1}}}_3 \, , 
\end{eqnarray}
\end{subequations}
can be derived from Eq.~(\ref{eq:Skyrme_int:Coord_representation2}),
where
\begin{subequations}
\begin{eqnarray}
\delta_{\xi_1\xi_2} 
& \equiv & \delta(\vec{r}_1 -\vec{r}_2) \, \delta_{\sigma_1 \sigma_2} \, 
           \delta_{q_1 q_2}\, , \\
\int \! \rmd \xi  
&\equiv & \int \! \rmd^3r \sum_{\sigma = \pm 1/2} \sum_{q = \pm 1/2} \, .
\end{eqnarray}
\end{subequations}

%
%
\subsection{Delta and gradient operators}
\label{sec:gradients_op_def}

\subsubsection{Delta operators}
\label{delta}

The delta operator $\hat{\delta}^r_{ij}$ describes an interaction between 
nucleons $i$ and $j$ located at the same position. Its two-body and 
three-body matrix elements in coordinate representation are given by
\begin{subequations}
\begin{eqnarray}
\langle \xi_1 \xi_2 | \hat{\delta}^r_{12} | \xi_3 \xi_4 \rangle 
& = & \langle \xi_1 \xi_2 | \xi_3 \xi_4 \rangle  \, 
      \delta(\vec{r}^{ }_3 - \vec{r}^{ }_4) \, , \\
\langle \xi_1 \xi_2 \xi_3 | \hat{\delta}^r_{12}  | \xi_4 \xi_5 \xi_6 \rangle 
& = & \langle \xi_1 \xi_2 \xi_3 | \xi_4 \xi_5 \xi_6 \rangle \, 
      \delta(\vec{r}^{ }_4 - \vec{r}^{ }_5)  \, , \\
\langle \xi_1 \xi_2 \xi_3 |  \hat{\delta}^r_{13}  | \xi_4 \xi_5 \xi_6 \rangle 
& = & \langle \xi_1 \xi_2 \xi_3 | \xi_4 \xi_5 \xi_6 \rangle \, 
      \delta(\vec{r}^{ }_4 - \vec{r}^{ }_6)  \,\,,\\
\langle \xi_1 \xi_2 \xi_3 |  \hat{\delta}^r_{23}  | \xi_4 \xi_5 \xi_6 \rangle 
& = & \langle \xi_1 \xi_2 \xi_3 | \xi_4 \xi_5 \xi_6 \rangle \, 
      \delta(\vec{r}^{ }_5 - \vec{r}^{ }_6)  \, .
\end{eqnarray}
\end{subequations}

%
%
\subsubsection{Gradient operators}
\label{gradients}

The one-body gradient  operator is introduced through matrix elements
 connecting coordinate and configuration basis states
\begin{equation}
\vec{\nabla}_{\vec{r}} \; \varphi_i (\xi)  
\equiv  \langle \xi | \hat{\vec{\nabla}} | i \rangle  
= \int \! \rmd\xi' \, 
  \langle \xi | \hat{\vec{\nabla}} | \xi' \rangle \, \varphi_i (\xi') \, . \label{nabla1}
\end{equation}
From the definition of its hermitian conjugate 
$\langle i | \hat{\vec{\nabla}}^{\dagger} | \xi \rangle
= \big[ \langle \xi | \hat{\vec{\nabla}} | i \rangle \big]^*$
it follows that
\begin{equation}
\vec{\nabla}_{\vec{r}} \; \varphi_i^* (\xi)  
= \langle i | \hat{\vec{\nabla}}^{\dagger} | \xi \rangle 
= \int \! \rmd\xi' \, \varphi_i^* (\xi') \, 
  \langle \xi' | \hat{\vec{\nabla}}^{\dagger} | \xi \rangle \, .  \label{nabla2}
\end{equation}
Matrix elements of the gradient operator and of its  hermitian conjugate  
in coordinate basis can deduce directly from Eqs.~(\ref{nabla1})
and~(\ref{nabla2})
\begin{subequations}
\begin{align}
\langle \xi_1 | \hat{\vec{\nabla}} | \xi_2 \rangle 
& = \langle \xi_1 | \xi_2 \rangle \, \vec{\nabla}_{\vec{r}_2} \, ,\\
\langle \xi_1 | \hat{\vec{\nabla}}^{\dagger} | \xi_2 \rangle 
& = \cev{\nabla}_{\vec{r}_1} \, \langle \xi_1 | \xi_2 \rangle  \, ,
\end{align}
\end{subequations}
where the convention used state that $\vec{\nabla}_{\vec{r}}$ acts on functions depending on $\vec{r}$ located to
its right whereas $\cev{\nabla}_{\vec{r}}$ acts on functions depending 
on $\vec{r}$ located to its left. The momentum operator 
$\hat{\vec{p}} \equiv -\mathrm{i} \hbar \hat{\vec{\nabla}}$ being hermitian, 
it follows trivially that $\langle i | \hat{\vec{\nabla}} | j \rangle
= -\langle i | \hat{\vec{\nabla}}^{\dagger} | j \rangle$ is anti-hermitian, 
such that
\begin{subequations}
\begin{align}
\langle \xi_1 | \hat{\vec{\nabla}} | \xi_2 \rangle 
&= \langle \xi_1 | \xi_2 \rangle \vec{\nabla}_{\vec{r}_2} 
  =-\cev{\nabla}_{\vec{r}_1} \langle \xi_1 | \xi_2 \rangle
\\
\langle \xi_1 | \hat{\vec{\nabla}}^{\dagger} | \xi_2 \rangle 
&= \cev{\nabla}_{\vec{r}_1} \langle \xi_1 | \xi_2 \rangle  
  =-\langle \xi_1 | \xi_2 \rangle \vec{\nabla}_{\vec{r}_2} 
\, ,
\end{align}
\end{subequations}

\subsubsection{Relative momentum operators}
\label{relativemomenta}

The gradient structure of the pseudo potential involves relative momentum operators associated with particles $i$ and $j$
\begin{equation}
\hat{\vec{k}}^{\,}_{ij} 
\equiv \frac{1}{2 \hbar} \, ( \hat{\vec{p}}_i - \hat{\vec{p}}_j ) 
= -\frac{{\mathrm i}}{2} \,
  \big( \hat{\vec{\nabla}}_{i} - \hat{\vec{\nabla}}_{j} \big) \, ,
\end{equation}
where $\hat{\vec{\nabla}}_{i}$ acts on particle $i$. Using  that $\hat{\vec{k}}^{\,}_{ij}$ is hermitian
$\hat{\vec{k}}^{\,}_{12} = \hat{\vec{k}}^{\,\dagger}_{12}$, we first provide two- and three-body matrix elements connecting coordinate and configuration basis states
\begin{subequations}
\label{eq:Skyrme_int:gradients_opertors}
\begin{align}
\langle \xi_1 \xi_2 | \hat{\vec{k}}^{\,}_{12} | ij \rangle   
&=   \big[ \vec{k}^{\,}_{\vec{r}^{\,}_1 \vec{r}^{\,}_2} \; \varphi_i (\xi_1) \,  \varphi_j (\xi_2) \big] \, , 
     \\
\langle ij | \hat{\vec{k}}^{\,}_{12} | \xi_1 \xi_2 \rangle   
&=   \big[ \vec{k}^{\,\ast}_{\vec{r}^{\,}_1 \vec{r}^{\,}_2} \; \varphi^\ast_i (\xi_1) \,  \varphi^\ast_j (\xi_2) \big] \, , \\
\langle \xi_1 \xi_2 \xi_3 | \hat{\vec{k}}^{\,}_{12} | ijk \rangle   
&=   \big[ \vec{k}^{\,}_{\vec{r}^{\,}_1 \vec{r}^{\,}_2} \; \varphi_i (\xi_1) \,  \varphi_j (\xi_2) \varphi_k (\xi_3) \big] \, , \\
\langle ijk | \hat{\vec{k}}^{\,}_{12} | \xi_1 \xi_2 \xi_3 \rangle   
&=   \big[ \vec{k}^{\,\ast}_{\vec{r}^{\,}_1 \vec{r}^{\,}_2} \; \varphi^\ast_i (\xi_1) \,  \varphi^\ast_j (\xi_2) \varphi^\ast_k (\xi_3) \big] \, ,  \\
\langle \xi_1 \xi_2 \xi_3 | \hat{\vec{k}}^{\,}_{13} | ijk \rangle   
&=   \big[ \vec{k}^{\,}_{\vec{r}^{\,}_1 \vec{r}^{\,}_3} \; \varphi_i (\xi_1) \,  \varphi_j (\xi_2) \varphi_k (\xi_3) \big] \, , \\
\langle ijk | \hat{\vec{k}}^{\,}_{13} | \xi_1 \xi_2 \xi_3 \rangle   
&=   \big[ \vec{k}^{\,\ast}_{\vec{r}^{\,}_1 \vec{r}^{\,}_3} \; \varphi^\ast_i (\xi_1) \,  \varphi^\ast_j (\xi_2) \varphi^\ast_k (\xi_3) \big] \, ,  \\
\langle \xi_1 \xi_2 \xi_3 | \hat{\vec{k}}^{\,}_{23} | ijk \rangle   
&=   \big[ \vec{k}^{\,}_{\vec{r}^{\,}_2 \vec{r}^{\,}_3} \; \varphi_i (\xi_1) \,  \varphi_j (\xi_2) \varphi_k (\xi_3) \big] \, , \\
\langle ijk | \hat{\vec{k}}^{\,}_{23} | \xi_1 \xi_2 \xi_3 \rangle   
&=   \big[ \vec{k}^{\,\ast}_{\vec{r}^{\,}_2 \vec{r}^{\,}_3} \; \varphi^\ast_i (\xi_1) \,  \varphi^\ast_j (\xi_2) \varphi^\ast_k (\xi_3) \big] \, ,
\end{align}
\end{subequations}
where 
\begin{equation}
\label{eq:Skyrme_int:gradients}
\vec{k}^{\,}_{\vec{r}^{\,}_i\vec{r}^{\,}_j}  
\equiv -\frac{{\mathrm i}}{2} 
      \big( \vec{\nabla}_{\vec{r}^{\,}_i} 
          - \vec{\nabla}_{\vec{r}^{\,}_j} \big) \, ,
\end{equation}
while $\vec{k}^{\,*}_{\vec{r}^{\,}_i \vec{r}^{\,}_j}$ denotes its complex 
conjugate. The brackets in Eq.~(\ref{eq:Skyrme_int:gradients_opertors}) 
indicate that $\vec{k}_{\vec{r}^{\,}_i\vec{r}^{\,}_j}$ acts only on the 
wave functions located inside. Matrix elements in the coordinate basis 
can be deduced to take the form
\begin{subequations}
\label{eq:Skyrme_int:gradients_opertors:2}
\begin{align}
\langle \xi_1 \xi_2 | \hat{\vec{k}}^{\,}_{12} | \xi_3 \xi_4 \rangle   
&=   \langle \xi_1 \xi_2 | \xi_3 \xi_4 \rangle \; \vec{k}^{\,}_{\vec{r}_3\vec{r}_4}  \label{eq:k12R}\\
 &=   \cev{k}^{\,\ast}_{\vec{r}_1\vec{r}_2}  \; \langle \xi_1 \xi_2 | \xi_3 \xi_4 \rangle  \, , \label{eq:k12L} \\
\langle \xi_1 \xi_2 \xi_3 | \hat{\vec{k}}^{\,}_{12} | \xi_4 \xi_5 \xi_6 \rangle   &=   \langle \xi_1 \xi_2 \xi_3 | \xi_4 \xi_5 \xi_6 \rangle \; \vec{k}^{\,}_{\vec{r}_4\vec{r}_5}  \\
 &=   \cev{k}^{\,\ast}_{\vec{r}_1\vec{r}_2}  \; \langle \xi_1 \xi_2 \xi_3 | \xi_4 \xi_5 \xi_6 \rangle  \, , \\
\langle \xi_1 \xi_2 \xi_3 | \hat{\vec{k}}^{\,}_{13} | \xi_4 \xi_5 \xi_6 \rangle   &=   \langle \xi_1 \xi_2 \xi_3 | \xi_4 \xi_5 \xi_6 \rangle \; \vec{k}^{\,}_{\vec{r}_4\vec{r}_6}  \\
 &=   \cev{k}^{\,\ast}_{\vec{r}_1\vec{r}_3}  \; \langle \xi_1 \xi_2 \xi_3 | \xi_4 \xi_5 \xi_6 \rangle  \, , \\
\langle \xi_1 \xi_2 \xi_3 | \hat{\vec{k}}^{\,}_{23} | \xi_4 \xi_5 \xi_6 \rangle   &=   \langle \xi_1 \xi_2 \xi_3 | \xi_4 \xi_5 \xi_6 \rangle \; \vec{k}^{\,}_{\vec{r}_5\vec{r}_6}  \\
 &=   \cev{k}^{\,\ast}_{\vec{r}_2\vec{r}_3}  \; \langle \xi_1 \xi_2 \xi_3 | \xi_4 \xi_5 \xi_6 \rangle  \, ,
\end{align}
\end{subequations}
where $\vec{k}^{\,}_{\vec{r}^{\,}_i\vec{r}^{\,}_j}$ acts on functions located
to its right while $\cev{k}^{\,*}_{\vec{r}^{\,}_i\vec{r}^{\,}_j}$ acts on 
functions located to its left. The quantum mechanical operator 
$\hat{\vec{k}}_{ij}$ has to be distinguished from its position-space matrix 
element $\vec{k}_{\vec{r}_3 \vec{r}_4}$ that is a differential operator on 
$\mathbb{R}^3 \otimes \mathbb{R}^3$. 

Thorough definitions of the matrix elements of elementary operators have 
been given above. Matrix elements of a {\it product} of such elementary 
operators can be computed in a pedestrian way by inserting as many 
completeness relations as necessary to invoke matrix elements of the 
elementary operators.\footnote{Taking short-cuts by "applying" operators 
sequentially on bras or kets rather than resorting to \textit{matrix elements}
 of elementary operators might sometimes lead to ambiguous computational 
steps.} Let us consider as an example two types of matrix elements that 
will have to be considered for the two-body Skyrme pseudo-potential. By 
virtue of Eqs.~(\ref{eq:k12L}) and~(\ref{eq:k12R}), and by inserting 
enough completeness relations on ${\cal H}_2$, one obtains
\begin{subequations}
\begin{eqnarray}
\lefteqn{
\langle \xi_1 \xi_2 | \hat{\delta}^r_{12} \hat{\vec{k}}^{\,2}_{12} | \xi_3 \xi_4 \rangle
} \nonumber \\ 
& = & \langle \xi_1 \xi_2 | \xi_3 \xi_4 \rangle \, 
      \delta(\vec{r}_3 - \vec{r}_4) \; \vec{k}^{\,2}_{\vec{r}_3\vec{r}_4}  
     \nonumber  \\
& = & \delta(\vec{r}_1 - \vec{r}_2) \; \cev{k}^{\,\ast 2}_{\vec{r}_1\vec{r}_2}
       \, \langle \xi_1 \xi_2 | \xi_3 \xi_4 \rangle  \, , 
     \\
\lefteqn{
\langle \xi_1 \xi_2 | \hat{\vec{k}}^{\,}_{12} \cdot 
\hat{\delta}^r_{12} \, \hat{\vec{k}}^{\,}_{12} | \xi_3 \xi_4 \rangle
} \nonumber \\
& = & \cev{k}^{\,\ast}_{\vec{r}_1\vec{r}_2} \, 
      \langle \xi_1 \xi_2 | \xi_3 \xi_4 \rangle \, 
      \delta(\vec{r}_3 - \vec{r}_4) \; \vec{k}^{\,}_{\vec{r}_3\vec{r}_4} 
\nonumber\\
& = & \langle \xi_1 \xi_2 | \xi_3 \xi_4 \rangle \, \vec{k}^{\,}_{\vec{r}_3\vec{r}_4}  \delta(\vec{r}_3 - \vec{r}_4) \; \vec{k}^{\,}_{\vec{r}_3\vec{r}_4}  \, \nonumber\\
& = &  \cev{k}^{\,\ast}_{\vec{r}_1\vec{r}_2}  \cdot \delta(\vec{r}_1 - \vec{r}_2) \; 
       \cev{k}^{\,\ast}_{\vec{r}_1\vec{r}_2}  \, 
       \langle \xi_1 \xi_2 | \xi_3 \xi_4 \rangle  \, . 
\end{eqnarray}
\end{subequations}
Any of the alternative formulae can be used when evaluating matrix elements 
of the interaction given that the resulting expressions can be related by 
one or several integrations by parts. Proceeding as above, one can easily 
show that gradient and delta operators do not commute
\begin{equation}
\hat{\vec{k}}^{\,}_{ij} \hat{\delta}^r_{ij} 
\neq \hat{\delta}^r_{ij} \hat{\vec{k}}^{\,}_{ij} 
= (\hat{\vec{k}}^{\,}_{ij} \hat{\delta}^r_{ij})^\dagger. 
\end{equation}

\subsection{Position, spin and isospin exchange operators}
\label{sec:Skyrme:ingredients:exchange}

Two-body coordinate-exchange operators will be used to formulate the 
pseudo-potential. Additionally, such operators are elementary building 
blocks of the antisymmetrizers that will enter the calculation of the EDF 
kernel, see Eq.~(\ref{eq:Skyrme_int:Energy2}). Applying the exchange 
operator \mbox{$\hat{P}_{ij} = \hat{P}_{ji}$}, that act on the coordinates 
of particles $i$ and $j$, with \mbox{$i \neq j$}, two- and three-body basis 
states are transformed according to
\begin{subequations}
\label{eq:Skyrme_int:exchanges}
\begin{align}
\hat{P}_{12} \, | \xi_3 \xi_4 \rangle 
& \equiv | \xi_4 \xi_3 \rangle \, , \\
\hat{P}_{12} \, | \xi_4 \xi_5 \xi_6 \rangle 
&\equiv | \xi_5 \xi_4 \xi_6  \rangle  \, , \\
\hat{P}_{13} \, | \xi_4 \xi_5 \xi_6 \rangle 
& \equiv | \xi_6 \xi_5 \xi_4 \rangle 
\, , \\
\hat{P}_{23} \, | \xi_4 \xi_5 \xi_6 \rangle 
& \equiv | \xi_4 \xi_6 \xi_5 \rangle 
\, ,
\end{align}
\end{subequations}
and similarly in any other basis representation. Of course, applying the 
same exchange operator twice gives back the original state. Furthermore, 
different exchange operators acting on the same space do in general not 
commute with one another, i.e.\
\begin{subequations}
\label{eq:Skyrme_int:double_exchange}
\begin{align}
\hat{P}_{12}\hat{P}_{13} 
&= \hat{P}_{13}\hat{P}_{23} = \hat{P}_{23}\hat{P}_{12}  \, , \\
\hat{P}_{12}\hat{P}_{23} 
&= \hat{P}_{23}\hat{P}_{13} = \hat{P}_{13}\hat{P}_{12}  \, .
\end{align}
\end{subequations}
In coordinate representation, exchange operators factorize into position-, 
spin- and isospin-exchange operators
$\hat{P}_{ij} \equiv \hat{P}^{r}_{ij} \, \hat{P}^{\sigma}_{ij} \, 
\hat{P}^{q}_{ij}$ that only exchange the corresponding coordinates, e.g.
\begin{subequations}
\label{eq:Skyrme_int:rsq_exchanges}
\begin{align}
\hat{P}^r_{12} \, | \xi_4 \xi_5 \xi_6 \rangle &\equiv | \vec{r}^{\,}_5 \sigma_4 q_4 ,  \vec{r}^{\,}_4 \sigma_5 q_5 ,  \vec{r}^{\,}_6 \sigma_6 q_6 \rangle  \,\,, \\
\hat{P}^\sigma_{12} \, | \xi_4 \xi_5 \xi_6 \rangle &\equiv | \vec{r}^{\,}_4 \sigma_5 q_4 ,  \vec{r}^{\,}_5 \sigma_4 q_5 ,  \vec{r}^{\,}_6 \sigma_6 q_6 \rangle  \,\,, \\
\hat{P}^q_{12} \, | \xi_4 \xi_5 \xi_6 \rangle &\equiv | \vec{r}^{\,}_4 \sigma_4 q_5 ,  \vec{r}^{\,}_5 \sigma_5 q_4 ,  \vec{r}^{\,}_6 \sigma_6 q_6 \rangle   \,\,.
\end{align}
\end{subequations}
Coordinate-exchange operators do not commute with relative momentum 
operators. One finds that 
\begin{subequations}
\begin{align}
\hat{\vec{k}}^{\,}_{ij} \; \hat{P}^r_{ij} 
& = \hat{P}^r_{ij} \; \hat{\vec{k}}^{\,}_{ji} 
  = - \hat{P}^r_{ij} \; \hat{\vec{k}}^{\,}_{ij} \, , 
\label{eq:Skyrme_int:anticommutator3} \\
\hat{\vec{k}}^{\,}_{ij} \; \hat{P}^r_{kj} 
& = \hat{P}^r_{kj} \; \hat{\vec{k}}^{\,}_{ik} \, ,\\
\hat{\vec{k}}^{\,}_{ij} \; \hat{P}^r_{ik} 
& = \hat{P}^r_{ik} \; \hat{\vec{k}}^{\,}_{kj} \, ,
\end{align}
\end{subequations}
i.e.\ in general, the commutation with a position-exchange operator changes 
the particle indices involved in the gradient operator. In the particular 
case where particle indices are the same in both operators they
anti-commute, see Eq.~(\ref{eq:Skyrme_int:anticommutator3}). These features 
can be established in a pedestrian way, e.g.
\begin{subequations}
\begin{eqnarray}
\langle \vec{r}_1 \vec{r}_2 \vec{r}_3 | \hat{\vec{k}}^{\,}_{12} \hat{P}^r_{12} | \vec{r}_4 \vec{r}_5 \vec{r}_6 \rangle 
&=& \, 
\langle \vec{r}_1 \vec{r}_2 \vec{r}_3 | \vec{r}_5 \vec{r}_4 \vec{r}_6 \rangle \vec{k}^{\,}_{\vec{r}_5 \vec{r}_4} 
 \\
&=& - 
\langle \vec{r}_1 \vec{r}_2 \vec{r}_3 | \hat{P}^r_{12} | \vec{r}_4 \vec{r}_5 \vec{r}_6 \rangle \vec{k}^{\,}_{\vec{r}_4 \vec{r}_5} 
\nonumber \\
&= & - \langle \vec{r}_1 \vec{r}_2 \vec{r}_3 | \hat{P}^r_{12} \hat{\vec{k}}^{\,}_{12} | \vec{r}_4 \vec{r}_5 \vec{r}_6 \rangle
\,\,\,,  \nonumber \\
\langle \vec{r}_1 \vec{r}_2 \vec{r}_3 | \hat{\vec{k}}^{\,}_{12} \hat{P}^r_{13} | \vec{r}_4 \vec{r}_5 \vec{r}_6 \rangle 
&=& \langle \vec{r}_1 \vec{r}_2 \vec{r}_3 | \vec{r}_6 \vec{r}_5 \vec{r}_4 \rangle \vec{k}^{\,}_{\vec{r}_6 \vec{r}_5} 
 \\
&=& \langle \vec{r}_1 \vec{r}_2 \vec{r}_3 |  \hat{P}^r_{13} \hat{\vec{k}}^{\,}_{32}  | \vec{r}_4 \vec{r}_5 \vec{r}_6 \rangle
 \nonumber \, .
\end{eqnarray}
\end{subequations}
This indicates that it may not be equivalent to have position-exchange 
operators located to the right or to the left of gradient operators in 
three-body potentials. Also, while it is always possible to replace 
$\hat{P}^r_{ij}$ directly by $\pm 1$ in the matrix elements of the two-body 
potential by virtue of Eq.~(\ref{eq:Skyrme_int:anticommutator3}), where 
the sign ultimately depends on the parity associated with the combination 
of gradient operators at play, this is in most cases not possible in matrix
elements of three-body operators. 

To evaluate matrix elements of the pseudo potential, it turns out to be 
useful to write spin-exchange operators in terms of spin Pauli matrices 
\cite{Nilsson}
\begin{equation}
\hat{P}^\sigma_{ij} 
= \frac{1}{2} (1 + \hat{\vec{\sigma}}_i \cdot \hat{\vec{\sigma}}_j) 
\, .
\end{equation}
Recalling that 
\begin{equation}
\hat{\sigma}_{i,\mu} \hat{\sigma}_{i,\nu} 
= \delta_{\mu\nu} \, 1 
  + {\mathrm i} \sum_\kappa \epsilon_{\mu\nu\kappa} \hat{\sigma}_{i,\kappa} 
\, ,
\end{equation}
when both Pauli matrices act on the same particle $i$, with 
\mbox{$\mu, \nu, \kappa \in \{ x, y, z \}$} and 
$\epsilon_{\mu\nu\kappa}$ denoting the Levi-Civita symbol, 
the product of two spin-exchange operators can be expressed as
\begin{subequations}
\label{eq:Skyrme_int:Double_ex_spin}
\begin{align}
\hat{P}^\sigma_{12} \hat{P}^\sigma_{13} 
= \frac{1}{4} \Big( 1 + \hat{\vec{\sigma}}_1 \cdot \hat{\vec{\sigma}}_2
&
                  + \hat{\vec{\sigma}}_1 \cdot \hat{\vec{\sigma}}_3
                  + \hat{\vec{\sigma}}_2 \cdot \hat{\vec{\sigma}}_3 
\\
& + {\mathrm i} \sum_{\mu\nu\kappa} \epsilon_{\mu\nu\kappa} \,
    \hat{\sigma}_{1,\kappa} \, \hat{\sigma}_{2,\mu} \, \hat{\sigma}_{3,\nu}
   \Big) \,,
\nonumber \\
\hat{P}^\sigma_{12} \hat{P}^\sigma_{23} 
= \frac{1}{4} \Big( 1
  + \hat{\vec{\sigma}}_1 \cdot \hat{\vec{\sigma}}_2
&
  + \hat{\vec{\sigma}}_2 \cdot \hat{\vec{\sigma}}_3 
  + \hat{\vec{\sigma}}_1 \cdot \hat{\vec{\sigma}}_3
\\
& + {\mathrm i} \sum_{\mu\nu\kappa} 
    \epsilon_{\mu\nu\kappa} \, \hat{\sigma}_{1,\mu} \,
    \hat{\sigma}_{2,\kappa} \, \hat{\sigma}_{3,\nu}
    \Big) \,.
\nonumber 
\end{align} 
\end{subequations}

\subsection{Antisymmetrization operators}
\label{sec:Skyrme:antisymmetrizers}

Let us introduce $\hat{{\cal A}}_{12}$ and $\hat{{\cal A}}_{123}$ as the two- 
and three-body antisymmetrizers, respectively, under the form
\begin{subequations}
\label{eq:Skyrme_int:antisymmetrizers}
\begin{align}
\label{eq:Skyrme_int:antisymmetrizers:A_12} 
\hat{{\cal A}}_{12}             
& \equiv ( 1 - \hat{P}_{12} )  \, ,
          \\
\label{eq:Skyrme_int:antisymmetrizers:A_123} 
\hat{{\cal A}}_{123}           
& \equiv  \hat{{\cal A}}_{12} ( 1 - \hat{P}_{13} - \hat{P}_{23} )
      \\                             
& =  ( 1 - \hat{P}_{12} - \hat{P}_{13} - \hat{P}_{23}
         + \hat{P}_{12}\hat{P}_{13} + \hat{P}_{12}\hat{P}_{23} ) \, .  \nonumber 
\end{align}
We introduce $\hat{{\cal A}}^{12}_{123}$ as another useful combination of 
exchange operators
\begin{align}
\hat{{\cal A}}^{12}_{123}   
&\equiv  ( 1 - \hat{P}_{13} - \hat{P}_{23} ) \, .
\label{eq:Skyrme_int:antisymmetrizers:A^12_123} 
\end{align}
\end{subequations}
Basic properties of two-body exchange operators lead to
\begin{subequations}
\label{eq:Skyrme_int:exchange_A}
\begin{eqnarray}
\hat{P}_{12} \, \hat{{\cal A}}_{12} | i j \rangle &=& - \hat{{\cal A}}_{12} | i j \rangle 
\, , \\
\hat{P}_{12} \, \hat{{\cal A}}_{123} | i j k \rangle &=& - \hat{{\cal A}}_{123} | i j k \rangle 
\, , \\
\hat{P}_{13} \, \hat{{\cal A}}_{123} | i j k \rangle &=& - \hat{{\cal A}}_{123} | i j k \rangle 
\, , \\
\hat{P}_{23} \, \hat{{\cal A}}_{123} | i j k \rangle &=& - \hat{{\cal A}}_{123} | i j k \rangle 
\, , 
\end{eqnarray}
\end{subequations}
from which trivially follows that
\begin{subequations}
\label{eq:Skyrme_int:rsq_exchanges_A}
\begin{align}
\hat{P}^r_{ij} \, {\cal A}_{123} | \xi_4 \xi_5 \xi_6 \rangle 
&= - \hat{P}^\sigma_{ij} \, \hat{P}^q_{ij} \, {\cal A}_{123} | \xi_4 \xi_5 \xi_6 \rangle \,, \\ 
\hat{P}^\sigma_{ij} \, {\cal A}_{123} | \xi_4 \xi_5 \xi_6 \rangle 
&= - \hat{P}^r_{ij} \, \hat{P}^q_{ij} \, {\cal A}_{123}           | \xi_4 \xi_5 \xi_6 \rangle \, , \\
\hat{P}^q_{ij} \, {\cal A}_{123} | \xi_4 \xi_5 \xi_6 \rangle 
&= - \hat{P}^r_{ij} \, \hat{P}^\sigma_{ij} \, {\cal A}_{123} | \xi_4 \xi_5 \xi_6 \rangle \, .
\end{align}
\end{subequations}

\section{The energy density functional}
\label{sect:Skyrme_EDF}

Before coming to the construction of the pseudo potential itself, let us 
explain its use within the context of EDF calculations.

\subsection{Reference States}

The EDF method originates from the picture of a nucleus as an ensemble 
of quasi particles moving independently in their self-created average 
field. It relies on the use of product states of Bogoliubov quasi-particles 
\begin{equation}
\label{eq:Intro_met:product_state}
| \Phi \rangle 
= \mathcal{N}_\Phi \prod_\mu \beta_\mu | 0 \rangle \, ,
\end{equation}
where $\{ \beta^\dagger_\mu \}$ and $\{\beta_\mu\} $ denote 
quasi-particle creation and annihilation operators relating 
to an arbitrary one-body basis $\{a^{\dagger}_\alpha, a_\alpha \}$
through a unitary canonical transformation of Bogoliubov 
type~\cite{blaizotripka,ring80a}
\begin{subequations}
\begin{align}
\beta_\mu 
& = \sum_i \big( U^{\dagger}_{\mu i} a_i + V^{\dagger}_{\mu i} a^\dagger_i
           \big) \, , \\
\beta^{\dagger}_\mu 
& = \sum_i \big(  V_{i \mu } a_i +  U_{i \mu} a^\dagger_i \big) \, .
\end{align}
\end{subequations}
The factor $\mathcal{N}_\Phi$ in Eq.~(\ref{eq:Intro_met:product_state}) ensures the normalization \mbox{$\langle \Phi | \Phi \rangle = 1$} of the 
quasi-particle vacuum. 
%
%
\subsection{Pseudo-potential-based EDF kernel}

Within the pseudo-potential-based formulation of the EDF method, 
the energy kernel is derived from a pseudo Hamiltonian that reads, in
an arbitrary basis, as
\begin{subequations}
\label{eq:Intro_met:hamiltonian}
\begin{align}
\hat{H}_{\text{pseudo}} 
= & \sum_{ij} 
    a^\dagger_i \, t^{(1)}_{ij} \, a_j^{\,}  \\
  & + \frac{1}{2!} \sum_{ijkl} 
      a^\dagger_i \, a^\dagger_j \, 
      v^{(2)}_{ijkl} \, 
      a_l \, a_k \\
  & + \frac{1}{3!} \sum_{ijklmn}  
      a^\dagger_i \,  a^\dagger_j a^\dagger_k \, 
      v^{(3)}_{ijklmn} \, 
      a_n \,  a_m \,  a_l 
     \label{eq:Intro_met:hamiltonian:3body}
 \\
 & + \, \cdots \, , \nonumber
\end{align}
\end{subequations}
where 
\begin{subequations}
\begin{eqnarray}
t^{(1)}_{ij} & \equiv & \langle i | \hat{t} | j \rangle \, , \\
v^{(2)}_{ijkl} & \equiv &  \langle i j | \hat{v}_{12} | k l \rangle \, , \\
v^{(3)}_{ijklmn} & \equiv &  \langle i j k | \hat{v}_{123} | l m n \rangle \, , 
\end{eqnarray}
\end{subequations}
denote matrix elements of the effective one-body kinetic energy operator 
and non-antisymmetrized matrix elements of two-body, three-body 
\ldots (density-independent) pseudo-potentials. In the present work, we do 
limit ourselves to two- and three-body pseudo potentials, but the further 
extension of the formalism to four-body and higher operators is 
straigthforward, though cumbersome.

The corresponding SR, i.e.\ diagonal, EDF kernel is computed as
\begin{subequations}
\begin{eqnarray}
\label{eq:Intro_met:HFB_kernel}
E 
& \equiv & \langle \Phi | \hat{H}_{\text{pseudo}} | \Phi \rangle 
      \\
& = & E[\rho, \kappa, 
            \kappa^{\ast}] \, , 
\end{eqnarray}
\end{subequations}
and takes the form of a functional of one-body density matrices
\begin{subequations}
\label{eq:Skyrme_int:dens_matrix}
\begin{align}
\label{eq:Skyrme_int:rho}
\rho_{i j} 
& \equiv \langle \Phi | a^{\dagger}_{j} a_{i} | \Phi \rangle \, , 
         \\
\label{eq:Skyrme_int:kappa}
\kappa_{i j} 
& \equiv \langle \Phi | a_{j} a_{i} | \Phi \rangle \,  , 
         \\
\label{eq:Skyrme_int:kappa*}
\kappa^{\ast}_{i j} 
& \equiv \langle \Phi | a^{\dagger}_{i} a^{\dagger}_{j} | \Phi \rangle \, ,
\end{align}
\end{subequations}
by virtue of Wick's theorem~\cite{wick50a}. The normal density 
matrix is hermitian 
\mbox{$\rho_{i j} = \rho^{\ast}_{j i}$}, 
whereas the anomalous density matrix is skew symmetric 
\mbox{$\kappa _{i j} = -\kappa _{j i}$}. 

Multi-reference calculations invoke an extension of the SR EDF kernel to
 define the {\it off diagonal} kernel involving two different Bogoliubov 
states. As opposed to hybrid and general functionals~\cite{duguet10a}, 
such an extension is formally straightforward and unambiguous for a 
pseudo-potential-based parametrization. By virtue of the generalized 
(i.e.\ off-diagonal) Wick theorem~\cite{balian69a,ring80a,blaizotripka}, 
the off diagonal energy is obtained from $E[\rho, \kappa, \kappa^{\ast}]$ 
by replacing the density matrices of Eq.~(\ref{eq:Skyrme_int:dens_matrix}) 
with \textit{transition} (i.e.\ off-diagonal) density 
matrices~\cite{duguet13a}, and multiplying the entire EDF kernel with a
norm kernel.

\subsection{EDF kernel in a configuration basis}

When evaluating Eq.~(\ref{eq:Intro_met:HFB_kernel}), the resulting terms 
can be grouped according to their content in normal and anomalous density 
matrices
\begin{align}
\label{eq:Skyrme_int:Energy}
E[\rho, \kappa, \kappa^{\ast}]
& \equiv  E^{\rho} 
   + E^{\rho\rho} 
   + E^{\kappa\kappa} 
   + E^{\rho\rho\rho} 
   + E^{\kappa\kappa\rho} \, .
\end{align}
There are several equivalent possibilities how these can be expressed. We
will choose a form where each given product of density matrices appears only
once, and where the antisymmetrization is done explicitly in the matrix
elements by virtue of the antisymmetrization operators 
$\hat{{\cal A}}_{12}$, $\hat{{\cal A}}_{123}$ and $\hat{{\cal A}}^{12}_{123}$
of Eq.~(\ref{eq:Skyrme_int:antisymmetrizers})
\begin{subequations}
\label{eq:Skyrme_int:Energy2}
\begin{align}
E^{\rho}
&= \sum_{ij} 
\langle i | \hat{t} | j \rangle \rho _{ji}  \, ,
\label{eq:Skyrme_int:Energy:kinetic} 
\\
E^{\rho\rho}
&=\frac{1}{2} \sum_{ijkl} 
\langle i j | \hat{v}_{12} \hat{{\cal A}}_{12} | k l \rangle \, \rho _{ki} \, \rho _{lj}  \, ,
\label{eq:Skyrme_int:Energy:normal:2body}
\\
E^{\kappa\kappa} 
&= \frac{1}{2} \sum_{ijkl} \langle i j | \hat{v}_{12} | k l \rangle \, \kappa^{*}_{ij} \, \kappa _{kl} 
\label{eq:Skyrme_int:Energy:anormal:2body}
\\
&= \frac{1}{4} \sum_{ijkl} 
\langle i j | \hat{v}_{12} \hat{{\cal A}}_{12} | k l \rangle \, \kappa ^{*}_{ij} \, \kappa _{kl}   \, ,
\label{eq:Skyrme_int:Energy:anormal:2body2}
\\
E^{\rho\rho\rho} 
&= \frac{1}{6} \sum_{ijklmn} 
\langle i j k | \hat{v}_{123} \hat{{\cal A}}_{123} | l m n \rangle \, \rho _{li} \, \rho _{mj} \, \rho _{nk} \, ,
\label{eq:Skyrme_int:Energy:normal:3body} 
\\
E^{\kappa\kappa\rho}
&=
\frac{1}{6} \sum_{ijklmn} 
\langle i j k | \hat{{\cal A}}^{12}_{123} \hat{v}_{123} \hat{{\cal A}}^{12}_{123} | l m n \rangle \, \kappa ^{*}_{ij} \, \kappa _{lm}  \, \rho _{nk} 
\label{eq:Skyrme_int:Energy:anormal:3body} 
\\
&=
\frac{1}{2} \sum_{ijklmn} 
\langle i j k | \hat{v}_{123} \hat{{\cal A}}^{12}_{123} | l m n \rangle \, \kappa ^{*}_{ij} \, \kappa _{lm}  \, \rho _{nk} 
\label{eq:Skyrme_int:Energy:anormal:3body1} 
\\
&=
\frac{1}{4} \sum_{ijklmn} 
\langle i j k | \hat{v}_{123} \hat{{\cal A}}_{123} | l m n \rangle \, \kappa ^{*}_{ij} \, \kappa _{lm}  \, \rho _{nk} \, .
\label{eq:Skyrme_int:Energy:anormal:3body2} 
\end{align}
\end{subequations}
Exploiting relations~(\ref{eq:Skyrme_int:double_exchange}) and the 
cyclic nature of the particle trace in expressions containing normal 
density matrices only, e.g.
\begin{eqnarray}
\label{eq:Skyrme_int:cyclic:trace} 
\sum_{ijkl} \langle i j |  \hat{P}_{12} \hat{v}_{12} | k l \rangle \, 
\rho _{ki} \, \rho _{lj} = \sum_{ijkl} \langle i j |  \hat{v}_{12} \hat{P}_{12} | k l \rangle \, 
    \rho _{ki} \, \rho _{lj}  \, , 
\end{eqnarray}
as well as the skew symmetry of the pairing tensor, antisymmetrizers and 
exchange operators can be placed either to the left or to the right of the 
pseudo-potential according to what is shown in 
Eqs.~(\ref{eq:Skyrme_int:Energy2}).


\subsection{Symmetry under particle exchange}

Because we are dealing  with identical particles, pseudo-potential 
operators must be symmetric under the exchange of any pair of 
nucleons, i.e.\
\begin{subequations}
\label{symmetric}
\begin{align}
\hat{v}_{\overline{12}} 
& = \hat{v}_{\overline{21}} \, , \\
\hat{v}_{\overline{123}} 
& = \hat{v}_{\overline{213}} 
  = \hat{v}_{\overline{132}} 
  = \hat{v}_{\overline{321}} 
  = \hat{v}_{\overline{231}} 
  = \hat{v}_{\overline{312}} \, ,
\end{align}
\end{subequations}
where the bar over a certain set of particle indices indicates from here on 
the symmetry of the operator under any permutation within that set. In 
matrix elements, exchanging particles corresponds to exchanging the complete 
set of associated single-particle quantum numbers in \textit{both} the bra and 
the ket. This leads to symmetry properties of the kind
\begin{subequations}
\label{eq:exchange:symmetry:1}
\begin{align}
\langle ij | \hat{v}_{\overline{12}} | kl \rangle 
& = \langle ji | \hat{v}_{\overline{12}} | lk \rangle \, , 
    \\
\langle ijk | \hat{v}_{\overline{123}} | lmn \rangle
& = \langle kij | \hat{v}_{\overline{123}} | nlm \rangle \, .
\end{align}
\end{subequations}
When constructing a two-body hermitian operator out of delta and relative 
momentum operators, along with exchange operators, the symmetry of the
potential under particle exchange is automatically fulfilled. 
When the three-body potential is constructed from the same two-body building 
blocks, its symmetry under permutation of the particle indices is not as 
automatically fulfilled. This property, however, can always be enforced 
by constructing the pseudo potential as the sum of six permutations over 
the particle indices of a non-symmetric potential $\hat{v}'_{123}$. More 
convenient for our purpose is to take advantage of the fact that any fully 
symmetric operator $\hat{v}_{\overline{123}}$ can be decomposed into 
the sum of three parts that are symmetric under the exchange of two particles
\begin{align}
\label{eq:Skyrme_int:3body:decomp} 
\hat{v}_{\overline{123}}  
&\equiv \hat{v}_{\overline{12}3}+\hat{v}_{\overline{13}2}
   +\hat{v}_{\overline{23}1} \, .
\end{align}
We will use this property to build explicitly $\hat{v}_{\overline{12}3}$, 
with the other two parts being obtained through the application of two-body 
exchange operators
\begin{equation}
\label{eq:Skyrme_int:3body:decomp2}
\hat{v}_{\overline{123}} 
= \hat{v}_{\overline{12}3} 
  + \hat{P}_{23} \, \hat{v}_{\overline{12}3} \, \hat{P}_{23} 
  + \hat{P}_{13} \, \hat{v}_{\overline{12}3} \, \hat{P}_{13} 
\, .
\end{equation}
Invoking the skew symmetry of $\kappa$, 
Eq.~(\ref{eq:Skyrme_int:3body:decomp2}), allows one to rewrite trilinear 
contributions to the EDF kernel in terms of $\hat{v}_{\overline{12}3}$ 
only, i.e.\
\begin{subequations}
\label{eq:Skyrme_int:Energy3}
\begin{align}
E^{\rho\rho\rho} 
&= \frac{1}{2} \sum_{ijklmn} 
\langle i j k | \hat{v}_{\overline{12}3}  \hat{{\cal A}}_{123} | l m n \rangle \, \rho _{li} \, \rho _{mj} \, \rho _{nk} \, ,
\label{eq:Skyrme_int:Energy3:normal:3bodyA} 
\\
E^{\kappa\kappa\rho}
&= \frac{1}{2} \sum_{ijklmn} 
\langle i j k | \hat{{\cal A}}^{12}_{123} \hat{v}_{\overline{12}3}  \hat{{\cal A}}^{12}_{123} | l m n \rangle \, \kappa ^{*}_{ij} \, \kappa _{lm}  \, \rho _{nk} 
 \, .
\label{eq:Skyrme_int:Energy3:anormal:3bodyB} 
\end{align}
\end{subequations}

\section{Building the pseudo potential}
\label{sect:Skyrme_int}

In this section, we describe the set-up of  two- and three-body Skyrme-type 
pseudo potentials containing up to two gradient operators.

\subsection{Generic structure}

We build hermitian two- and three-body operators out of two-body delta, 
relative momentum and exchange operators such that symmetries listed in 
Sec.~\ref{introductoryremarks} are fulfilled. To the best of our knowledge, 
all earlier attempts to construct Skyrme-type  three-body contact interactions
were limited to operator structures where $\hat{v}_{\overline{12}3}$ is set 
up by inserting an additional delta function into a subset of the standard 
two-body Skyrme interaction~\cite{blaizot75a,liu75a,onishi78a,arima86a,zheng90a,waroquier76a,waroquier79a,waroquier83a,waroquier83b,liu91a,liu91b}. 
As will be seen below, this does not generate the most general set of 
three-body terms.

We separate two- and three-body pseudo-potentials into a sum of terms, 
that all are functions of the elementary two-body operators, i.e.\
\begin{subequations}
\label{eq:Skyrme_int:Int_init}
\begin{align} 
\label{eq:Skyrme_int:2body:Int_init}
\hat{v}_{\overline{12}} 
& \equiv \sum_{i} \, 
         \hat{v}^i_{\overline{12}}
         \left[\hat{P}^{\{t_i,x_i\}}_{12}, 
          \hat{\vec{k}}^{\,(\dagger)}_{12}, 
          \hat{\delta}^r_{12}\right] \, ,
         \\
\label{eq:Skyrme_int:3body:Int_init}
\hat{v}_{\overline{12}3} 
& \equiv \sum_{i} \, 
         \hat{v}^i_{\overline{12}3}\left[ 
          \hat{P}^{\{u_i,y_i\}}_{123}, 
          \hat{\vec{k}}^{\,(\dagger)}_{12}, 
          \hat{\vec{k}}^{\,(\dagger)}_{23}, 
          \hat{\vec{k}}^{\,(\dagger)}_{13}, 
          \hat{\delta}^r_{13}\hat{\delta}^r_{23}\right] \, .
\end{align}
\end{subequations}
The index $i$ labels the possible coordinate-space structures, i.e.\
terms with a different content in gradient operators. The number of such
terms is limited by the number of interacting nucleons (i.e.\ two and three 
in the present case) and the number of gradient operators allowed (i.e.\ up 
to two in the present case). Each of these can be combined with a large 
number of distinct combinations of two-body exchange operators, represented 
by $\hat{P}^{\{t_i,x_i\}}_{12}$ and $\hat{P}^{\{u_i,y_i\}}_{123}$.


\subsection{Structure in coordinate space}

Each function $\hat{v}^i_{\overline{12}}$
contains a set of parameters denoted as $t_i$ and $x_{ij}$, and each 
function $\hat{v}^i_{\overline{12}3}$ contains a set of parameters denoted 
as $u_i$ and $y_{ij}$. Parameters $t_i$ and $u_i$ represent the overall 
coupling strength of a given coordinate space operator, whereas $x_{ij}$ 
and $y_{ij}$ weigh the possible combinations of spin- and isospin-exchange 
operators, labelled by $j$. 

We first specify the dependence of functions $\hat{v}^i_{\overline{12}}$ 
and $\hat{v}^i_{\overline{12}3}$ on gradient operators and delta functions 
by forming all possible hermitian scalars. Using for now the most simple 
spin and isospin dependence under the form of unit operators in spin and 
isospin space, one obtains
\begin{widetext}
\begin{subequations}
\label{eq:Skyrme_int:Grad_init}
\begin{align}
\hat{v}^0_{\overline{12}} (\hat{\ensuremath{\mathbbm{1}}}_{2,\sigma q}, \hat{\delta}^r_{12})
\equiv & \,
\hat{\ensuremath{\mathbbm{1}}}_{2,\sigma q} \, \hat{\delta}^r_{12}
\label{eq:Skyrme_int:2body:Grad_init:nograd}
\,,   \\
\hat{v}^1_{\overline{12}}(\hat{\ensuremath{\mathbbm{1}}}_{2,\sigma q},\hat{\vec{k}}^{\,(\dagger)}_{12}, \hat{\delta}^r_{12})
\equiv & \, 
 \hat{\ensuremath{\mathbbm{1}}}_{2,\sigma q} \, \frac{1}{2}
\left( \hat{\vec{k}}^{\,\dagger}_{12} \cdot \hat{\vec{k}}^{\,\dagger}_{12} \, \hat{\delta}^r_{12}
\, + \, 
\hat{\delta}^r_{12} \, \hat{\vec{k}}^{\,}_{12} \cdot \hat{\vec{k}}^{\,}_{12}  \right)
\label{eq:Skyrme_int:2body:Grad_init:k12k12}
\,,  \\
\hat{v}^2_{\overline{12}}(\hat{\ensuremath{\mathbbm{1}}}_{2,\sigma q},\hat{\vec{k}}^{\,(\dagger)}_{12}, \hat{\delta}^r_{12})
\equiv & \, \hat{\ensuremath{\mathbbm{1}}}_{2,\sigma q} \,
\hat{\vec{k}}^{\,\dagger}_{12} \, \hat{\delta}^r_{12} \, \cdot \hat{\vec{k}}^{\,}_{12}
\label{eq:Skyrme_int:2body:Grad_init:kp12k12}
\, ,  \\
\intertext{for two-body terms and}
\hat{v}^0_{\overline{12}3}(\hat{\ensuremath{\mathbbm{1}}}_{3,\sigma q}, \hat{\delta}^r_{13}\hat{\delta}^r_{23})
\equiv & \, 
\hat{\ensuremath{\mathbbm{1}}}_{3,\sigma q} \, \hat{\delta}^r_{13}\hat{\delta}^r_{23} \, 
\label{eq:Skyrme_int:3body:Grad_init:nograd}
 \,,  \\
\hat{v}^1_{\overline{12}3}(\hat{\ensuremath{\mathbbm{1}}}_{3,\sigma q},\hat{\vec{k}}^{\,(\dagger)}_{12}, \hat{\delta}^r_{13}\hat{\delta}^r_{23})
\equiv & \, 
\hat{\ensuremath{\mathbbm{1}}}_{3,\sigma q}  \,\frac{1}{2} 
\left( \hat{\vec{k}}^{\,\dagger}_{12} \cdot \hat{\vec{k}}^{\,\dagger}_{12}  \, \hat{\delta}^r_{13}\hat{\delta}^r_{23} \,
\, + \, 
\hat{\delta}^r_{13}\hat{\delta}^r_{23} \, \hat{\vec{k}}^{\,}_{12} \cdot \hat{\vec{k}}^{\,}_{12}  \right)
 \label{eq:Skyrme_int:3body:Grad_init:k12k12} 
 \,,  \\
\hat{v}^2_{\overline{12}3}(\hat{\ensuremath{\mathbbm{1}}}_{3,\sigma q},\hat{\vec{k}}^{\,(\dagger)}_{12}, \hat{\delta}^r_{13}\hat{\delta}^r_{23})
\equiv & \, \hat{\ensuremath{\mathbbm{1}}}_{3,\sigma q} \,
\hat{\vec{k}}^{\,\dagger}_{12} \, \hat{\delta}^r_{13}\hat{\delta}^r_{23} \, \cdot \hat{\vec{k}}^{\,}_{12}
 \label{eq:Skyrme_int:3body:Grad_init:kp12k12} 
 \,,  \\
\hat{v}^3_{\overline{12}3}(\hat{\ensuremath{\mathbbm{1}}}_{3,\sigma q},\hat{\vec{k}}^{\,(\dagger)}_{23},\hat{\vec{k}}^{\,(\dagger)}_{13}, \hat{\delta}^r_{13}\hat{\delta}^r_{23})
\equiv & \, 
\hat{\ensuremath{\mathbbm{1}}}_{3,\sigma q} \, \frac{1}{2} 
\left(  \hat{\vec{k}}^{\,\dagger}_{23} \cdot \hat{\vec{k}}^{\,\dagger}_{13}  \, \hat{\delta}^r_{13}\hat{\delta}^r_{23}
\, + \, 
\hat{\delta}^r_{13}\hat{\delta}^r_{23} \, \hat{\vec{k}}^{\,}_{13} \cdot \hat{\vec{k}}^{\,}_{23} \right)
 \label{eq:Skyrme_int:3body:Grad_init:k23k13} 
 \,,  \\
\hat{v}^4_{\overline{12}3}(\hat{\ensuremath{\mathbbm{1}}}_{3,\sigma q},\hat{\vec{k}}^{\,(\dagger)}_{23},\hat{\vec{k}}^{\,(\dagger)}_{13}, \hat{\delta}^r_{13}\hat{\delta}^r_{23})
\equiv & \, 
\hat{\ensuremath{\mathbbm{1}}}_{3,\sigma q} \, \frac{1}{2} 
\left( 
\hat{\vec{k}}^{\,\dagger}_{13} \, \hat{\delta}^r_{13}\hat{\delta}^r_{23} \, \cdot \hat{\vec{k}}^{\,}_{23} 
\, + \, 
\hat{\vec{k}}^{\,\dagger}_{23} \, \hat{\delta}^r_{13}\hat{\delta}^r_{23} \, \cdot \hat{\vec{k}}^{\,}_{13} \right)
 \,, \label{eq:Skyrme_int:3body:Grad_init:kp23k13}
\end{align}
\end{subequations}
\end{widetext}
for three-body terms. The list of arguments has been reduced to those 
each function actually depends on.


\subsection{Structure in spin and isospin spaces}

As the next step, we deduce the most general operators 
$\hat{P}^{\{t_i,x_i\}}_{12}$ and $\hat{P}^{\{u_i,y_i\}}_{123}$ that accompanies
each term in Eq.~(\ref{eq:Skyrme_int:Grad_init}). For terms involving two 
hermitian conjugate contributions, one has to employ 
$\hat{P}^{\{t_i,x_i\}}_{12}$ or $\hat{P}^{\{u_i,y_i\}}_{123}$ for one and 
$\hat{P}^{\{t_i,x_i\}\dagger}_{12}$ and $\hat{P}^{\{u_i,y_i\}\dagger}_{123}$ 
for the other, such that the overall operator remains indeed hermitian.

\textit{A priori}, the most general form is given by the sum of two- and 
three-body terms obtained by multiplying position-, spin- and 
isospin-exchange operators in all possible ways. In the end, 
$\hat{P}^{\{t_i,x_i\}}_{12}$ and $\hat{P}^{\{u_i,y_i\}}_{123}$ can be 
expressed solely in terms of spin- and isospin-exchange operators by
 virtue of Eq.~(\ref{eq:Skyrme_int:rsq_exchanges_A}). While 
\mbox{$\hat{P}^{\{t_i,x_i\}}_{12} = \hat{P}^{\{t_i,x_i\}\,\dagger}_{12}$} 
derives from the hermiticity of exchange operators defined in 
Eq.~(\ref{eq:Skyrme_int:rsq_exchanges}), the same does not hold in 
general for $\hat{P}^{\{u_i,y_i\}}_{123}$, because products of exchange 
operators of the same type (i.e.\ space, spin or isospin) associated 
with different pairs of particles do not commute.

All terms in Eq.~(\ref{eq:Skyrme_int:Grad_init}) but those entering 
$\hat{v}^4_{\overline{12}3}$ are individually symmetric under the exchange 
of particles 1 and 2 such that they have to be joined by a spin-isospin 
operator that itself is symmetric under such an exchange. These considerations 
lead to the following general spin-isospin operators acting on ${\cal H}_2$ 
and ${\cal H}_3$ that are symmetric under the exchange of particles~1 and~2 
\begin{widetext}
\begin{subequations}
\begin{align}
\hat{P}^{\{t_i,x_i\}}_{\overline{12}} \equiv&
t_i
\Big( \hat{\ensuremath{\mathbbm{1}}}_{2} 
+ x_{i1} \hat{P}^{\sigma}_{12} 
+ x_{i2} \hat{P}^{q}_{12} 
+ x_{i3} \hat{P}^{\sigma}_{12} \hat{P}^{q}_{12} 
\Big)
 \, ,
\label{eq:Skyrme_int:2body:truc}
\\
\hat{P}^{\{u_i,y_i\}}_{\overline{12}3} \equiv&
u_i
\Big[ \hat{\ensuremath{\mathbbm{1}}}_{3} 
+ y_{i1} \hat{P}^{\sigma}_{12} 
+ y_{i2} \left( \hat{P}^{\sigma}_{13} + \hat{P}^{\sigma}_{23} \right)
+ y_{i3} \left( \hat{P}^{\sigma}_{12} \hat{P}^{\sigma}_{13} + \hat{P}^{\sigma}_{12} \hat{P}^{\sigma}_{23} \right)
+ y_{i10} \hat{P}^{q}_{12} 
\nonumber \\ &
+ y_{i11} \hat{P}^{\sigma}_{12} \hat{P}^{q}_{12} 
+ y_{i12} \left( \hat{P}^{\sigma}_{13} \hat{P}^{q}_{12} + \hat{P}^{\sigma}_{23} \hat{P}^{q}_{12} \right) 
+ y_{i13} \left( \hat{P}^{\sigma}_{12} \hat{P}^{\sigma}_{13} \hat{P}^{q}_{12} + \hat{P}^{\sigma}_{12} \hat{P}^{\sigma}_{23} \hat{P}^{q}_{12} \right) 
+ y_{i20} \left( \hat{P}^{q}_{13} + \hat{P}^{q}_{23} \right) 
\nonumber \\ &
+ y_{i21} \left( \hat{P}^{\sigma}_{12} \hat{P}^{q}_{13} + \hat{P}^{\sigma}_{12} \hat{P}^{q}_{23} \right) 
+ y_{i22} \left( \hat{P}^{\sigma}_{13} \hat{P}^{q}_{13} + \hat{P}^{\sigma}_{23} \hat{P}^{q}_{23} \right) 
+ y_{i23} \left( \hat{P}^{\sigma}_{13} \hat{P}^{q}_{23} + \hat{P}^{\sigma}_{23} \hat{P}^{q}_{13} \right) 
\nonumber \\ &
+ y_{i24} \left( \hat{P}^{\sigma}_{12} \hat{P}^{\sigma}_{13} \hat{P}^{q}_{13} + \hat{P}^{\sigma}_{12} \hat{P}^{\sigma}_{23} \hat{P}^{q}_{23} \right) 
+ y_{i25} \left( \hat{P}^{\sigma}_{12} \hat{P}^{\sigma}_{23} \hat{P}^{q}_{13} + \hat{P}^{\sigma}_{12} \hat{P}^{\sigma}_{13} \hat{P}^{q}_{23} \right) 
+ y_{i30} \left( \hat{P}^{q}_{12} \hat{P}^{q}_{13} + \hat{P}^{q}_{12} \hat{P}^{q}_{23} \right) 
\nonumber \\ &
+ y_{i31} \left( \hat{P}^{\sigma}_{12} \hat{P}^{q}_{12} \hat{P}^{q}_{13} + \hat{P}^{\sigma}_{12} \hat{P}^{q}_{12} \hat{P}^{q}_{23} \right) 
+ y_{i32} \left( \hat{P}^{\sigma}_{13} \hat{P}^{q}_{12} \hat{P}^{q}_{13} + \hat{P}^{\sigma}_{23} \hat{P}^{q}_{12} \hat{P}^{q}_{23} \right) 
+ y_{i33} \left( \hat{P}^{\sigma}_{13} \hat{P}^{q}_{12} \hat{P}^{q}_{23} + \hat{P}^{\sigma}_{23} \hat{P}^{q}_{12} \hat{P}^{q}_{13} \right) 
\nonumber \\ &
+ y_{i34} \left( \hat{P}^{\sigma}_{12} \hat{P}^{\sigma}_{13} \hat{P}^{q}_{12} \hat{P}^{q}_{13} + \hat{P}^{\sigma}_{12} \hat{P}^{\sigma}_{23} \hat{P}^{q}_{12} \hat{P}^{q}_{23} \right) 
+ y_{i35} \left( \hat{P}^{\sigma}_{12} \hat{P}^{\sigma}_{13} \hat{P}^{q}_{12} \hat{P}^{q}_{23} + \hat{P}^{\sigma}_{12} \hat{P}^{\sigma}_{23} \hat{P}^{q}_{12} \hat{P}^{q}_{13} \right) 
\Big]
 \, .
\label{eq:Skyrme_int:3body:Px123}
\end{align}
\end{subequations}
As for terms entering $\hat{v}^4_{\overline{12}3}$, we have to introduce 
two other functions of spin- and isospin-exchange operators, the first of 
which only depends on particles~1 and~2, whereas the second one depends on 
all particles and is not symmetric under the exchange of particles~1 and~2
\begin{subequations}
\label{eq:Skyrme_int:Px3}
\begin{align}
\hat{P}^{\{u_i,y_i\}}_{123,a} 
&\equiv u_i
   \Big( \hat{\ensuremath{\mathbbm{1}}}_{3} 
        + y_{i1} \hat{P}^{\sigma}_{12} 
        + y_{i2} \hat{P}^{q}_{12} 
        + y_{i3} \hat{P}^{\sigma}_{12} \hat{P}^{q}_{12} 
   \Big)
 \, ,
\label{eq:Skyrme_int:3body:Px12:kp23k13}
\\
\hat{P}^{\{u_i,y_i\}}_{123,b} 
&\equiv u_i 
   \Big(  y_{i2} \hat{P}^{\sigma}_{13} 
        + y_{i3} \hat{P}^{\sigma}_{12} \hat{P}^{\sigma}_{13}
        + y_{i12} \hat{P}^{\sigma}_{13} \hat{P}^{q}_{12} 
        + y_{i13} \hat{P}^{\sigma}_{12} \hat{P}^{\sigma}_{13} \hat{P}^{q}_{12} 
        + y_{i20} \hat{P}^{q}_{13} 
        + y_{i21} \hat{P}^{\sigma}_{12} \hat{P}^{q}_{13} 
        + y_{i22} \hat{P}^{\sigma}_{13} \hat{P}^{q}_{13} 
\nonumber \\ &
        + y_{i23} \hat{P}^{\sigma}_{13} \hat{P}^{q}_{23} 
        + y_{i24} \hat{P}^{\sigma}_{12} \hat{P}^{\sigma}_{13} \hat{P}^{q}_{13} 
        + y_{i25} \hat{P}^{\sigma}_{12} \hat{P}^{\sigma}_{23} \hat{P}^{q}_{13} 
        + y_{i30} \hat{P}^{q}_{12} \hat{P}^{q}_{13} 
        + y_{i31} \hat{P}^{\sigma}_{12} \hat{P}^{q}_{12} \hat{P}^{q}_{13} 
        \nonumber \\ &
        + y_{i32} \hat{P}^{\sigma}_{13} \hat{P}^{q}_{12} \hat{P}^{q}_{13} 
        + y_{i33} \hat{P}^{\sigma}_{13} \hat{P}^{q}_{12} \hat{P}^{q}_{23} 
        + y_{i34} \hat{P}^{\sigma}_{12} \hat{P}^{\sigma}_{13} \hat{P}^{q}_{12} \hat{P}^{q}_{13} 
        + y_{i35} \hat{P}^{\sigma}_{12} \hat{P}^{\sigma}_{13} \hat{P}^{q}_{12} \hat{P}^{q}_{23} 
    \Big)
 \, .
\label{eq:Skyrme_int:3body:Px13:kp23k13}
\end{align}
\end{subequations}
\end{widetext}
It has to be noted that 
\begin{equation}
\hat{P}^{\{u_i,y_i\}}_{\overline{12}3}
=   \hat{P}^{\{u_i,y_i\}}_{123,a} 
  + \hat{P}^{\{u_i,y_i\}}_{123,b}
  + \hat{P}^{\{u_i,y_i\}}_{213,b} \, ,
\end{equation} 
i.e.\ operators $\hat{P}^{\{u_i,y_i\}}_{123,a}$ and 
$\hat{P}^{\{u_i,y_i\}}_{123,b}$ are nothing but sub-parts of 
$\hat{P}^{\{u_i,y_i\}}_{\overline{12}3}$. Also, one has that 
\mbox{$\hat{P}^{\{u_i,y_i\}}_{123,a}=\hat{P}^{\{u_i,y_i\}\,\dagger}_{123,a}$}.

Whenever a given term displays good parity under the exchange of the spatial
coordinates of particles $i$ and $j$, one can replace $\hat{P}^r_{ij}$ by 
$\pm1$ in its matrix elements \textit{a priori}. For instance, it can be
easily seen that $\hat{v}^2_{\overline{12}}$ changes its sign under particle
exchange; hence, it has negative parity. As a consequence, one can make the 
replacement
\mbox{$\hat{v}^2_{\overline{12}} \hat{P}^r_{12} = \hat{v}^2_{\overline{12}}$} 
in its matrix elements. Whenever such property can be exploited, one can 
use Eq.~(\ref{eq:Skyrme_int:rsq_exchanges_A}) to re-express $\hat{P}^q_{ij}$ 
in terms of $\hat{P}^\sigma_{ij}$. This can be done in 
Eqs.~(\ref{eq:Skyrme_int:2body:Grad_init:nograd}), 
(\ref{eq:Skyrme_int:2body:Grad_init:k12k12}), 
(\ref{eq:Skyrme_int:2body:Grad_init:kp12k12}), 
(\ref{eq:Skyrme_int:3body:Grad_init:k12k12}) 
and~(\ref{eq:Skyrme_int:3body:Grad_init:kp12k12}) for $\hat{P}^q_{12}$, as 
well as in Eq.~(\ref{eq:Skyrme_int:3body:Grad_init:nograd}) for 
$\hat{P}^q_{12}$, $\hat{P}^q_{13}$ and $\hat{P}^q_{23}$. In the end,
these considerations bring $\hat{P}^{\{t_i,x_i\}}_{\overline{12}}$ and 
$\hat{P}^{\{u_i,y_i\}}_{\overline{12}3}$ into the simpler form
\begin{widetext}
\begin{subequations}
\label{eq:Skyrme_int:Px2}
\begin{eqnarray}
\hat{P}^{\{t_i,x_i\}}_{\overline{12},\alpha} 
& = & t_i
      \big( \hat{\ensuremath{\mathbbm{1}}}_{2} 
            + x_{i} \hat{P}^{\sigma}_{12} 
      \big) \, ,
\label{eq:Skyrme_int:2body:Px12}
\\
\hat{P}^{\{u_i,y_i\}}_{\overline{12}3,\alpha} 
& = & u_i
      \big[ \hat{\ensuremath{\mathbbm{1}}}_{3} 
      + y_{i1} \hat{P}^{\sigma}_{12} 
      + y_{i2} \big( \hat{P}^{\sigma}_{13} + \hat{P}^{\sigma}_{23} \big)
      + y_{i3} \big( \hat{P}^{\sigma}_{12} \hat{P}^{\sigma}_{13} 
                      + \hat{P}^{\sigma}_{12} \hat{P}^{\sigma}_{23} \big) 
      \big] \, ,
\label{eq:Skyrme_int:3body:Px123,3}
\\
\hat{P}^{\{u_i,y_i\}}_{\overline{12}3,\beta} 
& = & u_i
      \big[ \hat{\ensuremath{\mathbbm{1}}}_{3} 
      + y_{i1} \hat{P}^{\sigma}_{12} 
      + y_{i2} \big( \hat{P}^{\sigma}_{13} + \hat{P}^{\sigma}_{23} \big)
      + y_{i3} \big( \hat{P}^{\sigma}_{12} \hat{P}^{\sigma}_{13} + \hat{P}^{\sigma}_{12} \hat{P}^{\sigma}_{23} \big)
      + y_{i20} \big( \hat{P}^{q}_{13} + \hat{P}^{q}_{23} \big) 
\nonumber \\ 
&   & + y_{i21} \big( \hat{P}^{\sigma}_{12} \hat{P}^{q}_{13} + \hat{P}^{\sigma}_{12} \hat{P}^{q}_{23} \big) 
+ y_{i22} \big( \hat{P}^{\sigma}_{13} \hat{P}^{q}_{13} + \hat{P}^{\sigma}_{23} \hat{P}^{q}_{23} \big) 
+ y_{i23} \big( \hat{P}^{\sigma}_{13} \hat{P}^{q}_{23} + \hat{P}^{\sigma}_{23} \hat{P}^{q}_{13} \big) 
\nonumber \\ 
&   & + y_{i24} \big( \hat{P}^{\sigma}_{12} \hat{P}^{\sigma}_{13} \hat{P}^{q}_{13} + \hat{P}^{\sigma}_{12} \hat{P}^{\sigma}_{23} \hat{P}^{q}_{23} \big) 
+ y_{i25} \big( \hat{P}^{\sigma}_{12} \hat{P}^{\sigma}_{23} \hat{P}^{q}_{13} + \hat{P}^{\sigma}_{12} \hat{P}^{\sigma}_{13} \hat{P}^{q}_{23} \big) 
\Big]
 \, .
\label{eq:Skyrme_int:3body:Px123,2}
\end{eqnarray}
\end{subequations}
No such reductions, however, are possible in 
Eqs.~(\ref{eq:Skyrme_int:3body:Grad_init:k23k13}) - 
(\ref{eq:Skyrme_int:3body:Grad_init:kp23k13}). In the end, the complete 
exploitation of the symmetry relations listed above leads to 
the following set of the most general possible structures
\begin{subequations}
\label{eq:Skyrme_int:Grad}
\begin{align}
\hat{v}^0_{\overline{12}}
\equiv & \, 
\hat{P}^{\{t_0,x_0\}}_{\overline{12},\alpha}
\hat{\delta}^r_{12}
\label{eq:Skyrme_int:2body:Grad:nograd}
\,,  \\
\hat{v}^1_{\overline{12}}
\equiv & \, 
\frac{1}{2} \hat{P}^{\{t_1,x_1\}}_{\overline{12},\alpha}
\left[ \hat{\vec{k}}^{\,\dagger}_{12} \cdot \hat{\vec{k}}^{\,\dagger}_{12}  \, \hat{\delta}^r_{12}
\, + \, 
\hat{\delta}^r_{12} \, \hat{\vec{k}}^{\,}_{12} \cdot \hat{\vec{k}}^{\,}_{12}  \right]
\label{eq:Skyrme_int:2body:Grad:k12k12}
\,,  \\
\hat{v}^2_{\overline{12}}
\equiv & \, 
\hat{P}^{\{t_2,x_2\}}_{\overline{12},\alpha}
\hat{\vec{k}}^{\,\dagger}_{12} \, \hat{\delta}^r_{12} \, \cdot \hat{\vec{k}}^{\,}_{12}
\label{eq:Skyrme_int:2body:Grad:kp12k12}
\, ,
\intertext{for two-body terms and}
\hat{v}^0_{\overline{12}3}
\equiv & \, 
\frac{1}{2}  
\left[ \hat{P}^{\{u_0,y_0\},\dagger}_{\overline{12}3,\alpha}
+
\hat{P}^{\{u_0,y_0\}}_{\overline{12}3,\alpha}  \right] \; \hat{\delta}^r_{13}\hat{\delta}^r_{23}
\label{eq:Skyrme_int:3body:Grad:nograd}
 \,,  \\
\hat{v}^1_{\overline{12}3}
\equiv & \, 
\frac{1}{2}  
\left[ \hat{P}^{\{u_1,y_1\},\dagger}_{\overline{12}3,\beta} \, \hat{\vec{k}}^{\,\dagger}_{12} \cdot \hat{\vec{k}}^{\,\dagger}_{12} \, \hat{\delta}^r_{13}\hat{\delta}^r_{23}
\, + \, 
\hat{P}^{\{u_1,y_1\}}_{\overline{12}3,\beta} \, \hat{\delta}^r_{13}\hat{\delta}^r_{23} \, \hat{\vec{k}}^{\,}_{12} \cdot \hat{\vec{k}}^{\,}_{12}  \right]
 \label{eq:Skyrme_int:3body:Grad:k12k12} 
 \,,  \\
\hat{v}^2_{\overline{12}3}
\equiv & \, 
\frac{1}{2} 
\left[ \hat{P}^{\{u_2,y_2\},\dagger}_{\overline{12}3,\beta} \, 
\, + \, 
\hat{P}^{\{u_2,y_2\}}_{\overline{12}3,\beta} \right] \,  \hat{\vec{k}}^{\,\dagger}_{12} \, \hat{\delta}^r_{13}\hat{\delta}^r_{23} \, \cdot \hat{\vec{k}}^{\,}_{12}
 \label{eq:Skyrme_int:3body:Grad:kp12k12} 
 \,,  \\
 \hat{v}^3_{\overline{12}3}
\equiv & \, 
\frac{1}{2} 
\left[  \hat{P}^{\{u_3,y_3\},\dagger}_{\overline{12}3} \, \hat{\vec{k}}^{\,\dagger}_{23} \cdot \hat{\vec{k}}^{\,\dagger}_{13}  \, \hat{\delta}^r_{13}\hat{\delta}^r_{23}
\, + \, 
\hat{P}^{\{u_3,y_3\}}_{\overline{12}3} \, \hat{\delta}^r_{13}\hat{\delta}^r_{23} \, \hat{\vec{k}}^{\,}_{13} \cdot \hat{\vec{k}}^{\,}_{23} \right]
 \label{eq:Skyrme_int:3body:Grad:k23k13} 
 \,,  \\
\hat{v}^4_{\overline{12}3}
\equiv & \, 
\frac{1}{2} 
\hat{P}^{\{u_4,y_4\}}_{123,a}
\Big[\hat{\vec{k}}^{\,\dagger}_{13} \, \hat{\delta}^r_{13}\hat{\delta}^r_{23} \, \cdot \hat{\vec{k}}^{\,}_{23}
\, + \, 
\hat{\vec{k}}^{\,\dagger}_{23} \, \hat{\delta}^r_{13}\hat{\delta}^r_{23} \, \cdot \hat{\vec{k}}^{\,}_{13}
\Big]
 \nonumber \\
  \, + \, & \,
 \frac{1}{2} 
\Big[ 
\hat{P}^{\{u_4,y_{41}\},\dagger}_{123,b} \,
\, + \, 
\hat{P}^{\{u_4,y_{41}\}}_{213,b} \,
\Big]
\hat{\vec{k}}^{\,\dagger}_{23} \, \hat{\delta}^r_{13}\hat{\delta}^r_{23} \, \cdot \hat{\vec{k}}^{\,}_{13} 
  \, + \,
 \frac{1}{2} 
\Big[ 
\hat{P}^{\{u_4,y_{41}\},\dagger}_{213,b} \,
\, + \, 
\hat{P}^{\{u_4,y_{41}\}}_{123,b} \,
\Big]
\hat{\vec{k}}^{\,\dagger}_{13} \, \hat{\delta}^r_{13}\hat{\delta}^r_{23} \, \cdot \hat{\vec{k}}^{\,}_{23}
  \nonumber \\
\, + \,  & \,
 \frac{1}{2} 
\Big[ 
\hat{P}^{\{u_4,y_{42}\},\dagger}_{213,b} \,
\, + \, 
\hat{P}^{\{u_4,y_{42}\}}_{123,b} \,
\Big]
\hat{\vec{k}}^{\,\dagger}_{23} \, \hat{\delta}^r_{13}\hat{\delta}^r_{23} \, \cdot \hat{\vec{k}}^{\,}_{13} 
\, + \, 
 \frac{1}{2} 
\Big[ 
\hat{P}^{\{u_4,y_{42}\},\dagger}_{123,b} \,
 \, + \, 
\hat{P}^{\{u_4,y_{42}\}}_{213,b} \,
\Big]
\hat{\vec{k}}^{\,\dagger}_{13} \, \hat{\delta}^r_{13}\hat{\delta}^r_{23} \, \cdot \hat{\vec{k}}^{\,}_{23} 
   \, , \label{eq:Skyrme_int:3body:Grad:kp23k13}
\end{align}
\end{subequations}
\end{widetext}
for three-body terms. At this stage, $\hat{v}_{\overline{12}}$
is defined out of six coupling constants, whereas $\hat{v}_{\overline{12}3}$
includes altogether about 70 parameters. The two-body terms correspond 
already to the final form of Skyrme's standard central two-body vertex. For 
the three-body pseudo potential, however, it can be
expected that many terms are in fact linearly dependent. Further redundancies 
among these terms, however, become increasingly  difficult to detect, and
we do not attempt to find them by hand. Instead, the task is carried out by a
formal algebra code that constructs first the complete energy functional 
deriving from Eq.~(\ref{eq:Skyrme_int:Grad}) and then analyses the 
correlations between the original terms in the pseudo potential.

%
%
\section{Deriving the EDF kernel}
\label{sec:general:EDF}

We now provide the full expression of the EDF kernel obtained from the two- 
and three-body pseudo potentials given in Eq.~(\ref{eq:Skyrme_int:Grad}). 
The mathematical steps actually taken by the numerical code to derive the 
results are sketched in Appendix~\ref{sect:howto}. Correlations among the 
terms in the original pseudo potentials are then identified using a singular 
value decomposition (SVD). This allows us to deduce a set of linearly
independent central three-body Skyrme-like pseudo potentials.


\subsection{Ingredients of the EDF kernel}


\subsubsection{Density matrices}
 
In coordinate representation, the normal and anomalous one-body density 
matrices read
\begin{subequations}
\label{eq:Skyrme_int:intro:nonlocdensity}
\begin{align}
\rho (\vec{r} \sigma q ,\vec{r}\,' \sigma ' q') 
&\equiv \langle \Phi | a^{\dagger}_{\vec{r}\,' \sigma ' q'} a_{\vec{r} \sigma q}
        | \Phi \rangle  
\nonumber \\
& = \sum_{ij} \varphi^{\ast}_{j} (\vec{r}\,' \sigma ' q') \, \varphi^{\,}_{i} (\vec{r} \sigma q)  \, \rho_{ij} \, , 
\label{eq:Skyrme_int:intro:nonlocdensity:rho}
\\
\kappa (\vec{r} \sigma q ,\vec{r}\,' \sigma ' q') 
&\equiv 
\langle \Phi | a_{\vec{r}\,' \sigma ' q'} a_{\vec{r} \sigma q} | \Phi \rangle
\nonumber \\
&= \sum_{ij} \varphi_{j} (\vec{r}\,' \sigma ' q') \, \varphi^{\,}_{i} (\vec{r} \sigma q) \,  \kappa_{ij} \, , 
\label{eq:Skyrme_int:intro:nonlocdensity:kappa}
\end{align}
\end{subequations}
We assume pure proton and neutron density matrices, such that
\mbox{$\rho (\vec{r} \sigma q,\vec{r}\,' \sigma ' q') 
= \kappa (\vec{r} \sigma q,\vec{r}\,' \sigma ' q') = 0$} when 
\mbox{$q \neq q'$}. As already noted, the normal density matrix is 
hermitian, i.e.\ 
\mbox{$\rho (\vec{r} \sigma q,\vec{r}\,' \sigma ' q') 
= \rho^* (\vec{r}\,' \sigma ' q',\vec{r} \sigma q)$}, whereas the pair 
tensor is skew-symmetric, i.e.\ 
$\kappa (\vec{r} \sigma q,\vec{r}\,' \sigma ' q') 
= - \kappa(\vec{r}\,' \sigma ' q',\vec{r} \sigma q)$.


\subsection{Non-local densities}
\label{sect:nonlocaldensgoodq}

When assuming pure proton and neutron single-particle states, the most
straightforward representation of the densities is obtained in terms of 
proton and neutron densities. In this case, non-local normal and 
anomalous densities take the form
\begin{subequations}
\label{eq:Skyrme_int:nonlocdensities}
\begin{align} 
\label{eq:Skyrme_int:nonlocdensities:rho}
\rho_q (\vec{r}  ,\vec{r}\,' )
& \equiv \sum_{\sigma} \rho (\vec{r} \sigma q ,\vec{r}\,' \sigma q) \, , 
          \\
\label{eq:Skyrme_int:nonlocdensities:s}
s_{q, \nu} (\vec{r}  ,\vec{r}\,' )
& \equiv \sum_{\sigma' \sigma} \rho (\vec{r} \sigma q,\vec{r}\,' \sigma ' q) \,
         \langle \sigma ' | \hat{\sigma}_{\nu} | \sigma \rangle \, , 
         \\
\label{eq:Skyrme_int:nonlocdensities:trho}
\tilde{\rho}_q (\vec{r}  ,\vec{r}\,' )
& \equiv \sum_{\sigma} 2\bar{\sigma} 
        \kappa (\vec{r} \sigma q,\vec{r}\,' \bar{\sigma} q) \, , 
        \\
\label{eq:Skyrme_int:nonlocdensities:ts}
\tilde{s}_{q, \nu} (\vec{r}  ,\vec{r}\,' )
& \equiv \sum_{\sigma' \sigma}  2 \bar{\sigma}' 
         \kappa (\vec{r} \sigma q ,\vec{r}\,' \bar{\sigma}' q) \, 
         \langle \sigma ' | \hat{\sigma}_{\nu} | \sigma \rangle \,  , 
\end{align}
One further introduces kinetic densities
\begin{align}
\label{eq:Skyrme_int:nonlocdensities:tau}
\tau_q (\vec{r}  ,\vec{r}\,' )
& \equiv \sum_\mu \nabla_{\vec{r}, \mu} \, \nabla_{\vec{r}\,'\!, \mu} \, 
         \rho_q (\vec{r}  ,\vec{r}\,' ) \, ,
          \\
T_{q, \nu} (\vec{r}  ,\vec{r}\,' )
& \equiv \sum_\mu \nabla_{\vec{r}, \mu} \, \nabla_{\vec{r}\,'\!, \mu} \, 
         s_{q, \nu} (\vec{r}  ,\vec{r}\,' ) \, 
 \,  , \\
\tilde{\tau}_q (\vec{r}  ,\vec{r}\,' )
& \equiv \sum_\mu \nabla_{\vec{r}, \mu} \, \nabla_{\vec{r}\,'\!, \mu} \, 
         \tilde{\rho}_q (\vec{r}  ,\vec{r}\,' )
 \,  , \\
\tilde{T}_{q, \nu} (\vec{r}  ,\vec{r}\,' ) 
& \equiv \sum_\mu \nabla_{\vec{r}, \mu} \, \nabla_{\vec{r}\,'\!, \mu} \, 
         \tilde{s}_{q, \nu} (\vec{r}  ,\vec{r}\,' ) \,  ,
\end{align}
and currents
\begin{align} 
j_{q,\mu} (\vec{r}  ,\vec{r}\,' ) 
& \equiv - \frac{{\mathrm i}}{2} \big( \nabla_{\vec{r}, \mu} \, -  \, \nabla_{\vec{r}\,'\!, \mu} \big)  \, \rho_q (\vec{r}  ,\vec{r}\,' ) 
 \, , \\
J_{q, \mu \nu} (\vec{r}  ,\vec{r}\,' ), 
& \equiv - \frac{{\mathrm i}}{2} \big( \nabla_{\vec{r}, \mu} \, -  \, \nabla_{\vec{r}\,'\!, \mu} \big) \, s_{q, \nu} (\vec{r}  ,\vec{r}\,' ) \, , 
         \\
\tilde{\jmath}_{q,\mu} (\vec{r}  ,\vec{r}\,' ) 
& \equiv - \frac{{\mathrm i}}{2} \big( \nabla_{\vec{r}, \mu} \, -  \, \nabla_{\vec{r}\,'\!, \mu} \big)  \, \tilde{\rho}_q (\vec{r}  ,\vec{r}\,' )
 \, , \\
\label{eq:Skyrme_int:nonlocdensities:Jtilde}
\tilde{J}_{q, \mu \nu} (\vec{r}  ,\vec{r}\,' )
& \equiv - \frac{{\mathrm i}}{2} \big( \nabla_{\vec{r}, \mu} \, -  \, \nabla_{\vec{r}\,'\!, \mu} \big) \, \tilde{s}_{q, \nu} (\vec{r}  ,\vec{r}\,' ) \, 
 \, ,
\end{align}
\end{subequations}
where \mbox{$\bar{\sigma} = - \sigma$}. Greek 
indices taking values $x$, $y$, or $z$ refer to cartesian components of 
spatial vectors and tensors. Densities without Greek index such as $\rho$ and 
$\tilde{\rho}$ are scalars. Densities in
Eq.~(\ref{eq:Skyrme_int:nonlocdensities})
denotes non-local matter, spin, pair, pair-spin, kinetic, spin-kinetic, 
pair-kinetic, pair-spin-kinetic, current, spin-current, pair-current and 
pair-spin-current densities for a given nucleon species $q$, respectively. 

The particular definition of the pairing densities 
$\tilde{\rho}_q (\vec{r}  ,\vec{r}\,' )$ and 
$\tilde{s}_{q, \nu} (\vec{r} ,\vec{r}\,' )$ involves the time-reversal 
of coordinates $\vec{r}\,' \sigma' q'$, which is done to provide a compact 
representation of the EDF kernel derived from contact interactions in terms 
of a local densities, which cannot be achieved in terms of $\kappa$ due 
to its being skew symmetric~\cite{dobaczewski84a,dobaczewski95a}.

The non-local pair density is symmetric under coordinate exchange, 
whereas the pair-spin density is skew-symmetric
\begin{subequations}
\label{eq:Skyrme_int:nonlocdensities:symm}
\begin{eqnarray}
\tilde{\rho}_q (\vec{r}\,'  ,\vec{r} )
& = & + \tilde{\rho}_q (\vec{r}  ,\vec{r}\,' ) \, ,
      \\
\tilde{s}_{q, \nu} (\vec{r}\,'  ,\vec{r} )
& = & - \tilde{s}_{q, \nu} (\vec{r} ,\vec{r}\,' ) \, .
\end{eqnarray}
\end{subequations}
Instead of constructing them from the pair tensor
(\ref{eq:Skyrme_int:intro:nonlocdensity:kappa}), the pair densities could 
alternatively also be derived from a pair density matrix defined 
as~\cite{dobaczewski84a,dobaczewski95a}
\begin{equation}
\tilde{\rho}(\vec{r} \sigma q, \vec{r}\,' \sigma' q')
\equiv 2 \sigma' \, \kappa (\vec{r} \sigma q, \vec{r}\,' \sigma' q')
\, .
\end{equation}
The full normal and pair density matrices can be expressed in terms of
the non-local densities, which is equivalent to expanding a complex 
\mbox{$2 \times 2$} matrix in spin space over the unit matrix and 
Pauli matrices, which together form a complete basis of that space,
\begin{subequations}
\label{eq:density:matrices:su2:development}
\begin{eqnarray}
\lefteqn{
\rho (\vec{r} \sigma q, \vec{r}\,' \sigma' q')
} \nonumber \\
& = & \frac{1}{2} \, \big[
       \rho_q ( \vec{r}, \vec{r}\,' ) \delta_{\sigma \sigma'}
      + \vec{s}_{q} ( \vec{r}, \vec{r}\,' ) \cdot 
       \langle \sigma | \hat{\vec{\sigma}} | \sigma' \rangle
      \big] \, \delta_{qq'} \, ,
\\
\lefteqn{
\tilde{\rho}(\vec{r} \sigma q, \vec{r}\,' \sigma' q')
} \nonumber \\
& = &  \frac{1}{2} \, \big[
       \tilde{\rho}_q ( \vec{r}, \vec{r}\,' ) \delta_{\sigma \sigma'}
      + \tilde{\vec{s}}_{q} ( \vec{r}, \vec{r}\,' ) \cdot 
       \langle \sigma | \hat{\vec{\sigma}} | \sigma' \rangle
     \big] \, \delta_{qq'} \, .
\end{eqnarray}
\end{subequations}


\subsection{Local densities}
\label{localdensgoodq}

Ultimately, the expression of the EDF kernel invokes local densities obtained from the non-local ones through
\begin{eqnarray}
\label{eq:Skyrme_int:locdensities}
\mathcal{P}_q(\vec{r}) 
& \equiv & \mathcal{P}_q (\vec{r},\vec{r}) \, , \\
\tilde{\mathcal{P}}_q(\vec{r}) 
& \equiv & \tilde{\mathcal{P}}_q (\vec{r},\vec{r}) \, .
\end{eqnarray}
where $\mathcal{P}_q$ and $\tilde{\mathcal{P}}_q$ represent any of the 
normal or anomalous densities in Eq.~(\ref{eq:Skyrme_int:nonlocdensities}).
The local pair densities $\tilde{s}_{q, \nu}(\vec{r})$, 
$\tilde{T}_{q, \nu}(\vec{r})$ and $\tilde{j}_{q, \mu}(\vec{r})$ turn out 
to be zero 
\begin{eqnarray}
\tilde{s}_{q, \nu}(\vec{r})
& = & \tilde{T}_{q, \nu}(\vec{r})
  =   \tilde{j}_{q, \mu}(\vec{r})
  = 0 \, ,
\end{eqnarray}
because the corresponding non-local densities 
(\ref{eq:Skyrme_int:nonlocdensities}) are skew-symmetric under exchange of 
$\vec{r}$ and $\vec{r}\,'$, Eq.~(\ref{eq:Skyrme_int:nonlocdensities:symm}).

Single-reference EDF calculations based on time-reversal-breaking 
quasiparticle vacua, Eq.~(\ref{eq:Intro_met:product_state}), involve 
pair densities that are in general complex. The energy being real, the 
EDF kernel necessarily contains also their complex conjugate deriving from
$\tilde{\rho}^*_q (\vec{r},\vec{r}\,')
= \sum_{\sigma} 2 \bar{\sigma} \, 
        \kappa^* (\vec{r} \sigma q, \vec{r}\,' \bar{\sigma} q)
= \sum_{\sigma} 2 \bar{\sigma} \,
  \langle \Phi | a^\dagger_{\vec{r} \sigma q } a^\dagger_{\vec{r}\,' \bar{\sigma} q} 
  | \Phi \rangle
$ etc.


\subsection{Isoscalar and isovector densities}
\label{sect:densgoodT}
\label{localdensiso}

We now recouple the density matrices to isoscalars and isovectors, which
allow for a more transparent representation of the physics contained in
an energy functional
\begin{subequations}
\label{eq:Skyrme_int:nonlocdensities:IsoSV}
\begin{align} 
\label{eq:Skyrme_int:nonlocdensities:rhoIsoS}
\rho_0 (\vec{r}  ,\vec{r}\,' ) 
& \equiv \sum_{\sigma q} \rho (\vec{r} \sigma q ,\vec{r}\,' \sigma q) \, ,
   \\
\label{eq:Skyrme_int:nonlocdensities:rhoIsoV}
\rho_{1,\frak{a}} (\vec{r}  ,\vec{r}\,' ) 
& \equiv \sum_{\sigma q' q} \rho (\vec{r} \sigma q ,\vec{r}\,' \sigma q')  \,
         \tau_{\frak{a},q'q} \, , 
         \\  
\label{eq:Skyrme_int:nonlocdensities:sIsoS}
s_{0, \nu} (\vec{r}  ,\vec{r}\,' )
& \equiv  \sum_{\sigma' \sigma q} 
       \rho (\vec{r} \sigma q,\vec{r}\,' \sigma ' q)  \, 
       {\sigma}_{\nu ,\sigma' \sigma} \, ,
    \\
\label{eq:Skyrme_int:nonlocdensities:sIsoV}
s_{1,\frak{a}, \nu} (\vec{r}  ,\vec{r}\,' )
& \equiv \sum_{\sigma' \sigma q' q} 
         \rho (\vec{r} \sigma q,\vec{r}\,' \sigma ' q')  \,
         {\sigma}_{\nu ,\sigma' \sigma} \, {\tau}_{\frak{a},q'q} \, , 
         \\
\label{eq:Skyrme_int:nonlocdensities:trhoIsoS}
\breve{\rho}_0 (\vec{r}  ,\vec{r}\,' )
& \equiv \sum_{\sigma q} 4\bar{\sigma}\bar{q} \, 
         \kappa (\vec{r} \sigma q,\vec{r}\,' \bar{\sigma} \bar{q}) \, ,
     \\
\label{eq:Skyrme_int:nonlocdensities:trhoIsoV}
\breve{\rho}_{1,\frak{a}} (\vec{r}  ,\vec{r}\,' )
& \equiv \sum_{\sigma q' q} 4\bar{\sigma}\bar{q}' \, 
         \kappa (\vec{r} \sigma q,\vec{r}\,' \bar{\sigma} \bar{q}')  \, 
         {\tau}_{\frak{a},q'q} \, ,
      \\
\label{eq:Skyrme_int:nonlocdensities:tsIsoS}
\breve{s}_{0, \nu} (\vec{r}  ,\vec{r}\,' )
& \equiv \sum_{\sigma' \sigma q}  4\bar{\sigma}' \bar{q} \, 
         \kappa (\vec{r} \sigma q ,\vec{r}\,' \bar{\sigma}' \bar{q})  \,
         {\sigma}_{\nu ,\sigma' \sigma} \, ,
   \\
\label{eq:Skyrme_int:nonlocdensities:tsIsoV}
\breve{s}_{1,\frak{a}, \nu} (\vec{r}  ,\vec{r}\,' )
& \equiv \sum_{\sigma' \sigma q' q}  \!\!\! 4 \bar{\sigma}'\bar{q}' \, 
         \kappa (\vec{r} \sigma q ,\vec{r}\,' \bar{\sigma}' \bar{q}') \,
         {\sigma}_{\nu ,\sigma' \sigma} \, {\tau}_{\frak{a},q'q}  
   \, , 
\end{align}
\end{subequations}
where \mbox{${\sigma}_{\nu ,\sigma' \sigma} 
\equiv \langle \sigma ' | \hat{\sigma}_{\nu} | \sigma \rangle$} and 
\mbox{${\tau}_{\frak{a},q'q} 
\equiv \langle q' | \hat{\tau}_{\frak{a}} | q \rangle$} are matrix elements 
of spin and isospin Pauli matrices, respectively, whereas $\bar{q}\equiv -q$. 
Densities with index $0$ are isoscalar, whereas densities with index 
$(1,\frak{a})$ are cartesian components of isovector densities, with the 
index in fractur labeling its components $\frak{a} \in \{1,2,3\}$.

In this representation, local densitities are obtained in the same manner
as above in Eq.~(\ref{eq:Skyrme_int:locdensities}).

When limiting oneself to pure proton and neutron density matrices as done 
here, it follows that~\cite{perlinska04a,rohozinski10a}
\begin{itemize}
\item 
the first ($\frak{a} = 1$) and second ($\frak{a} = 2$) isovector components 
of all normal densities are zero,
\item 
all isoscalar pairing densities are zero,
\item
the third component ($\frak{a} = 3$) of all isovector pairing densities is 
zero.
\end{itemize}
For the sake of compact notation, normal isovector densities will be written 
without reference to the isospin component
\begin{equation}
\rho_{1} 
\equiv \rho_{1,3}
\end{equation}
in what follows. For pair densities, however, the index for the third
isospin component has to be kept.

\subsection{Link between the two representations}

The definition of the pair densities in isoscalar/isovector representation
labelled by a by "breve", such as $\breve{\rho}_t(\vec{r})$, differs from 
the ones labelled by a "tilde", such as $\tilde{\rho}_q(\vec{r})$, by a 
transformation in isospin that is the homologue of the transformation in spin 
space that leads from $\kappa(\vec{r} \sigma q, \vec{r}\,' \sigma' q')$ to 
$\tilde{\rho}(\vec{r} \sigma q, \vec{r}\,' \sigma' q')$~\cite{perlinska04a,rohozinski10a}. 
Both transformations are performed in order to obtain local pair densities 
from the skew-symmetric $\kappa(\vec{r} \sigma q, \vec{r}\,' \sigma' q')$. 
As long as densities are represented in proton-neutron representation, only 
the transformation in spin space is needed, whereas the recoupling  of the 
local densities in isospin space requires also the transformation in isospin 
space. For any normal density ${\cal P}$, the transformation between the two 
representations is given by
\begin{subequations}
\label{eq:Skyrme_int:isoscalar_isovector}
\begin{align}
{\cal P}_n 
&\equiv  \frac{1}{2} \, \big( {\cal P}_0 + {\cal P}_1 \big) \, , \\
{\cal P}_p
&\equiv  \frac{1}{2} \, \big( {\cal P}_0 - {\cal P}_1 \big) \, .
\end{align}
\end{subequations}
The first $\big( \breve{{\cal P}}_{1,1} \big)$ and second 
$\big( \breve{{\cal P}}_{1,2} \big)$ 
isovector components of the pairing densities are related to neutron and 
proton pairing densities $\tilde{{\cal P}}_q$, \mbox{$q \in \{n,p\}$}, 
according to \cite{perlinska04a}
\begin{subequations}
\label{eq:Skyrme_int:isoscalar_isovector_pairing}
\begin{align}
\tilde{{\cal P}}_n 
& \equiv \frac{1}{2} \, \big( \breve{{\cal P}}_{1,1} + {\mathrm{i}} \breve{{\cal P}}_{1,2} \big) \, , \\
\tilde{{\cal P}}_p 
&\equiv  \frac{1}{2} \, \big( \breve{{\cal P}}_{1,1} - {\mathrm{i}} \breve{{\cal P}}_{1,2} \big) \, .
\end{align}
\end{subequations}
%
%
%

\subsection{Deriving the energy functional}

The analytical derivation of the EDF kernel from the three-body
pseudo-potential considered here is more cumbersome than for the usual 
two-body Skyrme pseudo-potential. The main reason relates to the large 
number of terms obtained by multiplying the antisymmetrizer $\mathcal{A}_{123}$
with the exchange operators of Eqs.~(\ref{eq:Skyrme_int:Px3}), 
(\ref{eq:Skyrme_int:3body:Px12:kp23k13}) and~(\ref{eq:Skyrme_int:Px2}). 
Still, the intrinsic difficulty of calculating each individual term is the 
same and, for most of them, the evaluation can be done in the same manner 
as for the two-body Skyrme interaction. A slight complication arises in 
a small number of terms where position-exchange operators cannot be directly 
replaced by $\pm 1$ in the matrix elements. For those terms, one has to pay 
additional attention on which non-local density the gradient operators 
do act. 

In the end, the main challenge is to handle the sheer number of terms to be 
evaluated. The numerical code that performs the necessary algebraic 
manipulations is based on shape recognition and has been written as a 
Unix shell script~\cite{sadoudithese}. In order to present how the 
calculation proceeds in the code, Appendix~\ref{sect:howto} lists the steps 
to be taken to reduce the matrix elements of the pseudo potential to an
EDF kernel that depends on local densities only for a few typical terms 
arising from the three-body pseudo potential.


\subsection{Redundant terms in the pseudo potential}

Having derived the energy functional from the two- and three-body pseudo 
potentials, we are now looking for strict correlations between its terms. 
The analysis is performed at the level of the EDF kernel, i.e.\ examining
whether the energy functional deriving from different potential terms 
are linearly dependent. To do so, we apply the singular 
value decomposition to the matrix relating the coupling constants 
multiplying each term in the EDF kernel to the set of parameters entering 
the pseudo potential. Whenever such a correlation is identified, the 
number of independent terms in the original pseudo potential is reduced.  

\begin{table}[t!]
\begin{center}
\caption{\label{3bodyOut:corr:simple}
Equivalent terms in the three-body pseudo potential of 
Eq.~(\ref{eq:Skyrme_int:Grad}).
}
\begin{tabular}{rcl}
\hline \hline \noalign{\smallskip}
term & & correlated terms \\
\noalign{\smallskip}  \hline \noalign{\smallskip}
$u_0$  
& $\leftarrow$ & $u_0 y_{03}$ 
\\
$u_0 y_{01}$
& $\leftarrow$ & $u_0 y_{02}$ 
 \\
$u_1$  
& $\leftarrow$ & $u_1 y_{13},u_3 y_{30},u_3 y_{33},u_1 y_{121},u_1 y_{122}$,
   \\
&            & $u_3 y_{321},u_3 y_{322},u_1 y_{123},u_3 y_{323}$  
 \\
$u_1 y_{11}$  
& $\leftarrow$ & $u_1 y_{12},u_3 y_{31},u_3 y_{32},u_1 y_{120},u_1 y_{125}$,
   \\
&            & $u_3 y_{320},u_3 y_{325},u_1 y_{124},u_3 y_{324} $
 \\
$u_2 $ 
& $\leftarrow$ & $u_4,u_4 y_{43},u_5 y_{522},u_6 y_{622},u_2 y_{222}$,
   \\
&            & $u_5 y_{534},u_6 y_{634}$ 
 \\
$u_2 y_{21}$  
& $\leftarrow$ & $u_5 y_{52},u_5 y_{513},u_6 y_{62},u_5 y_{520}$,
   \\
&            & $u_6 y_{620},u_2 y_{225},u_5 y_{525},u_5 y_{531}$,
   \\
&            & $u_6 y_{624},u_2 y_{224},u_6 y_{632},u_6 y_{633}$ 
 \\
$u_2 y_{22}$ 
& $\leftarrow$ & $u_4 y_{42},u_6 y_{613},u_5 y_{532},u_5 y_{533}$ 
 \\
$u_2 y_{23}$ 
& $\leftarrow$ & $u_5 y_{512},u_6 y_{612},u_5 y_{530},u_5 y_{535}$ 
 \\
$u_2 y_{220}$  
& $\leftarrow$ & $u_4 y_{41},u_5 y_{524},u_6 y_{625},u_6 y_{631}$ 
 \\
$u_2 y_{221}$  
& $\leftarrow$ & $u_5 y_{53},u_5 y_{521},u_6 y_{621},u_6 y_{635}$ 
 \\
$u_2 y_{223}$  
& $\leftarrow$ & $y_{63},u_5 y_{523},u_6 y_{623},u_6 y_{630}$ \\
\noalign{\smallskip} \hline \hline
\end{tabular}
\end{center}
\end{table}

\begin{table}[t!]
\begin{center}
\caption{\label{3bodyOut:SVD}
Correlations between terms of the pseudo-potential
Eq.~(\ref{eq:Skyrme_int:Grad}). See text.
}
\begin{tabular}{r r r r }
\hline \hline \noalign{\smallskip}
term          & $u_2$ &$u_2 y_{21}$ & $u_2 y_{22}$ \\
\noalign{\smallskip}  \hline \noalign{\smallskip}
$u_2 y_{23} =$  & $-1$ & $+1$ & $+1$ \\
$u_2 y_{220}=$ &      & $+1$ & $-1$ \\
$u_2 y_{221}=$ & $-1$ & $+2$ & $-1$ \\
$u_2 y_{223}=$ & $+2$ & $-3$ &      \\
\noalign{\smallskip} \hline \hline
\end{tabular}
\end{center}
\end{table}

Two-body terms in Eq.~(\ref{eq:Skyrme_int:Grad}) serve as a consistency 
check for the procedure. In this case, the formal algebra code gives the 
well-known energy functional of 
Refs.~\cite{dobaczewski84a,perlinska04a,rohozinski10a}, and the correlation 
analysis confirms that all terms are linearly independent as expected. 

Most of the redundancies in the three-body pseudo-potential are easily 
identified given that the functional obtained from one term is often 
directly proportional to the functional derived from another term; see 
Table~\ref{3bodyOut:corr:simple} for the identification of such strict 
proportionality. For a smaller number of terms, however, only the SVD can 
reveal their more intricate interdependency; see Table~\ref{3bodyOut:SVD}. 
For example, the energy functional obtained from the term with parameter 
$u_2 y_{221}$ in Eq.~(\ref{eq:Skyrme_int:Grad}) equals the sum of the 
energy functionals obtained from the terms with parameters $u_2$, 
$u_2 y_{21}$ and $u_2 y_{22}$ with relative weights $-1$, $+2$ and $-1$, 
respectively. The two terms containing a single spin or isospin exchange 
operator in $\hat{P}^{\{u_0,y_0\}}_{\overline{12}3,\alpha}$,
Eq.~(\ref{eq:Skyrme_int:3body:Grad:nograd}), give an energy functional 
that is zero. The term with simultaneous spin and isospin exchange (or, 
equivalently, a position exchange) in 
Eq.~(\ref{eq:Skyrme_int:3body:Grad:nograd}) provides the same energy 
functional as the term without exchange operator, such that the three-body 
term without gradient is in the end defined by a single free parameter, 
as expected from earlier studies~\cite{waroquier76a,stringari78a}. Last 
but not least, it turns out that all gradient terms of 
Eqs.~(\ref{eq:Skyrme_int:3body:Grad:k23k13}) 
and~(\ref{eq:Skyrme_int:3body:Grad:kp23k13}) are fully correlated to 
those in Eqs.~(\ref{eq:Skyrme_int:3body:Grad:k12k12}) 
and~(\ref{eq:Skyrme_int:3body:Grad:kp12k12}), respectively.

\begin{table}[b!]
\begin{center}
\caption{\label{3bodyOut:earlierwork}
Three-body terms in Eq.~(\ref{eq:Skyrme_int:int:final}) labelled by
their parameters that have been considered in earlier work as
indicated by $+$.
}
\begin{tabular}{rcccccc}
\hline \hline \noalign{\smallskip}
Ref.                & $u_0$ & $u_1$ & $u_1 y_1$ & $u_2$ & $u_2 y_{21}$ & $u_2 y_{22}$ \\
\noalign{\smallskip}  \hline \noalign{\smallskip}
\cite{blaizot75a}   & $+$   & $+$   & $-$       & $-$   & $-$          & $-$ \\
\cite{liu75a}       & $+$   & $+$   & $-$       & $-$   & $-$          & $-$ \\
\cite{onishi78a}    & $+$   & $-$   & $-$       & $+$   & $+$          & $-$ \\
\cite{waroquier83a} & $+$   & $+$   & $-$       & $-$   & $-$          & $-$ \\
\cite{arima86a}     & $+$   & $+$   & $-$       & $+$   & $+$          & $-$ \\
\cite{zheng90a}     & $+$   & $+$   & $-$       & $-$   & $-$          & $-$ \\
\cite{liu91a}       & $+$   & $+$   & $+$       & $+$   & $+$          & $-$ \\
\noalign{\smallskip}
\hline \hline
\end{tabular}
\end{center}
\end{table}


\subsection{Final form of the pseudo potential}

The irreducible set of central three-body operators containing two 
gradients is not unique as there are many equivalent possibilities to 
select independent terms. For consistency with the standard representation 
of the central part of the two-body Skyrme interaction
\begin{subequations}
\label{eq:Skyrme_int:int:final}
\begin{eqnarray}
\hat{v}_{\overline{12}} 
& = & t_0 \big( \hat{\ensuremath{\mathbbm{1}}}_{2}  
                + x_0 \hat{P}^{\sigma}_{12} \big) \,
      \hat{\delta}^r_{12} \label{eq:Skyrme_int:2body_int:t0}  \\
&   & + \frac{t_1}{2} \big( \hat{\ensuremath{\mathbbm{1}}}_{2}  
       + x_1 \hat{P}^{\sigma}_{12} \big)
      \big(  \hat{\vec{k}}^{\,\dagger \, 2}_{12} \;\hat{\delta}^r_{12}
            + \hat{\delta}^r_{12} \; \hat{\vec{k}}^{\,2}_{12} \big) 
       \label{eq:Skyrme_int:2body_int:t1}    \\
&   & + t_2 \big( \hat{\ensuremath{\mathbbm{1}}}_{2}  
                   + x_2 \hat{P}^{\sigma}_{12} \big) \, 
        \hat{\vec{k}}^{\,\dagger}_{12} \; \hat{\delta}^r_{12} \cdot 
        \hat{\vec{k}}^{\,}_{12}  \label{eq:Skyrme_int:2body_int:t2} \, ,
\end{eqnarray}
we choose a form that contains only spin-exchange operators. This leads to
\begin{eqnarray}
\lefteqn{\hat{v}_{\overline{12}3} 
} \nonumber \\
& = & u_0 \; \,\hat{\delta}^r_{13} \hat{\delta}^r_{23} 
      \label{eq:Skyrme_int:3body_int:final:u0} \\
&   & + \frac{u_1}{2} \big( \hat{\ensuremath{\mathbbm{1}}}_{3} 
                              + y_1 \hat{P}^{\sigma}_{12} \big) \, 
      \big( \hat{\vec{k}}^{\,\dagger \, 2}_{12} \,\hat{\delta}^r_{13} 
             \hat{\delta}^r_{23} + \hat{\delta}^r_{13} \hat{\delta}^r_{23} \, 
              \hat{\vec{k}}^{\,2}_{12} \big) 
      \label{eq:Skyrme_int:3body_int:final:u1} \\
&   & +u_2 \big( \hat{\ensuremath{\mathbbm{1}}}_{3} 
      + y_{21} \hat{P}^{\sigma}_{12} \big) \, \hat{\vec{k}}^{\,\dagger}_{12}
        \; \hat{\delta}^r_{13} \hat{\delta}^r_{23} \, \cdot 
       \hat{\vec{k}}^{\,}_{12} 
     \label{eq:Skyrme_int:3body_int:final:u2} \\
&   & + u_2 \, y_{22} \big(\hat{P}^{\sigma}_{13} + 
                   \hat{P}^{\sigma}_{23}\big) \, 
      \hat{\vec{k}}^{\,\dagger}_{12} \; \hat{\delta}^r_{13} \hat{\delta}^r_{23} \, \cdot \hat{\vec{k}}^{\,}_{12} \, ,
\label{eq:Skyrme_int:3body_int:final:u2:2} 
\end{eqnarray}
\end{subequations}
where $u_0$, $u_1$, $y_1$, $u_2$,
$y_{21}$, and $y_{22}$ denotes the final set of free parameters complementing $t_0$, $x_0$, $t_1$, $x_1$, $t_2$ and $x_2$. There are altogether twelve parameters for the central terms. 

The complexity  of the final three-body pseudo-potential 
is tremendously reduced compared to the original expression, 
Eq.~(\ref{eq:Skyrme_int:Grad}). As a matter of fact, its spatial content 
can be obtained by inserting a mere delta operator $\hat{\delta}^r_{23}$
into  the central two-body terms. Still, the spin-isospin structure in
Eq.~(\ref{eq:Skyrme_int:3body_int:final:u2:2}) is richer than that of the 
corresponding two-body operator of Eq.~(\ref{eq:Skyrme_int:2body_int:t2}),
such that the form of the three-body pseudo potential could not have been 
completely guessed by analogy with the two-body terms.

Compared to three-body contact potentials with two gradients 
studied in the past, our final form contains one term, 
Eq.~(\ref{eq:Skyrme_int:3body_int:final:u2:2}), that, to the best of our 
knowledge, has never been considered before; see \ 
Table~\ref{3bodyOut:earlierwork}. As a matter of fact, most published work
considered an even smaller subset. Most importantly, these earlier works 
\cite{blaizot75a,liu75a,onishi78a,waroquier83a,arima86a,zheng90a,liu91a}
 discussed only nuclear matter and/or spherical nuclei and ignored
the possibility of pairing correlations, such that they present only 
incomplete expressions for the energy functional and the corresponding
one-body fields.

\begin{table}[t!]
\begin{center}
\caption{\label{tab:Skyrme_int:2bodyEDF:coeff}
Coefficients of the normal part of the bilinear EDF kernel,
Eq.~(\ref{eq:Skyrme_int:2bodyEDF:normal}), as a function of the 
parameters of the pseudo-potential of 
Eqs.~(\ref{eq:Skyrme_int:2body_int:t0})-(\ref{eq:Skyrme_int:2body_int:t2}). 
Missing entries are zero.}  
\begin{tabular}{r c c c c c c }
\hline \hline \noalign{\smallskip}
&$t_0$&$t_0x_0$&$t_1$&$t_1x_1$&$t_2$&$t_2x_2$ \\
\noalign{\smallskip}  \hline \noalign{\smallskip}
$A^\rho_0\;=$&$+\frac{3}{8}$&&&&&\\[0.3mm]
$A^\rho_1\;=$&$-\frac{1}{8}$&$-\frac{1}{4}$&&&&\\[0.3mm]
$A^s_0\;=$&$-\frac{1}{8}$&$+\frac{1}{4}$&&&&\\[0.3mm]
$A^s_1\;=$&$-\frac{1}{8}$&&&&&\\[0.3mm]
$A^\tau_0\;=$&&&$+\frac{3}{16}$&&$+\frac{5}{16}$&$+\frac{1}{4}$\\[0.3mm]
$A^\tau_1\;=$&&&$-\frac{1}{16}$&$-\frac{1}{8}$&$+\frac{1}{16}$&$+\frac{1}{8}$\\[0.3mm]
$A^T_0\;=$&&&$-\frac{1}{16}$&$+\frac{1}{8}$&$+\frac{1}{16}$&$+\frac{1}{8}$\\[0.3mm]
$A^T_1\;=$&&&$-\frac{1}{16}$&&$+\frac{1}{16}$&\\[0.3mm]
$A^{\nabla\rho}_0\;=$&&&$+\frac{9}{64}$&&$-\frac{5}{64}$&$-\frac{1}{16}$\\[0.3mm]
$A^{\nabla\rho}_1\;=$&&&$-\frac{3}{64}$&$-\frac{3}{32}$&$-\frac{1}{64}$&$-\frac{1}{32}$\\[0.3mm]
$A^{\nabla s}_0\;=$&&&$-\frac{3}{64}$&$+\frac{3}{32}$&$-\frac{1}{64}$&$-\frac{1}{32}$\\[0.3mm]
$A^{\nabla s}_1\;=$&&&$-\frac{3}{64}$&&$-\frac{1}{64}$&\\[0.3mm]
$A^j_0\;=$&&&$-\frac{3}{16}$&&$-\frac{5}{16}$&$-\frac{1}{4}$\\[0.3mm]
$A^j_1\;=$&&&$+\frac{1}{16}$&$+\frac{1}{8}$&$-\frac{1}{16}$&$-\frac{1}{8}$\\[0.3mm]
$A^J_0\;=$&&&$+\frac{1}{16}$&$-\frac{1}{8}$&$-\frac{1}{16}$&$-\frac{1}{8}$\\[0.3mm]
$A^J_1\;=$&&&$+\frac{1}{16}$&&$-\frac{1}{16}$&\\
\noalign{\smallskip} \hline \hline
\end{tabular}
\end{center}
\end{table}

\begin{table}[t!]
\caption{\label{tab:Skyrme_int:2bodyEDF:coeffP}
Same as Table~\ref{tab:Skyrme_int:2bodyEDF:coeff}, but 
for the anomalous part of the bilinear EDF kernel, 
Eq.~(\ref{eq:Skyrme_int:2bodyEDF:anormal}).
}  
\begin{center}
\begin{tabular}{r c c c c c c}
\hline \hline \noalign{\smallskip}
 &$t_0$&$t_0x_0$&$t_1$&$t_1x_1$&$t_2$&$t_2x_2$\\
\noalign{\smallskip}  \hline \noalign{\smallskip}
$A^{\breve{\rho}}=$&$+\frac{1}{8}$&$-\frac{1}{8}$&&&&\\[0.3mm]
$A^{\breve{\tau}^*}=$&&&$+\frac{1}{16}$&$-\frac{1}{16}$&&\\[0.3mm]
$A^{\breve{\tau}}=$&&&$+\frac{1}{16}$&$-\frac{1}{16}$&&\\[0.3mm]
$A^{\nabla \breve{\rho}}=$&&&$+\frac{1}{32}$&$-\frac{1}{32}$&&\\[0.3mm]
$A^{\breve{J}}=$&&&&&$+\frac{1}{8}$&$+\frac{1}{8}$\\
\noalign{\smallskip} \hline \hline
\end{tabular}
\end{center}
\end{table}


\subsection{Energy functional}
\label{sec:Skyrme_int:EDF:Result}

Starting from a Skyrme-like pseudo-potential, each term of the resulting 
energy functional can be expressed as the integral over a local energy 
density , i.e.\
\begin{subequations}
\label{energydensities}
\begin{eqnarray}
E^\rho 
&\equiv & \int \! d^3r \; {\cal E}^\rho(\vec{r}) \, , \\
E^{\rho \rho} 
&\equiv& \int \! d^3r \; {\cal E}^{\rho \rho}(\vec{r}) \,  , \\
E^{\kappa \kappa} 
&\equiv& \int \! d^3r \; {\cal E}^{\kappa \kappa}(\vec{r}) \, , \\
E^{\rho \rho \rho} 
&\equiv& \int \! d^3r \; {\cal E}^{\rho \rho \rho}(\vec{r}) \,  , \\
E^{\kappa \kappa \rho} 
&\equiv& \int \! d^3r \; {\cal E}^{\kappa \kappa \rho}(\vec{r}) \, .
\end{eqnarray}
\end{subequations}
We give now the energy functional in a representation using isoscalar
and isovector densities. Its representation in terms of proton and neutron 
densities can be found in Appendix~\ref{sec:Skyrme_int:EDF:np}.


\subsubsection{Linear part}

Omitting the argument $\vec{r}$ of the local densities for brevity, 
the linear energy density associated with the effective one-body kinetic 
energy operator is given by
\begin{equation}
\label{eq:Skyrme_int:1bodyEDF}
{\cal E}^{\rho}
= \frac{\hbar^2}{2m} \tau_0 \, .
\end{equation}

\subsubsection{Bilinear part}

The normal part of the bilinear energy density is well known and reads 
\begin{eqnarray}
\label{eq:Skyrme_int:2bodyEDF:normal}
{\cal E}^{\rho \rho}
& = & \sum_{t=0,1}\Big[ 
       A^\rho_t \, \rho_t\rho_t
     + A^\tau_t \, \rho_t\tau_t
     + A^{\nabla\rho}_t (\vec{\nabla}\rho_t) \cdot (\vec{\nabla}\rho_t) 
      \nonumber \\ 
&   & +\sum_{\mu\nu} A^J_t J_{t,\mu \nu} J_{t,\mu\nu}
      + A^s_t \, \vec{s}_t \cdot \vec{s}_t
      + A^T_t \, \vec{s}_t \cdot \vec{T}_t
      \nonumber \\ 
&   & 
      + A^j_t \, \vec{j}_t\cdot\vec{j}_t
      + \sum_{\mu\nu} A^{\nabla s}_t (\nabla_\mu s_{t,\nu}) (\nabla_\mu s_{t,\nu}) 
      \Big] \, , 
\end{eqnarray}
whereas its anomalous part is given by
\begin{eqnarray}
\label{eq:Skyrme_int:2bodyEDF:anormal}
{\cal E}^{\kappa \kappa} 
& = & \sum_{\frak{a}=1,2} \Big[
      A^{\breve{\rho}} \breve{\rho}^*_{1,\frak{a}} \breve{\rho}_{1,\frak{a}}   
     +A^{\breve{\tau}^*} \breve{\tau}^*_{1,\frak{a}} \breve{\rho}_{1,\frak{a}}
     +A^{\breve{\tau}} \breve{\tau}_{1,\frak{a}} \breve{\rho}^*_{1,\frak{a}}   
     \nonumber \\
&   & + A^{\nabla \breve{\rho}} (\vec{\nabla} \breve{\rho}^*_{1,\frak{a}} ) 
        \cdot (\vec{\nabla} \breve{\rho}_{1,\frak{a}})   
      +\sum_{\mu\nu} A^{\breve{J}} \breve{J}^*_{1,\frak{a}, \mu \nu} 
                                         \breve{J}_{1,\frak{a}, \mu \nu}  
      \Big]
      \, .
     \nonumber \\
\end{eqnarray}
The relations between the coupling constants of the energy functional
and the pseudo-potential parameters are listed in 
Tables~\ref{tab:Skyrme_int:2bodyEDF:coeff} 
and~\ref{tab:Skyrme_int:2bodyEDF:coeffP}.


\subsubsection{Trilinear part}

The normal part of the trilinear energy density reads
\begin{widetext}
\begin{eqnarray}
\label{eq:Skyrme_int:3bodyEDF:normal}
{\cal E}^{\rho \rho \rho}
& = & \sum_{t=0,1}
      \Big\{ 
   B^\rho_t    \,        \rho_t \rho_t \rho_0 
 +B^s_t          \,                    \vec{s}_t\cdot\vec{s}_t \rho_0 
 +B^\tau_t     \,       \rho_t\tau_t \rho_0
 +B^{\tau s}_t        \,           \tau_t  \vec{s}_t \cdot \vec{s}_0                      
 +B^T_t           \,                 \vec{s}_t\cdot\vec{T}_t  \rho_0                           
 +B^T_{t\bar{t}}       \;              \vec{s}_t\cdot\vec{T}_{\bar{t}}  \rho_1                    
\nonumber \\ 
&  & 
 +B^{\nabla\rho}_t   \,  (\vec{\nabla}\rho_t)\cdot(\vec{\nabla}\rho_t) \rho_0
 +B^{j}_t            \,                \vec{j}_t\cdot\vec{j}_t  \rho_0                          
 +\sum_{\mu\nu} 
\big[ 
    B^{\nabla s}_t    \,   (\nabla_{\mu}s_{t,\nu}) (\nabla_{\mu}s_{t,\nu}) \rho_0
 +B^{\nabla\rho s}_t      \,       (\nabla_{\mu}\rho_t) (\nabla_{\mu}s_{t,\nu}) \,  s_{0,\nu}  
\nonumber \\ 
&   &
 +B^{\nabla\rho s}_{t\bar{t}}   \,  (\nabla_{\mu}\rho_t) (\nabla_{\mu}s_{\bar{t},\nu})   s_{1,\nu}
 +B^J_t  \, J_{t,\mu \nu} J_{t,\mu\nu} \rho_0
 +B^{Js}_t       \,               j_{t,\mu}J_{t,\mu\nu}   s_{0,\nu}                       
 +B^{Js}_{t\bar{t}}    \,           j_{t,\mu}J_{\bar{t},\mu\nu}  s_{1,\nu} 
 \big]
\nonumber \\ 
&   & 
+\sum_{\mu \nu \lambda \kappa} \epsilon_{\nu \lambda \kappa} 
\big[
  B^{\nabla s J}_t  \,  (\nabla_{\mu}s_{t,\nu}) J_{t,\mu\lambda} s_{0,\kappa} 
+ B^{\nabla s J}_{t\bar{t}}  \,  (\nabla_{\mu}s_{t,\nu}) J_{\bar{t},\mu\lambda} s_{1,\kappa}
 \big]
\Big\}                      
  +B^s_{10}         \,                  \vec{s}_1 \cdot \vec{s}_0  \rho_1                    
 +B^\tau_{10}    \,   \rho_1\tau_0  \rho_1     
 +B^{\tau s}_{10}        \,        \tau_{0}   \vec{s}_1  \cdot \vec{s}_1                                               
\nonumber \\ 
&   & 
 +B^{\nabla\rho}_{10} \, (\vec{\nabla}\rho_1)\cdot(\vec{\nabla}\rho_0)  \rho_1 
 + \sum_{\mu\nu}  B^{\nabla s}_{10}       \,          (\nabla_{\mu}s_{1,\nu}) (\nabla_{\mu}s_{0,\nu})  \rho_1  
 +B^{j}_{10}                \,        \vec{j}_1\cdot\vec{j}_0   \rho_1                       
  +\sum_{\mu\nu} B^J_{10} \,  J_{1,\mu \nu} J_{0,\mu\nu}  \rho_1
\, ,
\end{eqnarray}
whereas its anomalous part is given by
\begin{eqnarray}
\label{eq:Skyrme_int:3bodyEDF:anormal} 
{\cal E}^{\kappa \kappa \rho} 
& = &\sum_{\frak{a}=1,2}
\bigg\{
B^{\breve{\rho}}_0 \breve{\rho}^*_{1,\frak{a}} \breve{\rho}_{1,\frak{a}} \rho_{0}   
+ B^{\breve{\tau}^*}_0 \breve{\tau}^*_{1,\frak{a}} \breve{\rho}_{1,\frak{a}} \rho_{0}   
+ B^{\breve{\tau}}_0 \breve{\rho}^*_{1,\frak{a}} \breve{\tau}_{1,\frak{a}} \rho_{0}   
+ B^{\breve{\rho} \tau}_0 \breve{\rho}^*_{1,\frak{a}} \breve{\rho}_{1,\frak{a}} \tau_{0}  
+ B^{\nabla \breve{\rho}}_0 (\vec{\nabla} \breve{\rho}^*_{1,\frak{a}}) \cdot (\vec{\nabla} \breve{\rho}_{1,\frak{a}}) \rho_{0}   
 \nonumber \\ 
&   &
+ B^{\nabla \breve{\rho}^* \breve{\rho}}_0 (\vec{\nabla} \breve{\rho}^*_{1,\frak{a}}) \breve{\rho}_{1,\frak{a}} \cdot (\vec{\nabla} \rho_{0}) 
+ B^{\breve{\rho}^*  \nabla \breve{\rho}}_0 \breve{\rho}^*_{1,\frak{a}} (\vec{\nabla} \breve{\rho}_{1,\frak{a}}) \cdot (\vec{\nabla} \rho_{0}) 
+ {\mathrm i} B^{\nabla \breve{\rho}^* j}_0 (\vec{\nabla} \breve{\rho}^*_{1,\frak{a}}) \breve{\rho}_{1,\frak{a}}  \cdot \vec{j}_{0} 
+ {\mathrm i} B^{\nabla \breve{\rho} j}_0 \breve{\rho}^*_{1,\frak{a}} (\vec{\nabla} \breve{\rho}_{1,\frak{a}}) \cdot \vec{j}_{0} 
 \nonumber  \\ 
&  &
+ \sum_{\mu \nu} \big[ 
B^{\breve{J}}_0 \breve{J}^*_{1,\frak{a}, \mu \nu} \breve{J}_{1,\frak{a}, \mu \nu} \rho_{0}   
+ B^{\breve{J}^* \breve{\rho}}_0 \breve{J}^*_{1,\frak{a}, \mu \nu} \breve{\rho}_{1,\frak{a}}  J_{0, \mu \nu} 
+ B^{\breve{\rho}^* \breve{J}}_0 \breve{\rho}^*_{1,\frak{a}}  \breve{J}_{1,\frak{a}, \mu \nu} J_{0, \mu \nu} 
+ {\mathrm i} B^{\nabla \breve{\rho}^* \breve{J}}_0 (\nabla_\mu \breve{\rho}^*_{1,\frak{a}}) \breve{J}_{1,\frak{a}, \mu \nu} s_{0, \nu}   
 \nonumber  \\ 
&  &
+ {\mathrm i} B^{\breve{J}^* \nabla \breve{\rho}}_0 \breve{J}^*_{1,\frak{a}, \mu \nu} (\nabla_\mu \breve{\rho}_{1,\frak{a}}) s_{0, \nu}   
+ {\mathrm i} B^{\breve{J}^* \nabla s}_0 \breve{J}^*_{1,\frak{a}, \mu \nu} \breve{\rho}_{1,\frak{a}}  (\nabla_\mu s_{0, \nu}) 
+ {\mathrm i} B^{\breve{J} \nabla s}_0 \breve{\rho}^*_{1,\frak{a}}  \breve{J}_{1,\frak{a}, \mu \nu}  (\nabla_\mu s_{0, \nu})
\big]
 \nonumber  \\ 
&  &
+ \sum_{\mu \nu \lambda \kappa} \epsilon_{\nu \lambda \kappa}  \big[ 
\,  {\mathrm i} B^{\breve{J}^2 s}_0 \breve{J}^*_{1,\frak{a}, \mu \nu} \breve{J}_{1,\frak{a}, \mu \lambda} s_{0, \kappa}  
\big]
\bigg\}
+ \! \sum_{\frak{a},\frak{b}=1,2} \sum_{\frak{c}=3} \epsilon_{\frak{a}\frak{b}\frak{c}} 
\bigg\{
{\mathrm i} B^{\breve{\rho}}_1 \breve{\rho}^*_{1,\frak{a}} \breve{\rho}_{1,\frak{b}} \rho_{1,\frak{c}}   
+ {\mathrm i} B^{\breve{\tau}^*}_1 \breve{\tau}^*_{1,\frak{a}} \breve{\rho}_{1,\frak{b}} \rho_{1,\frak{c}}   
+ {\mathrm i} B^{\breve{\tau}}_1 \breve{\rho}^*_{1,\frak{a}} \breve{\tau}_{1,\frak{b}} \rho_{1,\frak{c}}   
 \nonumber  \\ 
&  &
+ {\mathrm i} B^{\breve{\rho} \tau}_1 \breve{\rho}^*_{1,\frak{a}} \breve{\rho}_{1,\frak{b}}  \tau_{1,\frak{c}} 
+ {\mathrm i} B^{\nabla \breve{\rho}}_1 (\vec{\nabla} \breve{\rho}^*_{1,\frak{a}}) \cdot (\vec{\nabla} \breve{\rho}_{1,\frak{b}}) \rho_{1,\frak{c}}   
+ {\mathrm i} B^{\nabla \breve{\rho}^* \breve{\rho}}_1 (\vec{\nabla} \breve{\rho}^*_{1,\frak{a}}) \breve{\rho}_{1,\frak{b}} \cdot (\vec{\nabla} \rho_{1,\frak{c}})
+ {\mathrm i} B^{\breve{\rho}^* \nabla \breve{\rho}}_1 \breve{\rho}^*_{1,\frak{a}}  (\vec{\nabla} \breve{\rho}_{1,\frak{b}}) \cdot (\vec{\nabla} \rho_{1,\frak{c}} )
 \nonumber \\
&   & 
+ B^{\nabla \breve{\rho}^* j}_1 (\vec{\nabla} \breve{\rho}^*_{1,\frak{a}}) \breve{\rho}_{1,\frak{b}}  \cdot \vec{j}_{1,\frak{c}}   
+ B^{\nabla \breve{\rho} j}_1  \breve{\rho}^*_{1,\frak{a}}  (\vec{\nabla} \breve{\rho}_{1,\frak{b}}) \cdot \vec{j}_{1,\frak{c}}   
+ \sum_{\mu \nu} \bigg[ 
{\mathrm i} B^{\breve{J}}_1 \breve{J}^*_{1,\frak{a}, \mu \nu} \breve{J}_{1,\frak{b}, \mu \nu} \rho_{1,\frak{c}}   
+ {\mathrm i} B^{\breve{J}^* \breve{\rho}}_1 \breve{J}^*_{1,\frak{a}, \mu \nu} \breve{\rho}_{1,\frak{b}}  J_{1,\frak{c}, \mu \nu}  
 \nonumber  \\ 
&  &
+ {\mathrm i} B^{\breve{\rho}^* \breve{J}}_1 \breve{\rho}^*_{1,\frak{a}}  \breve{J}_{1,\frak{b}, \mu \nu} J_{1,\frak{c}, \mu \nu}  
+ B^{\nabla \breve{\rho}^* \breve{J}}_1  (\nabla_\mu \breve{\rho}^*_{1,\frak{a}}) \breve{J}_{1,\frak{b}, \mu \nu} s_{1,\frak{c}, \nu}   
+ B^{\breve{J}^* \nabla \breve{\rho}}_1 \breve{J}^*_{1,\frak{a}, \mu \nu} (\nabla_\mu \breve{\rho}_{1,\frak{b}}) s_{1,\frak{c}, \nu}   
 \nonumber  \\ 
&  &
+ B^{\breve{J}^* \nabla s}_1 \breve{J}^*_{1,\frak{a}, \mu \nu} \breve{\rho}_{1,\frak{b}}  (\nabla_\mu s_{1,\frak{c}, \nu})
+ B^{\breve{J} \nabla s}_1 \breve{\rho}^*_{1,\frak{a}}  \breve{J}_{1,\frak{b}, \mu \nu}  (\nabla_\mu s_{1,\frak{c}, \nu})
\bigg]
+ \sum_{\mu \nu \lambda \kappa} \epsilon_{\nu \lambda \kappa}  \bigg[ 
\, B^{\breve{J}^2 s}_1 \breve{J}^*_{1,\frak{a}, \mu \nu} \breve{J}_{1,\frak{b}, \mu \lambda} s_{1,\frak{c}, \kappa} \,  
\bigg]
\bigg\}
\, .
\end{eqnarray}
\end{widetext}
Sums over Greek indices run over $x$, $y$, and $z$ components of spatial 
vectors, whereas sums over indices in fractur are over isovector components. 
In the normal part of the trilinear EDF, the notation $\bar{t}$ is such that 
$\bar{t} = 1$ (0) whenever $t=0$ (1). Coupling constants are related to 
pseudo-potential parameters according to 
Tables~\ref{tab:Skyrme_int:3bodyEDF:coeff} 
and~\ref{tab:Skyrme_int:3bodyEDF:coeffP}.

\begin{table}[t!]
\begin{center}
\caption{\label{tab:Skyrme_int:3bodyEDF:coeff}
Same as Tab.~\ref{tab:Skyrme_int:2bodyEDF:coeff} for the normal 
part of the trilinear EDF, Eq.~(\ref{eq:Skyrme_int:3bodyEDF:normal}).
}  
\begin{tabular}{r c c c c c c }
\hline \hline \noalign{\smallskip}
&$u_0$&$u_1$&$u_1y_1$&$u_2$&$u_2y_{21}$&$u_2y_{22}$\\
\noalign{\smallskip}  \hline \noalign{\smallskip}
$B^\rho_0\;=$&$+\frac{3}{16}$&&&&&\\[0.3mm]
$B^\rho_1\;=$&$-\frac{3}{16}$&&&&&\\[0.3mm]
$B^\tau_0\;=$&&$+\frac{3}{32}$&&$+\frac{15}{64}$&$+\frac{3}{16}$&$+\frac{3}{32}$\\[0.3mm]
$B^\tau_{10}\;=$&&$-\frac{1}{32}$&$+\frac{1}{32}$&$-\frac{5}{64}$&$-\frac{1}{16}$&$-\frac{7}{32}$\\[0.3mm]
$B^\tau_1\;=$&&$-\frac{1}{16}$&$-\frac{1}{32}$&$+\frac{1}{32}$&$+\frac{1}{16}$&$-\frac{1}{16}$\\[0.3mm]
$B^{\nabla\rho}_0\;=$&&$+\frac{15}{128}$&&$-\frac{15}{256}$&$-\frac{3}{64}$&$-\frac{3}{128}$\\[0.3mm]
$B^{\nabla\rho}_{10}\;=$&&$-\frac{5}{64}$&$+\frac{1}{32}$&$+\frac{5}{128}$&$+\frac{1}{32}$&$+\frac{7}{64}$\\[0.3mm]
$B^{\nabla\rho}_1\;=$&&$-\frac{5}{128}$&$-\frac{1}{32}$&$-\frac{7}{256}$&$-\frac{1}{32}$&$-\frac{5}{128}$\\[0.3mm]
$B^J_0\;=$&&$+\frac{1}{32}$&$-\frac{1}{16}$&$-\frac{7}{64}$&$-\frac{1}{8}$&$+\frac{1}{32}$\\[0.3mm]
$B^J_{10}\;=$&&$-\frac{1}{16}$&$+\frac{1}{16}$&$+\frac{1}{32}$&&$+\frac{3}{16}$\\[0.3mm]
$B^J_1\;=$&&$+\frac{1}{32}$&&$-\frac{7}{64}$&$-\frac{1}{16}$&$-\frac{1}{32}$\\[0.3mm]
$B^s_0\;=$&$-\frac{3}{16}$&&&&&\\[0.3mm]
$B^s_{10}\;=$&$+\frac{3}{8}$&&&&&\\[0.3mm]
$B^s_1\;=$&$-\frac{3}{16}$&&&&&\\[0.3mm]
$B^T_0\;=$&&$-\frac{1}{16}$&$+\frac{1}{32}$&$+\frac{1}{32}$&$+\frac{1}{16}$&$+\frac{1}{8}$\\[0.3mm]
$B^T_{10}\;=$&&$+\frac{1}{16}$&$-\frac{1}{32}$&$-\frac{1}{32}$&$-\frac{1}{16}$&$-\frac{1}{8}$\\[0.3mm]
$B^T_{01}\;=$&&$+\frac{1}{16}$&&$-\frac{1}{32}$&&\\[0.3mm]
$B^T_1\;=$&&$-\frac{1}{16}$&&$+\frac{1}{32}$&&\\[0.3mm]
$B^{\tau s}_0\;=$&&$-\frac{1}{32}$&$-\frac{1}{32}$&$-\frac{5}{64}$&$-\frac{1}{16}$&$+\frac{5}{32}$\\[0.3mm]
$B^{\tau s}_{10}\;=$&&$-\frac{1}{32}$&&$-\frac{5}{64}$&$-\frac{1}{16}$&$-\frac{1}{32}$\\[0.3mm]
$B^{\tau s}_1\;=$&&$+\frac{1}{16}$&$+\frac{1}{32}$&$-\frac{1}{32}$&$-\frac{1}{16}$&$+\frac{1}{16}$\\[0.3mm]
$B^{\nabla s}_0\;=$&&$-\frac{5}{128}$&$+\frac{1}{32}$&$-\frac{7}{256}$&$-\frac{1}{32}$&$+\frac{1}{128}$\\[0.3mm]
$B^{\nabla s}_{10}\;=$&&$+\frac{5}{64}$&$-\frac{1}{32}$&$+\frac{1}{128}$&&$+\frac{3}{64}$\\[0.3mm]
$B^{\nabla s}_1\;=$&&$-\frac{5}{128}$&&$-\frac{7}{256}$&$-\frac{1}{64}$&$-\frac{1}{128}$\\[0.3mm]
$B^{\nabla\rho s}_0\;=$&&$-\frac{5}{64}$&$-\frac{1}{32}$&$+\frac{5}{128}$&$+\frac{1}{32}$&$-\frac{5}{64}$\\[0.3mm]
$B^{\nabla\rho s}_{01}\;=$&&$-\frac{5}{64}$&&$+\frac{5}{128}$&$+\frac{1}{32}$&$+\frac{1}{64}$\\[0.3mm]
$B^{\nabla\rho s}_{10}\;=$&&$+\frac{5}{64}$&&$+\frac{1}{128}$&$+\frac{1}{32}$&$+\frac{1}{64}$\\[0.3mm]
$B^{\nabla\rho s}_1\;=$&&$+\frac{5}{64}$&$+\frac{1}{32}$&$+\frac{1}{128}$&&$-\frac{3}{64}$\\[0.3mm]
$B^{Js}_0\;=$&&$+\frac{1}{16}$&$+\frac{1}{16}$&$+\frac{5}{32}$&$+\frac{1}{8}$&$-\frac{5}{16}$\\[0.3mm]
$B^{Js}_{01}\;=$&&$+\frac{1}{16}$&&$+\frac{5}{32}$&$+\frac{1}{8}$&$+\frac{1}{16}$\\[0.3mm]
$B^{Js}_{10}\;=$&&$-\frac{1}{16}$&&$+\frac{1}{32}$&$+\frac{1}{8}$&$+\frac{1}{16}$\\[0.3mm]
$B^{Js}_1\;=$&&$-\frac{1}{16}$&$-\frac{1}{16}$&$+\frac{1}{32}$&&$-\frac{3}{16}$\\[0.3mm]
$B^{\nabla sJ}_{0}\;=$&&&&$-\frac{3}{64}$&$-\frac{3}{32}$&$+\frac{3}{32}$\\[0.3mm]
$B^{\nabla sJ}_{01}\;=$&&&$+\frac{1}{16}$&$-\frac{3}{64}$&$-\frac{1}{32}$&$+\frac{1}{32}$\\[0.3mm]
$B^{\nabla sJ}_{10}\;=$&&&$-\frac{1}{32}$&$-\frac{3}{64}$&$-\frac{1}{32}$&$+\frac{1}{32}$\\[0.3mm]
$B^{\nabla sJ}_{1}\;=$&&&$-\frac{1}{32}$&$-\frac{3}{64}$&$-\frac{1}{32}$&$+\frac{1}{32}$\\[0.3mm]
$B^j_0\;=$&&$-\frac{3}{32}$&&$-\frac{15}{64}$&$-\frac{3}{16}$&$-\frac{3}{32}$\\[0.3mm]
$B^j_{10}\;=$&&$+\frac{1}{16}$&$-\frac{1}{16}$&$+\frac{5}{32}$&$+\frac{1}{8}$&$+\frac{7}{16}$\\[0.3mm]
$B^j_1\;=$&&$+\frac{1}{32}$&$+\frac{1}{16}$&$-\frac{7}{64}$&$-\frac{1}{8}$&$-\frac{5}{32}$\\[1.5mm]
\hline \hline
\end{tabular}
\end{center}
\end{table}

\begin{table}[t!]
\begin{center}
\caption{\label{tab:Skyrme_int:3bodyEDF:coeffP}
Same as Tab.~\ref{tab:Skyrme_int:2bodyEDF:coeff}, but for the anomalous 
part of the trilinear EDF kernel, Eq.~(\ref{eq:Skyrme_int:3bodyEDF:anormal}).
}
\begin{tabular}{r c c c c c c }
\hline \hline \noalign{\smallskip}
 &$u_0$&$u_1$&$u_1y_1$&$u_2$&$u_2y_{21}$&$u_2y_{22}$\\
\noalign{\smallskip}  \hline \noalign{\smallskip}
$B^{\breve{\rho}}_0=$&$+\frac{3}{16}$&&&&&\\[0.3mm]
$B^{\breve{\tau}^*}_0=$&&$+\frac{3}{64}$&$-\frac{3}{128}$&&&\\[0.3mm]
$B^{\breve{\tau}}_0=$&&$+\frac{3}{64}$&$-\frac{3}{128}$&&&\\[0.3mm]
$B^{\breve{\rho} \tau}_0=$&&$+\frac{1}{32}$&$+\frac{1}{64}$&$+\frac{5}{64}$&$+\frac{1}{16}$&$-\frac{1}{16}$\\[0.3mm]
$B^{\nabla \breve{\rho}}_0=$&&$+\frac{1}{32}$&$-\frac{1}{128}$&$+\frac{5}{256}$&$+\frac{1}{64}$&$-\frac{1}{64}$\\[0.3mm]
$B^{\nabla \breve{\rho}^* \breve{\rho}}_0=$&&$+\frac{5}{128}$&$+\frac{1}{128}$&$-\frac{5}{256}$&$-\frac{1}{64}$&$+\frac{1}{64}$\\[0.3mm]
$B^{\breve{\rho}^*  \nabla \breve{\rho}}_0=$&&$+\frac{5}{128}$&$+\frac{1}{128}$&$-\frac{5}{256}$&$-\frac{1}{64}$&$+\frac{1}{64}$\\[0.3mm]
$B^{\nabla \breve{\rho}^* j}_0=$&&$-\frac{1}{64}$&$-\frac{1}{128}$&$-\frac{5}{128}$&$-\frac{1}{32}$&$+\frac{1}{32}$\\[0.3mm]
$B^{\nabla \breve{\rho} j}_0=$&&$+\frac{1}{64}$&$+\frac{1}{128}$&$+\frac{5}{128}$&$+\frac{1}{32}$&$-\frac{1}{32}$\\[0.3mm]
$B^{\breve{J}}_0=$&&&&$+\frac{9}{64}$&$+\frac{1}{8}$&$+\frac{1}{16}$\\[0.3mm]
$B^{\breve{J}^* \breve{\rho}}_0=$&&&$-\frac{1}{64}$&$-\frac{3}{64}$&$-\frac{1}{16}$&$+\frac{1}{16}$\\[0.3mm]
$B^{\breve{\rho}^* \breve{J}}_0=$&&&$-\frac{1}{64}$&$-\frac{3}{64}$&$-\frac{1}{16}$&$+\frac{1}{16}$\\[0.3mm]
$B^{\nabla \breve{\rho}^* \breve{J}}_0=$&&&$+\frac{1}{128}$&$+\frac{3}{128}$&$+\frac{1}{32}$&$-\frac{1}{32}$\\[0.3mm]
$B^{\breve{J}^* \nabla \breve{\rho}}_0=$&&&$-\frac{1}{128}$&$-\frac{3}{128}$&$-\frac{1}{32}$&$+\frac{1}{32}$\\[0.3mm]
$B^{\breve{J}^* \nabla s}_0=$&&&$-\frac{1}{64}$&$+\frac{3}{128}$&$+\frac{1}{32}$&$-\frac{1}{32}$\\[0.3mm]
$B^{\breve{J} \nabla s}_0=$&&&$+\frac{1}{64}$&$-\frac{3}{128}$&$-\frac{1}{32}$&$+\frac{1}{32}$\\[0.3mm]
$B^{\breve{J}^2 s}_0=$&&&&$+\frac{3}{64}$&$+\frac{1}{32}$&$-\frac{1}{8}$\\[0.3mm]
$B^{\breve{\rho}}_1=$&$-\frac{3}{16}$&&&&&\\[0.3mm]
$B^{\breve{\tau}^*}_1=$&&$-\frac{3}{64}$&$+\frac{3}{128}$&&&\\[0.3mm]
$B^{\breve{\tau}}_1=$&&$-\frac{3}{64}$&$+\frac{3}{128}$&&&\\[0.3mm]
$B^{\breve{\rho} \tau}_1=$&&$-\frac{1}{32}$&$-\frac{1}{64}$&$+\frac{1}{64}$&$+\frac{1}{32}$&$-\frac{1}{32}$\\[0.3mm]
$B^{\nabla \breve{\rho}}_1=$&&$-\frac{1}{32}$&$+\frac{1}{128}$&$+\frac{1}{256}$&$+\frac{1}{128}$&$-\frac{1}{128}$\\[0.3mm]
$B^{\nabla \breve{\rho}^* \breve{\rho}}_1=$&&$-\frac{5}{128}$&$-\frac{1}{128}$&$-\frac{1}{256}$&$-\frac{1}{128}$&$+\frac{1}{128}$\\[0.3mm]
$B^{\breve{\rho}^* \nabla \breve{\rho}}_1=$&&$-\frac{5}{128}$&$-\frac{1}{128}$&$-\frac{1}{256}$&$-\frac{1}{128}$&$+\frac{1}{128}$\\[0.3mm]
$B^{\nabla \breve{\rho}^* j}_1=$&&$-\frac{1}{64}$&$-\frac{1}{128}$&$+\frac{1}{128}$&$+\frac{1}{64}$&$-\frac{1}{64}$\\[0.3mm]
$B^{\nabla \breve{\rho} j}_1=$&&$+\frac{1}{64}$&$+\frac{1}{128}$&$-\frac{1}{128}$&$-\frac{1}{64}$&$+\frac{1}{64}$\\[0.3mm]
$B^{\breve{J}}_1=$&&&&$-\frac{3}{64}$&$-\frac{1}{32}$&$-\frac{5}{32}$\\[0.3mm]
$B^{\breve{J}^* \breve{\rho}}_1=$&&&$+\frac{1}{64}$&$-\frac{3}{64}$&$-\frac{1}{32}$&$+\frac{1}{32}$\\[0.3mm]
$B^{\breve{\rho}^* \breve{J}}_1=$&&&$+\frac{1}{64}$&$-\frac{3}{64}$&$-\frac{1}{32}$&$+\frac{1}{32}$\\[0.3mm]
$B^{\nabla \breve{\rho}^* \breve{J}}_1=$&&&$+\frac{1}{128}$&$-\frac{3}{128}$&$-\frac{1}{64}$&$+\frac{1}{64}$\\[0.3mm]
$B^{\breve{J}^* \nabla \breve{\rho}}_1=$&&&$-\frac{1}{128}$&$+\frac{3}{128}$&$+\frac{1}{64}$&$-\frac{1}{64}$\\[0.3mm]
$B^{\breve{J}^* \nabla s}_1=$&&&$-\frac{1}{64}$&$-\frac{3}{128}$&$-\frac{1}{64}$&$+\frac{1}{64}$\\[0.3mm]
$B^{\breve{J} \nabla s}_1=$&&&$+\frac{1}{64}$&$+\frac{3}{128}$&$+\frac{1}{64}$&$-\frac{1}{64}$\\[0.3mm]
$B^{\breve{J}^2 s}_1=$&&&&$-\frac{3}{64}$&$-\frac{1}{16}$&$-\frac{1}{32}$\\[1.5mm]
\hline \hline
\end{tabular}
\end{center}
\end{table}


\subsubsection{Discussion}

A few further comments on the structure of bilinear and trilinear
contributions to the EDF kernel are in order.

In the case of pure proton and neutron density matrices, as considered here, 
only a pairing functional of isovector character 
remains~\cite{perlinska04a,rohozinski10a}, as all isoscalar pair densities 
are zero, see Sect.~\ref{sect:densgoodT}. The generic isospin structure of 
the terms containing isovector densities is
\begin{subequations}
\label{eq:Skyrme_int:isoscalar_isovector_pairing:restriction}
\begin{align}
\label{eq:Skyrme_int:isoscalar_isovector_pairing:restriction:1}
\sum_{\frak{a}} \mathcal{P}_{1,\frak{a}} \, \mathcal{P}'_{1,\frak{a}} 
& = \mathcal{P}_{1,3}\, \mathcal{P}'_{1,3} \, , \\
\label{eq:Skyrme_int:isoscalar_isovector_pairing:restriction:2}
\sum_{\frak{a}} \breve{\mathcal{P}}^{*}_{1,\frak{a}} \, 
\breve{\mathcal{P}}'_{1,\frak{a}} 
&=   \breve{\mathcal{P}}^{*}_{1,1} \, \breve{\mathcal{P}}'_{1,1} 
   + \breve{\mathcal{P}}^{*}_{1,2} \, \breve{\mathcal{P}}'_{1,2} \,, \\
\label{eq:Skyrme_int:isoscalar_isovector_pairing:restriction:3}
\sum_{\frak{a}} \mathcal{P}_{1,\frak{a}} \, \mathcal{P}'_{1,\frak{a}} \, 
\mathcal{P}''_{0}
& = \mathcal{P}_{1,3}\, \mathcal{P}'_{1,3} \, \mathcal{P}''_{0} \, , \\
\label{eq:Skyrme_int:isoscalar_isovector_pairing:restriction:4}
\sum_{\frak{a}} \breve{\mathcal{P}}^{*}_{1,\frak{a}} \, 
\breve{\mathcal{P}}'_{1,\frak{a}} \, \mathcal{P}''_{0} 
&= \big(  \breve{\mathcal{P}}^{*}_{1,1} \, \breve{\mathcal{P}}'_{1,1} 
        + \breve{\mathcal{P}}^{*}_{1,2} \, \breve{\mathcal{P}}'_{1,2} 
   \big) \, \mathcal{P}''_{0} \, , \\
\label{eq:Skyrme_int:isoscalar_isovector_pairing:restriction:5}
\sum_{\frak{a}\frak{b}\frak{c}} 
\epsilon_{\frak{a}\frak{b}\frak{c}} 
\breve{\mathcal{P}}^{*}_{1,\frak{a}} \,
\breve{\mathcal{P}}'_{1,\frak{b}} \,
\mathcal{P}''_{1,\frak{c}} 
&= \big( \breve{\mathcal{P}}^{*}_{1,1} \, \breve{\mathcal{P}}'_{1,2} 
       - \breve{\mathcal{P}}^{*}_{1,2} \, \breve{\mathcal{P}}'_{1,1} 
    \big) \,
    \mathcal{P}''_{1,3} \, .
\end{align}
\end{subequations}
Equations~(\ref{eq:Skyrme_int:isoscalar_isovector_pairing:restriction:1}) to 
(\ref{eq:Skyrme_int:isoscalar_isovector_pairing:restriction:4})
correspond to scalar products of two isovectors coupled to
isospin zero (and which might be multiplied by a normal isoscalar density),
whereas Eq.~(\ref{eq:Skyrme_int:isoscalar_isovector_pairing:restriction:5}) 
displays a triple product of three isovectors that are thereby also 
coupled to an isoscalar.

As all pair densities are in general complex, all terms containing two 
different pair densities $\breve{\mathcal{P}}^{*}$ and $\breve{\mathcal{P}}'$
take the form $\breve{\mathcal{P}}^{*} \, \breve{\mathcal{P}}' 
+ \breve{\mathcal{P}} \, \breve{\mathcal{P}}^{\prime *}$ in order for the 
functional kernel to be real.

The bilinear part of the functional does not contain all possible
combinations of local densities compatible with spatial symmetries
\cite{doba96b,perlinska04a}. Indeed, some of those combinations only emerge 
in the functional derived from spin-orbit and tensor 
forces~\cite{lesinski07a,hellemans12a}. The same applies to the trilinear 
part of the functional. We postpone the discussion of spin-orbit and tensor 
terms to a future publication~\cite{sadoudi:LS+tensor}.

There are two equivalent ways of writing the terms with derivatives of local 
densities in the bilinear part of the EDF, i.e.\ the third and last term 
in Eq.~(\ref{eq:Skyrme_int:2bodyEDF:normal}) and the last term in 
Eq.~(\ref{eq:Skyrme_int:2bodyEDF:anormal}), that differ from each other by 
an integration by parts~\cite{doba96b}. Usually, these terms are expressed 
in terms of 
Laplacians~\cite{bender03a,doba96b,bender09b,lesinski07a,hellemans12a}. 
Trilinear terms, however, do not offer such a freedom. For internal 
consistency, we thus define associated terms in the bilinear part of 
the EDF in terms of first derivatives of local densities, at variance with 
most of the literature.

Trilinear terms ${\cal E}^{\rho \rho \rho}$ and 
${\cal E}^{\kappa \kappa \rho}$ display a much more complex structure 
than what would be obtained by adding a mere density dependence to the
coupling constants entering ${\cal E}^{\rho \rho}$ and 
${\cal E}^{\kappa \kappa}$. Although one can find trilinear terms that do 
have the structure of terms appearing in the bilinear part of the functional 
times $\rho_0(\vec{r})$, relative weights between isoscalar and isovector 
terms, or between time-even, time-odd and pairing terms, are not the same as 
in their bilinear counterparts. This is a consequence of Pauli's exclusion 
principle that is fully preserved for energy functionals deriving from a 
three-body pseudo-potential, but violated for functionals deriving from 
density-dependent two-body interactions; see the discussion in 
Refs.~\cite{stringari78a,waroquier83a} regarding terms without gradients.

Keeping the coupling constants consistent with 
Tables~\ref{tab:Skyrme_int:2bodyEDF:coeff}, 
\ref{tab:Skyrme_int:2bodyEDF:coeffP}, \ref{tab:Skyrme_int:3bodyEDF:coeff}, 
and~\ref{tab:Skyrme_int:3bodyEDF:coeffP}, 
the energy functional~(\ref{eq:Skyrme_int:1bodyEDF}) - 
(\ref{eq:Skyrme_int:3bodyEDF:anormal}) is invariant under arbitrary local 
gauge transformations, see Appendix~\ref{sec:Skyrme_int:Gauge}. This property 
indicates the local character of the underlying pseudo 
potential~\cite{blaizotripka}. Galilean invariance, which is a necessary
requirement for interactions to be meaningfully used in dynamical calculations
such as time-dependent HF, is one special case of the more general invariance 
under arbitrary gauge transformations and therefore automatically fulfilled.

The first critical check that the formal algebra code is proceeding 
correctly is provided by the fact that well-known results (EDF, one-body 
fields, infinite matter properties) associated with the central part of 
the (density-independent) two-body Skyrme interaction are recovered. 
For the trilinear part, two non-trivial, though indirect, tests give us 
further confidence, i.e.\ (i) the local gauge invariance of the EDF kernel 
is exactly fulfilled as mentioned above and (ii) the numerical code that 
tracks both the real and imaginary parts of the EDF kernel computes the
latter to be strictly zero for each hermitian piece of the pseudo potential.

Properties of homogeneous nuclear matter, both symmetric and asymmetric in 
isospin and/or spin, along with Landau parameters, are discussed in 
in Appendix~\ref{sect:INM}. Additionally, the energy functional in the 
often used proton-neutron representation  is provided in 
Appendix~\ref{sec:Skyrme_int:EDF:np}, whereas the expressions for the 
associated one-body fields entering the Hartree-Fock-Bogoliubov equations 
of motion are listed in Appendix~\ref{Skyrme_int:fields}.

%
\section{Conclusions and outlook}
\label{sect:concl}

We have constructed the most general central Skyrme-type three-body 
pseudo-potential containing up to two derivatives, derived the corresponding 
energy density functional (i.e.\ time-even and time-odd contributions to 
the normal part of the EDF along with the complete anomalous part) and 
one-body fields as well as computed an extensive set of infinite nuclear 
matter properties. Our objective is to build EDF parametrizations that 
derive strictly from (density-independent) two- and three-body Skyrme-like 
pseudo potentials as required for spuriousity-free multi-reference 
calculations.

The main observations and conclusions of the present work are that
\begin{itemize}
\item
The central three-body pseudo potential is defined out of six independent 
parameters in total. Combined with the central part of the two-body Skyrme 
pseudo-potential, this leads to a total of twelve parameters priori
to considering spin-orbit and tensor terms. 

\item
The structure of some of the three-body terms containing gradients cannot 
be conjectured by just inserting an additional delta function into a 
two-body Skyrme interaction of standard form. 

\item
The EDF trilinear kernel possesses a much more complex structure than 
the functional resulting from a density-dependent two-body vertex, in 
particular as far as time-odd and pairing parts are concerned. 

\end{itemize}
The main points for future studies are
\begin{itemize}

\item 
The discussion regarding three-body spin-orbit and tensor pseudo 
potentials constructed along the same lines as here will be given 
elsewhere~\cite{sadoudi:LS+tensor}.  These do not contribute to bulk 
properties of non-polarized nuclear matter, but can be used to fine-tune 
the nucleon-number dependence of the shell structure with more freedom 
than when using two-body spin-orbit and tensor interactions 
only~\cite{lesinski07a,bender09b}.

\item 
A first tentative adjustment of the parameters of the newly derived 
EDF kernel, complemented by the Coulomb interaction, is currently 
underway~\cite{bennaceur:fit}. The most important question to answer before 
proceeding further regards the capacity of the presently-developed 
pseudo-potential-based EDF to give a satisfying description of bulk 
properties of nuclei, including pairing correlations.

\item 
When combined with a Coulomb energy functional that contains exact 
exchange and pairing contributions, the pseudo-potential-based 
EDF constructed here can be used without ambiguities in multi-reference EDF 
calculations performing symmetry restoration and/or configuration mixing 
based on the generator coordinate method. 

\end{itemize}
%

%
%
\acknowledgments

Many inspiring discussions with K.~Bennaceur are gratefully acknowledged. 
This work has been supported by the Agence Nationale de la Recherche 
under grant no.\ ANR 2010 BLANC 0407 "NESQ".

%
%
\begin{appendix}

\section{Neutron-proton representation of the EDF}
\label{sec:Skyrme_int:EDF:np}

A widely used alternative to the representation of the EDF kernel
in terms of isoscalar and isovector densities presented in 
Sec.~\ref{sec:Skyrme_int:EDF:Result} is a representation in terms 
of proton and neutron densities.

\subsection{Energy density}


\subsubsection{Linear part}

The kinetic energy density is given by
\begin{equation}
\label{eq:1bodyEDF:np}
{\cal E}^{\rho} 
= \frac{\hbar^2}{2m} \sum_{q} \tau_{q}   \,.
\end{equation}

\begin{table}[t!]
\begin{center}
\caption{\label{tab:2bodyOut:npCoeff}
Coupling constants of the normal bilinear part of the EDF in 
neutron-proton representation, Eq.~(\ref{ErhorhoNP}), as a function of the 
pseudo-potential parameters of Eqs.~(\ref{eq:Skyrme_int:2body_int:t0}) to 
(\ref{eq:Skyrme_int:2body_int:t2}). Missing entries are zero.
}
\begin{tabular}{r c c c c c c }
\hline \hline \noalign{\smallskip}
&$t_0$&$t_0x_0$&$t_1$&$t_1x_1$&$t_2$&$t_2x_2$\\
\noalign{\smallskip}  \hline \noalign{\smallskip}
$A^{\rho_1 \rho_1}=$&$+\frac{1}{4}$&$-\frac{1}{4}$&&&&\\[0.3mm]
$A^{\rho_1 \rho_2}=$&$+\frac{1}{2}$&$+\frac{1}{4}$&&&&\\[0.3mm]
$A^{s_1 s_1}=$&$-\frac{1}{4}$&$+\frac{1}{4}$&&&&\\[0.3mm]
$A^{s_1 s_2}=$&&$+\frac{1}{4}$&&&&\\[0.3mm]
$A^{\tau_1 \rho_1}=$&&&$+\frac{1}{8}$&$-\frac{1}{8}$&$+\frac{3}{8}$&$+\frac{3}{8}$\\[0.3mm]
$A^{\tau_1 \rho_2}=$&&&$+\frac{1}{4}$&$+\frac{1}{8}$&$+\frac{1}{4}$&$+\frac{1}{8}$\\[0.3mm]
$A^{T_1 s_1}=$&&&$-\frac{1}{8}$&$+\frac{1}{8}$&$+\frac{1}{8}$&$+\frac{1}{8}$\\[0.3mm]
$A^{T_1 s_2}=$&&&&$+\frac{1}{8}$&&$+\frac{1}{8}$\\[0.3mm]
$A^{\nabla \rho_1 \nabla \rho_1}=$&&&$+\frac{3}{32}$&$-\frac{3}{32}$&$-\frac{3}{32}$&$-\frac{3}{32}$\\[0.3mm]
$A^{\nabla \rho_1 \nabla \rho_2}=$&&&$+\frac{3}{16}$&$+\frac{3}{32}$&$-\frac{1}{16}$&$-\frac{1}{32}$\\[0.3mm]
$A^{\nabla s_1 \nabla s_1}=$&&&$-\frac{3}{32}$&$+\frac{3}{32}$&$-\frac{1}{32}$&$-\frac{1}{32}$\\[0.3mm]
$A^{\nabla s_1 \nabla s_2}=$&&&&$+\frac{3}{32}$&&$-\frac{1}{32}$\\[0.3mm]
$A^{j_1 j_1}=$&&&$-\frac{1}{8}$&$+\frac{1}{8}$&$-\frac{3}{8}$&$-\frac{3}{8}$\\[0.3mm]
$A^{j_1 j_2}=$&&&$-\frac{1}{4}$&$-\frac{1}{8}$&$-\frac{1}{4}$&$-\frac{1}{8}$\\[0.3mm]
$A^{J_1 J_1}=$&&&$+\frac{1}{8}$&$-\frac{1}{8}$&$-\frac{1}{8}$&$-\frac{1}{8}$\\[0.3mm]
$A^{J_1 J_2}=$&&&&$-\frac{1}{8}$&&$-\frac{1}{8}$\\
\noalign{\smallskip} \hline \hline
\end{tabular}
\end{center}
\end{table}

\begin{table}[t!]
\begin{center}
\caption{\label{tab:2bodyOut:npCoeffP}
Same as Table~\ref{tab:2bodyOut:npCoeff}, but for the anomalous bilinear part 
of the EDF, Eq.~(\ref{EkappakappaNP}).
}
\begin{tabular}{r c c c c c c }
\hline \hline \noalign{\smallskip}
&$t_0$&$t_0x_0$&$t_1$&$t_1x_1$&$t_2$&$t_2x_2$ \\
\noalign{\smallskip}  \hline \noalign{\smallskip}
$A^{\tilde{\rho}^*_1 \tilde{\rho}_1}=$&$+\frac{1}{4}$&$-\frac{1}{4}$&&&&\\[0.3mm]
$A^{\tilde{\tau}^*_1 \tilde{\rho}_1}=$&&&$+\frac{1}{8}$&$-\frac{1}{8}$&&\\[0.3mm]
$A^{\tilde{\tau}_1 \tilde{\rho}^*_1}=$&&&$+\frac{1}{8}$&$-\frac{1}{8}$&&\\[0.3mm]
$A^{\nabla \tilde{\rho}^*_1 \nabla \tilde{\rho}_1}=$&&&$+\frac{1}{16}$&$-\frac{1}{16}$&&\\[0.3mm]
$A^{\tilde{J}^*_1 \tilde{J}_1}_1=$&&&&&$+\frac{1}{4}$&$+\frac{1}{4}$\\
\noalign{\smallskip} \hline \hline
\end{tabular}
\end{center}
\end{table}


\subsubsection{Bilinear part}

The normal part of the bilinear energy density reads 
\begin{align}
\label{ErhorhoNP}
{\cal E}^{\rho \rho} 
&= \sum_{q} \Big\{ \nonumber
  A^{\rho_1 \rho_1} \rho_{q} \rho_{q}   
+ A^{\rho_1 \rho_2} \rho_{q} \rho_{\bar{q}}   
+ A^{s_1 s_1} \vec{s}_{q} \cdot \vec{s}_{q}   
 \\ &\nonumber 
+ A^{s_1 s_2} \vec{s}_{q} \cdot \vec{s}_{\bar{q}}   
+ A^{\tau_1 \rho_1} \tau_{q} \rho_{q}   
+ A^{\tau_1 \rho_2} \tau_{q} \rho_{\bar{q}}   
 \\ &\nonumber 
+ A^{T_1 s_1} \vec{T}_{q} \cdot \vec{s}_{q}   
+ A^{T_1 s_2} \vec{T}_{q} \cdot \vec{s}_{\bar{q}}   
 \\ 
& \nonumber 
+ A^{\nabla \rho_1 \nabla \rho_1} (\vec{\nabla} \rho_{q}) \cdot (\vec{\nabla} \rho_{q})   
+ A^{\nabla \rho_1 \nabla \rho_2} (\vec{\nabla} \rho_{q}) \cdot (\vec{\nabla} \rho_{\bar{q}})   
 \\ 
& \nonumber 
+ \sum_{\mu\nu} \Big[
    A^{\nabla s_1 \nabla s_1} (\nabla_\mu s_{q, \nu}) (\nabla_\mu s_{q, \nu})   
 \\ 
& \nonumber 
+ A^{\nabla s_1 \nabla s_2} (\nabla_\mu s_{q, \nu}) (\nabla_\mu s_{\bar{q}, \nu})   
 \\ &\nonumber 
+ A^{J_1 J_1} J_{q, \mu \nu} J_{q, \mu \nu}   
+ A^{J_1 J_2} J_{q, \mu \nu} J_{\bar{q}, \mu \nu}   \Big]
 \\ &
+ A^{j_1 j_1} \vec{j}_{q} \cdot \vec{j}_{q}   
+ A^{j_1 j_2} \vec{j}_{q} \cdot \vec{j}_{\bar{q}}   
\Big\}
 \,,  
\end{align}
whereas its anomalous part takes the form
\begin{eqnarray}
\label{EkappakappaNP}
{\cal E}^{\kappa \kappa} 
& = &
\sum_{q} \Big[
  A^{\tilde{\rho}^*_1 \, \tilde{\rho}_1} \, \tilde{\rho}^*_{q} \tilde{\rho}_{q}   
+ A^{\tilde{\tau}^*_1 \, \tilde{\rho}_1} \, \tilde{\tau}^*_{q} \tilde{\rho}_{q}   
+ A^{\tilde{\tau}_1 \, \tilde{\rho}^*_1} \, \tilde{\tau}_{q} \tilde{\rho}^*_{q}   
  \nonumber \\ 
&  & + A^{\nabla \tilde{\rho}^*_1 \nabla \tilde{\rho}_1} 
  (\vec{\nabla} \tilde{\rho}^*_{q}) \cdot (\vec{\nabla} \tilde{\rho}_{q})
   \nonumber \\ 
&  & + \sum_{\mu\nu} 
   A^{\tilde{J}^*_1 \tilde{J}_1}_1 \, \tilde{J}^*_{q, \mu \nu} \tilde{J}_{q, \mu \nu}   
\Big] \, . 
\end{eqnarray}
Index $\bar{q}$ appearing in the sums over nucleon species $q$ denotes 
nucleons of the other kind \mbox{$\bar{q} \neq q$}. The relation between 
the parameters of the pseudo-potential and the coefficients of the energy 
functional are given in Tables~\ref{tab:2bodyOut:npCoeff} 
and~\ref{tab:2bodyOut:npCoeffP}. 


\subsubsection{Trilinear part}

The normal part of the trilinear energy density reads 
\begin{widetext}
\begin{eqnarray}
\label{ErhorhorhoNP}
{\cal E}^{\rho \rho \rho} 
&= & 
\sum_{q} \Big\{
 B^{\rho_1 \rho_1 \rho_2} \rho_{q} \rho_{q} \rho_{\bar{q}}   
+ B^{s_1 s_1 \rho_2} \vec{s}_{q} \cdot \vec{s}_{q} \rho_{\bar{q}}  
+ B^{\tau_1 \rho_1 \rho_2} \tau_{q} \rho_{q} \rho_{\bar{q}}   
+ B^{\tau_1 \rho_1 \rho_1} \tau_{q} \rho_{q} \rho_{q}  
+ B^{\tau_1 \rho_2 \rho_2} \tau_{q} \rho_{\bar{q}} \rho_{\bar{q}}   
+ B^{T_1 s_1 \rho_2} \vec{T}_{q} \cdot \vec{s}_{q} \rho_{\bar{q}}   
\nonumber  \\ 
& &
+ B^{T_1 s_2 \rho_1} \vec{T}_{q} \cdot \vec{s}_{\bar{q}} \rho_{q}   
+ B^{\tau_1 s_1 s_1} \tau_{q} \vec{s}_{q} \cdot \vec{s}_{q}   
+ B^{\tau_1 s_1 s_2} \tau_{q} \vec{s}_{q} \cdot \vec{s}_{\bar{q}}   
+ B^{\tau_1 s_2 s_2} \tau_{q} \vec{s}_{\bar{q}} \cdot \vec{s}_{\bar{q}}   
+ B^{\nabla \rho_1 \nabla \rho_1 \rho_1} (\vec{\nabla} \rho_{q}) \cdot (\vec{\nabla} \rho_{q}) \rho_{q}
\nonumber  \\ 
&  &
+ B^{\nabla \rho_1 \nabla \rho_1 \rho_2} (\vec{\nabla} \rho_{q}) \cdot (\vec{\nabla} \rho_{q}) \rho_{\bar{q}}   
+ B^{\nabla \rho_1 \nabla \rho_2 \rho_1} (\vec{\nabla} \rho_{q}) \cdot (\vec{\nabla} \rho_{\bar{q}}) \rho_{q}   
+ B^{j_1 j_1 \rho_1} \vec{j}_{q} \cdot \vec{j}_{q} \rho_{q}   
+ B^{j_1 j_1 \rho_2} \vec{j}_{q} \cdot \vec{j}_{q} \rho_{\bar{q}}   
+ B^{j_1 j_2 \rho_1} \vec{j}_{q} \cdot \vec{j}_{\bar{q}} \rho_{q}   
 \nonumber  \\ 
& & 
+ \sum_{\mu\nu} \Big[ 
B^{\nabla s_1 \nabla s_1 \rho_1} (\nabla_\mu s_{q, \nu}) (\nabla_\mu s_{q, \nu}) \rho_{q}   
+ B^{\nabla s_1 \nabla s_1 \rho_2} (\nabla_\mu s_{q, \nu}) (\nabla_\mu s_{q, \nu}) \rho_{\bar{q}}   
+ B^{\nabla s_1 \nabla s_2 \rho_1} (\nabla_\mu s_{q, \nu}) (\nabla_\mu s_{\bar{q}, \nu}) \rho_{q}   
 \nonumber  \\ 
& &
+ B^{\nabla \rho_1 \nabla s_1 s_1} (\nabla_\mu \rho_{q}) (\nabla_\mu s_{q, \nu}) s_{q, \nu}   
+ B^{\nabla \rho_1 \nabla s_1 s_2} (\nabla_\mu \rho_{q}) (\nabla_\mu s_{q, \nu}) s_{\bar{q}, \nu}   
+ B^{\nabla \rho_1 \nabla s_2 s_1} (\nabla_\mu \rho_{q}) (\nabla_\mu s_{\bar{q}, \nu}) s_{q, \nu}   
 \nonumber  \\ 
& &
+ B^{\nabla \rho_1 \nabla s_2 s_2} (\nabla_\mu \rho_{q}) (\nabla_\mu s_{\bar{q}, \nu}) s_{\bar{q}, \nu}   
+ B^{J_1 J_1 \rho_1} J_{q, \mu \nu} J_{q, \mu \nu} \rho_{q}   
+ B^{J_1 J_1 \rho_2} J_{q, \mu \nu} J_{q, \mu \nu} \rho_{\bar{q}}   
+ B^{J_1 J_2 \rho_1} J_{q, \mu \nu} J_{\bar{q}, \mu \nu} \rho_{q}   
 \nonumber  \\ 
& &
+ B^{j_1 J_1 s_1} j_{q,\mu} J_{q, \mu \nu} s_{q, \nu}   
+ B^{j_1 J_1 s_2} j_{q,\mu} J_{q, \mu \nu} s_{\bar{q}, \nu}   
+ B^{j_1 J_2 s_1} j_{q,\mu} J_{\bar{q}, \mu \nu} s_{q, \nu}   
+ B^{j_1 J_2 s_2} j_{q,\mu} J_{\bar{q}, \mu \nu} s_{\bar{q}, \nu}   
\Big]
 \nonumber  \\ 
& &
 + \sum_{\mu\nu\lambda\kappa} \, \epsilon_{\nu\lambda\kappa}  \, \Big[ 
   B^{\nabla s_1 J_1 s_1} (\nabla_\mu s_{q, \nu}) J_{q, \mu \lambda} s_{q, \kappa}  \,  
+ B^{\nabla s_1 J_1 s_2} (\nabla_\mu s_{q, \nu}) J_{q, \mu \lambda} s_{\bar{q}, \kappa}  \, 
+ B^{\nabla s_1 J_2 s_1} (\nabla_\mu s_{q, \nu}) J_{\bar{q}, \mu \lambda} s_{q, \kappa}  \, 
\nonumber  \\ 
& &
+ B^{\nabla s_1 J_2 s_2} (\nabla_\mu s_{q, \nu}) J_{\bar{q}, \mu \lambda} s_{\bar{q}, \kappa}  \, 
\Big]
\Big\}  \,,
\end{eqnarray}
whereas its anomalous part is given by
\begin{eqnarray}
\label{ErhokappakappaNP}
 {\cal E}^{\kappa \kappa \rho} 
&= & \sum_{q} \Big\{
  B^{\tilde{\rho}^*_1 \tilde{\rho}_1 \rho_2} \tilde{\rho}^*_{q} \tilde{\rho}_{q} \rho_{\bar{q}}   
+ B^{\tau_1 \tilde{\rho}^*_1 \tilde{\rho}_1} \tau_{q} \tilde{\rho}^*_{q} \tilde{\rho}_{q}   
+ B^{\tau_2 \tilde{\rho}^*_1 \tilde{\rho}_1} \tau_{\bar{q}} \tilde{\rho}^*_{q} \tilde{\rho}_{q}   
+ B^{\tilde{\tau}^*_1 \tilde{\rho}_1 \rho_2} \tilde{\tau}^*_{q} \tilde{\rho}_{q} \rho_{\bar{q}}   
+ B^{\tilde{\tau}_1 \tilde{\rho}^*_1 \rho_2} \tilde{\tau}_{q} \tilde{\rho}^*_{q} \rho_{\bar{q}}   
\nonumber \\ 
& &
+ B^{\nabla \tilde{\rho}^*_1 \nabla \tilde{\rho}_1 \rho_1} (\vec{\nabla} \tilde{\rho}^*_{q}) \cdot (\vec{\nabla} \tilde{\rho}_{q}) \rho_{q}   
+ B^{\nabla \tilde{\rho}^*_1 \nabla \tilde{\rho}_1 \rho_2} (\vec{\nabla} \tilde{\rho}^*_{q}) \cdot (\vec{\nabla} \tilde{\rho}_{q}) \rho_{\bar{q}}   
+ B^{\nabla \tilde{\rho}^*_1 \nabla \rho_1 \tilde{\rho}_1} (\vec{\nabla} \tilde{\rho}^*_{q}) \cdot (\vec{\nabla} \rho_{q}) \tilde{\rho}_{q}   
\nonumber  \\ 
& &
+ B^{\nabla \tilde{\rho}_1 \nabla \rho_1 \tilde{\rho}^*_1} (\vec{\nabla} \tilde{\rho}_{q}) \cdot (\vec{\nabla} \rho_{q}) \tilde{\rho}^*_{q}   
+ B^{\nabla \tilde{\rho}^*_1 \nabla \rho_2 \tilde{\rho}_1} (\vec{\nabla} \tilde{\rho}^*_{q}) \cdot (\vec{\nabla} \rho_{\bar{q}}) \tilde{\rho}_{q}   
+ B^{\nabla \tilde{\rho}_1 \nabla \rho_2 \tilde{\rho}^*_1} (\vec{\nabla} \tilde{\rho}_{q}) \cdot (\vec{\nabla} \rho_{\bar{q}}) \tilde{\rho}^*_{q}   
 \nonumber  \\ 
& &
+ {\mathrm i}B^{\nabla \tilde{\rho}^*_1 j_1 \tilde{\rho}_1} (\vec{\nabla} \tilde{\rho}^*_{q}) \cdot \vec{j}_{q,\mu} \tilde{\rho}_{q}   
+ {\mathrm i}B^{\nabla \tilde{\rho}^*_1 j_2 \tilde{\rho}_1} (\vec{\nabla} \tilde{\rho}^*_{q}) \cdot \vec{j}_{\bar{q}} \tilde{\rho}_{q}   
+ {\mathrm i}B^{\nabla \tilde{\rho}_1 j_1 \tilde{\rho}^*_1} (\vec{\nabla} \tilde{\rho}_{q}) \cdot \vec{j}_{q} \tilde{\rho}^*_{q}   
+ {\mathrm i}B^{\nabla \tilde{\rho}_1 j_2 \tilde{\rho}^*_1} (\vec{\nabla} \tilde{\rho}_{q}) \cdot \vec{j}_{\bar{q}} \tilde{\rho}^*_{q}   
 \nonumber  \\ 
& & 
+ \sum_{\mu\nu} \Big[
    B^{\tilde{J}^*_1 \tilde{J}_1 \rho_1} \tilde{J}^*_{q, \mu \nu} \tilde{J}_{q, \mu \nu} \rho_{q}   
+ B^{\tilde{J}^*_1 \tilde{J}_1 \rho_2} \tilde{J}^*_{q, \mu \nu} \tilde{J}_{q, \mu \nu} \rho_{\bar{q}}   
+ B^{\tilde{J}^*_1 J_1 \tilde{\rho}_1} \tilde{J}^*_{q, \mu \nu} J_{q, \mu \nu} \tilde{\rho}_{q}   
+ B^{\tilde{J}^*_1 J_2 \tilde{\rho}_1} \tilde{J}^*_{q, \mu \nu} J_{\bar{q}, \mu \nu} \tilde{\rho}_{q}   
\nonumber \\ 
& & 
+ B^{\tilde{J}_1 J_1 \tilde{\rho}^*_1} \tilde{J}_{q, \mu \nu} J_{q, \mu \nu} \tilde{\rho}^*_{q}   
+ B^{\tilde{J}_1 J_2 \tilde{\rho}^*_1} \tilde{J}_{q, \mu \nu} J_{\bar{q}, \mu \nu} \tilde{\rho}^*_{q}   
+ {\mathrm i}B^{\nabla \tilde{\rho}^*_1 \tilde{J}_1 s_1} (\nabla_\mu \tilde{\rho}^*_{q}) \tilde{J}_{q, \mu \nu} s_{q, \nu}   
+ {\mathrm i}B^{\nabla \tilde{\rho}^*_1 \tilde{J}_1 s_2} (\nabla_\mu \tilde{\rho}^*_{q}) \tilde{J}_{q, \mu \nu} s_{\bar{q}, \nu}   
\nonumber  \\ 
& & 
+ {\mathrm i}B^{\nabla \tilde{\rho}_1 \tilde{J}^*_1 s_1} (\nabla_\mu \tilde{\rho}_{q}) \tilde{J}^*_{q, \mu \nu} s_{q, \nu}   
+ {\mathrm i}B^{\nabla \tilde{\rho}_1 \tilde{J}^*_1 s_2} (\nabla_\mu \tilde{\rho}_{q}) \tilde{J}^*_{q, \mu \nu} s_{\bar{q}, \nu}   
+ {\mathrm i}B^{\nabla s_1 \tilde{J}^*_1 \tilde{\rho}_1} (\nabla_\mu s_{q, \nu}) \tilde{J}^*_{q, \mu \nu} \tilde{\rho}_{q}   
\nonumber  \\ 
& &
+ {\mathrm i}B^{\nabla s_2 \tilde{J}^*_1 \tilde{\rho}_1} (\nabla_\mu s_{\bar{q}, \nu}) \tilde{J}^*_{q, \mu \nu} \tilde{\rho}_{q}   
+ {\mathrm i}B^{\nabla s_1 \tilde{J}_1 \tilde{\rho}^*_1} (\nabla_\mu s_{q, \nu}) \tilde{J}_{q, \mu \nu} \tilde{\rho}^*_{q}   
+ {\mathrm i}B^{\nabla s_2 \tilde{J}_1 \tilde{\rho}^*_1} (\nabla_\mu s_{\bar{q}, \nu}) \tilde{J}_{q, \mu \nu} \tilde{\rho}^*_{q}   \Big]
\nonumber  \\ 
& &
 + \sum_{\mu\nu\lambda\kappa} \, \epsilon_{\nu\lambda\kappa}  \, \Big[ 
 {\mathrm i}B^{\tilde{J}^*_1 \tilde{J}_1 s_1} \tilde{J}^*_{q, \mu \nu} \tilde{J}_{q, \mu \lambda} s_{q, \kappa}  \, 
+ {\mathrm i}B^{\tilde{J}^*_1 \tilde{J}_1 s_2} \tilde{J}^*_{q, \mu \nu} \tilde{J}_{q, \mu \lambda} s_{\bar{q}, \kappa}  \,   
 \Big]
 \Big\}  \, .
\end{eqnarray}
\end{widetext}
The relations between the parameters of the pseudo-potential and the 
coefficients of the energy functional are listed in 
Tables~\ref{tab:3bodyOut:npCoeff} and~\ref{tab:3bodyOut:npCoeffP}.

\begin{table}[t!]
\begin{center}
\caption{\label{tab:3bodyOut:npCoeff}
Same as Table~\ref{tab:2bodyOut:npCoeff} for the normal part of the trilinear 
EDF kernel, Eq.~(\ref{ErhorhorhoNP}).
}
\begin{tabular}{r r r r r r r }
\hline \hline \noalign{\smallskip}
&$u_0$&$u_1$&$u_1y_1$&$u_2$&$u_2y_{21}$&$u_2y_{22}$\\
\noalign{\smallskip}  \hline \noalign{\smallskip}
$B^{\rho_1 \rho_1 \rho_2}=$&$+\frac{3}{4}$&&&&&\\[0.3mm]
$B^{s_1 s_1 \rho_2}=$&$-\frac{3}{4}$&&&&&\\[0.3mm]
$B^{\tau_1 \rho_1 \rho_1}=$&&&&$+\frac{3}{16}$&$+\frac{3}{16}$&$-\frac{3}{16}$\\[0.3mm]
$B^{\tau_1 \rho_1 \rho_2}=$&&$+\frac{1}{4}$&$-\frac{1}{16}$&$+\frac{5}{8}$&$+\frac{1}{2}$&$+\frac{5}{8}$\\[0.3mm]
$B^{\tau_1 \rho_2 \rho_2}=$&&$+\frac{1}{8}$&$+\frac{1}{16}$&$+\frac{1}{8}$&$+\frac{1}{16}$&$-\frac{1}{16}$\\[0.3mm]
$B^{T_1 s_1 \rho_2}=$&&$-\frac{1}{4}$&$+\frac{1}{16}$&$+\frac{1}{8}$&$+\frac{1}{8}$&$+\frac{1}{4}$\\[0.3mm]
$B^{T_1 s_2 \rho_1}=$&&&$+\frac{1}{16}$&&$+\frac{1}{8}$&$+\frac{1}{4}$\\[0.3mm]
$B^{\tau_1 s_1 s_1}=$&&&&$-\frac{3}{16}$&$-\frac{3}{16}$&$+\frac{3}{16}$\\[0.3mm]
$B^{\tau_1 s_1 s_2}=$&&&$-\frac{1}{16}$&&&$+\frac{3}{8}$\\[0.3mm]
$B^{\tau_1 s_2 s_2}=$&&$-\frac{1}{8}$&$-\frac{1}{16}$&$-\frac{1}{8}$&$-\frac{1}{16}$&$+\frac{1}{16}$\\[0.3mm]
$B^{\nabla \rho_1 \nabla \rho_1 \rho_1}=$&&&&$-\frac{3}{64}$&$-\frac{3}{64}$&$+\frac{3}{64}$\\[0.3mm]
$B^{\nabla \rho_1 \nabla \rho_1 \rho_2}=$&&$+\frac{5}{32}$&$-\frac{1}{16}$&$-\frac{1}{8}$&$-\frac{7}{64}$&$-\frac{11}{64}$\\[0.3mm]
$B^{\nabla \rho_1 \nabla \rho_2 \rho_1}=$&&$+\frac{5}{16}$&$+\frac{1}{16}$&$-\frac{1}{16}$&$-\frac{1}{32}$&$+\frac{1}{32}$\\[0.3mm]
$B^{\nabla s_1 \nabla s_1 \rho_1}=$&&&&$-\frac{3}{64}$&$-\frac{3}{64}$&$+\frac{3}{64}$\\[0.3mm]
$B^{\nabla s_1 \nabla s_1 \rho_2}=$&&$-\frac{5}{32}$&$+\frac{1}{16}$&$-\frac{1}{16}$&$-\frac{3}{64}$&$-\frac{3}{64}$\\[0.3mm]
$B^{\nabla s_1 \nabla s_2 \rho_1}=$&&&$+\frac{1}{16}$&&$-\frac{1}{32}$&$+\frac{1}{32}$\\[0.3mm]
$B^{\nabla \rho_1 \nabla s_1 s_1}=$&&&&$+\frac{3}{32}$&$+\frac{3}{32}$&$-\frac{3}{32}$\\[0.3mm]
$B^{\nabla \rho_1 \nabla s_1 s_2}=$&&&&&$-\frac{1}{32}$&$-\frac{5}{32}$\\[0.3mm]
$B^{\nabla \rho_1 \nabla s_2 s_1}=$&&&$-\frac{1}{16}$&&$+\frac{1}{32}$&$-\frac{1}{32}$\\[0.3mm]
$B^{\nabla \rho_1 \nabla s_2 s_2}=$&&$-\frac{5}{16}$&$-\frac{1}{16}$&$+\frac{1}{16}$&$+\frac{1}{32}$&$-\frac{1}{32}$\\[0.3mm]
$B^{j_1 j_1 \rho_1}=$&&&&$-\frac{3}{16}$&$-\frac{3}{16}$&$+\frac{3}{16}$\\[0.3mm]
$B^{j_1 j_1 \rho_2}=$&&$-\frac{1}{8}$&$+\frac{1}{8}$&$-\frac{1}{2}$&$-\frac{7}{16}$&$-\frac{11}{16}$\\[0.3mm]
$B^{j_1 j_2 \rho_1}=$&&$-\frac{1}{4}$&$-\frac{1}{8}$&$-\frac{1}{4}$&$-\frac{1}{8}$&$+\frac{1}{8}$\\[0.3mm]
$B^{J_1 J_1 \rho_1}=$&&&&$-\frac{3}{16}$&$-\frac{3}{16}$&$+\frac{3}{16}$\\[0.3mm]
$B^{J_1 J_1 \rho_2}=$&&$+\frac{1}{8}$&$-\frac{1}{8}$&$-\frac{1}{4}$&$-\frac{3}{16}$&$-\frac{3}{16}$\\[0.3mm]
$B^{J_1 J_2 \rho_1}=$&&&$-\frac{1}{8}$&&$-\frac{1}{8}$&$+\frac{1}{8}$\\[0.3mm]
$B^{j_1 J_1 s_1}=$&&&&$+\frac{3}{8}$&$+\frac{3}{8}$&$-\frac{3}{8}$\\[0.3mm]
$B^{j_1 J_1 s_2}=$&&&&&$-\frac{1}{8}$&$-\frac{5}{8}$\\[0.3mm]
$B^{j_1 J_2 s_1}=$&&&$+\frac{1}{8}$&&$+\frac{1}{8}$&$-\frac{1}{8}$\\[0.3mm]
$B^{j_1 J_2 s_2}=$&&$+\frac{1}{4}$&$+\frac{1}{8}$&$+\frac{1}{4}$&$+\frac{1}{8}$&$-\frac{1}{8}$\\[0.3mm]
$B^{\nabla s_1 J_1 s_1}=$&&&&$-\frac{3}{16}$&$-\frac{3}{16}$&$+\frac{3}{16}$\\[0.3mm]
$B^{\nabla s_1 J_1 s_2}=$&&&$-\frac{1}{16}$&&$-\frac{1}{16}$&$+\frac{1}{16}$\\[0.3mm]
$B^{\nabla s_1 J_2 s_1}=$&&&$-\frac{1}{16}$&&$-\frac{1}{16}$&$+\frac{1}{16}$\\[0.3mm]
$B^{\nabla s_1 J_2 s_2}=$&&&$+\frac{1}{8}$&&$-\frac{1}{16}$&$+\frac{1}{16}$\\
\noalign{\smallskip} \hline \hline
\end{tabular}
\end{center}
\end{table}

\begin{table}[t!]
\begin{center}
\caption{\label{tab:3bodyOut:npCoeffP}
Same as Table~\ref{tab:2bodyOut:npCoeff}, but for the anomalous part 
of the trilinear EDF kernel, Eq.~(\ref{ErhokappakappaNP}).
}
\begin{tabular}{r r r r r r r }
\hline \hline \noalign{\smallskip}
&$u_0$&$u_1$&$u_1y_1$&$u_2$&$u_2y_{21}$&$u_2y_{22}$\\
\noalign{\smallskip}  \hline \noalign{\smallskip}
$B^{\tilde{\rho}^*_1 \tilde{\rho}_1 \rho_2}=$&$+\frac{3}{4}$&&&&&\\[0.3mm]
$B^{\tilde{\tau}^*_1 \tilde{\rho}_1 \rho_2}=$&&$+\frac{3}{16}$&$-\frac{3}{32}$&&&\\[0.3mm]
$B^{\tilde{\tau}_1 \tilde{\rho}^*_1 \rho_2}=$&&$+\frac{3}{16}$&$-\frac{3}{32}$&&&\\[0.3mm]
$B^{\tau_1 \tilde{\rho}^*_1 \tilde{\rho}_1}=$&&&&$+\frac{3}{16}$&$+\frac{3}{16}$&$-\frac{3}{16}$\\[0.3mm]
$B^{\tau_2 \tilde{\rho}^*_1 \tilde{\rho}_1}=$&&$+\frac{1}{8}$&$+\frac{1}{16}$&$+\frac{1}{8}$&$+\frac{1}{16}$&$-\frac{1}{16}$\\[0.3mm]
$B^{\nabla \tilde{\rho}^*_1 \nabla \tilde{\rho}_1 \rho_1}=$&&&&$+\frac{3}{64}$&$+\frac{3}{64}$&$-\frac{3}{64}$\\[0.3mm]
$B^{\nabla \tilde{\rho}^*_1 \nabla \tilde{\rho}_1 \rho_2}=$&&$+\frac{1}{8}$&$-\frac{1}{32}$&$+\frac{1}{32}$&$+\frac{1}{64}$&$-\frac{1}{64}$\\[0.3mm]
$B^{\nabla \tilde{\rho}^*_1 \nabla \rho_1 \tilde{\rho}_1}=$&&&&$-\frac{3}{64}$&$-\frac{3}{64}$&$+\frac{3}{64}$\\[0.3mm]
$B^{\nabla \tilde{\rho}^*_1 \nabla \rho_2 \tilde{\rho}_1}=$&&$+\frac{5}{32}$&$+\frac{1}{32}$&$-\frac{1}{32}$&$-\frac{1}{64}$&$+\frac{1}{64}$\\[0.3mm]
$B^{\nabla \tilde{\rho}_1 \nabla \rho_1 \tilde{\rho}^*_1}=$&&&&$-\frac{3}{64}$&$-\frac{3}{64}$&$+\frac{3}{64}$\\[0.3mm]
$B^{\nabla \tilde{\rho}_1 \nabla \rho_2 \tilde{\rho}^*_1}=$&&$+\frac{5}{32}$&$+\frac{1}{32}$&$-\frac{1}{32}$&$-\frac{1}{64}$&$+\frac{1}{64}$\\[0.3mm]
$B^{\tilde{J}^*_1 \tilde{J}_1 \rho_1}=$&&&&$+\frac{3}{16}$&$+\frac{3}{16}$&$-\frac{3}{16}$\\[0.3mm]
$B^{\tilde{J}^*_1 \tilde{J}_1 \rho_2}=$&&&&$+\frac{3}{8}$&$+\frac{5}{16}$&$+\frac{7}{16}$\\[0.3mm]
$B^{\tilde{J}^*_1 J_1 \tilde{\rho}_1}=$&&&&$-\frac{3}{16}$&$-\frac{3}{16}$&$+\frac{3}{16}$\\[0.3mm]
$B^{\tilde{J}^*_1 J_2 \tilde{\rho}_1}=$&&&$-\frac{1}{16}$&&$-\frac{1}{16}$&$+\frac{1}{16}$\\[0.3mm]
$B^{\tilde{J}_1 J_1 \tilde{\rho}^*_1}=$&&&&$-\frac{3}{16}$&$-\frac{3}{16}$&$+\frac{3}{16}$\\[0.3mm]
$B^{\tilde{J}_1 J_2 \tilde{\rho}^*_1}=$&&&$-\frac{1}{16}$&&$-\frac{1}{16}$&$+\frac{1}{16}$\\[0.3mm]
$B^{\tilde{J}^*_1 \tilde{J}_1 s_1}=$&&&&$+\frac{3}{16}$&$+\frac{3}{16}$&$-\frac{3}{16}$\\[0.3mm]
$B^{\tilde{J}^*_1 \tilde{J}_1 s_2}=$&&&&&$-\frac{1}{16}$&$-\frac{5}{16}$\\[0.3mm]
$B^{\nabla \tilde{\rho}^*_1 \tilde{J}_1 s_1}=$&&&&$+\frac{3}{32}$&$+\frac{3}{32}$&$-\frac{3}{32}$\\[0.3mm]
$B^{\nabla \tilde{\rho}^*_1 \tilde{J}_1 s_2}=$&&&$+\frac{1}{32}$&&$+\frac{1}{32}$&$-\frac{1}{32}$\\[0.3mm]
$B^{\nabla \tilde{\rho}^*_1 j_1 \tilde{\rho}_1}=$&&&&$-\frac{3}{32}$&$-\frac{3}{32}$&$+\frac{3}{32}$\\[0.3mm]
$B^{\nabla \tilde{\rho}^*_1 j_2 \tilde{\rho}_1}=$&&$-\frac{1}{16}$&$-\frac{1}{32}$&$-\frac{1}{16}$&$-\frac{1}{32}$&$+\frac{1}{32}$\\[0.3mm]
$B^{\nabla \tilde{\rho}_1 \tilde{J}^*_1 s_1}=$&&&&$-\frac{3}{32}$&$-\frac{3}{32}$&$+\frac{3}{32}$\\[0.3mm]
$B^{\nabla \tilde{\rho}_1 \tilde{J}^*_1 s_2}=$&&&$-\frac{1}{32}$&&$-\frac{1}{32}$&$+\frac{1}{32}$\\[0.3mm]
$B^{\nabla s_1 \tilde{J}^*_1 \tilde{\rho}_1}=$&&&&$+\frac{3}{32}$&$+\frac{3}{32}$&$-\frac{3}{32}$\\[0.3mm]
$B^{\nabla s_2 \tilde{J}^*_1 \tilde{\rho}_1}=$&&&$-\frac{1}{16}$&&$+\frac{1}{32}$&$-\frac{1}{32}$\\[0.3mm]
$B^{\nabla \tilde{\rho}_1 j_1 \tilde{\rho}^*_1}=$&&&&$+\frac{3}{32}$&$+\frac{3}{32}$&$-\frac{3}{32}$\\[0.3mm]
$B^{\nabla \tilde{\rho}_1 j_2 \tilde{\rho}^*_1}=$&&$+\frac{1}{16}$&$+\frac{1}{32}$&$+\frac{1}{16}$&$+\frac{1}{32}$&$-\frac{1}{32}$\\[0.3mm]
$B^{\nabla s_1 \tilde{J}_1 \tilde{\rho}^*_1}=$&&&&$-\frac{3}{32}$&$-\frac{3}{32}$&$+\frac{3}{32}$\\[0.3mm]
$B^{\nabla s_2 \tilde{J}_1 \tilde{\rho}^*_1}=$&&&$+\frac{1}{16}$&&$-\frac{1}{32}$&$+\frac{1}{32}$\\
\noalign{\smallskip} \hline \hline
\end{tabular}
\end{center}
\end{table}

%

%
\section{Infinite nuclear matter}
\label{sect:INM}

\subsection{General definitions}

A first insight into the physics described by a given energy functional is 
provided by the analysis of the model system of homogeneous infinite nuclear 
matter (INM), where the Coulomb interaction is neglected. Although it is an 
idealized system, INM has relevance to the study of several real systems, 
e.g.\ the physics of neutron stars or the dynamics of supernovae explosions. 
In this context, one is first and foremost interested in computing the 
equation of state (EOS) of such a system, i.e.\ its energy per nucleon as 
a function of its density. Below, pairing correlations are omitted as they 
little impact bulk properties such as the EOS. However, one should note that 
pairing properties, such as pairing gaps, of INM are of importance to the 
physics of neutron stars, see e.g.\ Ref.~\cite{Chamel:2012br}.

Infinite nuclear matter being translationally invariant, it is convenient 
to use a plane wave basis
\begin{equation}
\label{eq:INM:Intro:WF}
\langle \vec{r} \sigma q | \vec{k} \sigma' q' \rangle 
= \varphi_{\vec{k} \sigma' q'}(\vec{r} \sigma q)  
= \left( 2\pi \right)^{-\frac{3}{2}} \, 
  \exp({\mathrm i} \vec{k} \cdot \vec{r}) \, 
  \delta_{\sigma\sigma'} \, \delta_{qq'} \, ,
\end{equation}
where $q\sigma=\{n\uparrow,n\downarrow,p\uparrow,p\downarrow\}$ labels
proton/neutron states with spin up/down. Neglecting pairing, the SR state 
reduces to a Slater determinant obtained by filling individual orbitals 
$\varphi_{\vec{k}\sigma' q'}(\vec{r} \sigma q)$ up to the Fermi momentum, 
i.e.\ the normal density matrix is diagonal in the plane-wave basis and 
equal to 1 for states characterized by $|\vec{k}| \leq k_{F,q\sigma}$ and 
0 otherwise, where $k_{F,q\sigma}$ denotes the spin- and isospin-dependent 
Fermi momentum. In doing so, we make the usual assumption that the 
respective Fermi surface is spherical for each particle species and spin 
direction.

When calculating densities for each of the combinations $q \sigma$,
the sum over basis states $(i,j)$ in Eq.~(\ref{eq:Skyrme_int:locdensities}) 
becomes an integral over the Fermi spheres of radius $k_{F,q\sigma}$,
leading to expressions for the matter and kinetic densities of the form
\begin{subequations}
\begin{align}
\label{eq:INM:Intro:rho:npud}
\rho_{q\sigma}     
&=  \frac{1}{6\pi^2} k_{F,q\sigma}^3 \, ,
     \\
\label{eq:INM:Intro:kin:npud}
\tau_{q\sigma} 
&= \frac{3}{20} \; \frac{2}{3\pi^2} \; k_{F,q\sigma}^5 \, .
\end{align}
\end{subequations}
With the choice of a Fermi surface centered at \mbox{$\vec{k} = 0$}, 
current densities vanish \mbox{$\vec{j}_{q\sigma} = 0$}. Also, all 
gradients of local densities are zero \mbox{$\nabla_\nu \rho_{q\sigma} = 0$} 
by construction, as are the pair densities.

The densities (\ref{eq:INM:Intro:rho:npud}) for the four different 
combinations of spin and particle species can be recoupled to 
scalar-isoscalar $\rho_0$, scalar-isovector $\rho_1$, vector-isoscalar 
$s_0$ and vector-isovector $s_1$ densities~\cite{bender02a}
\begin{subequations}
\label{eq:INM:rho01s01}
\begin{align}
\rho_0 
& = \rho_{n\uparrow} + \rho_{n\downarrow} + \rho_{p\uparrow} + \rho_{p\downarrow} \, , \\
\rho_1 
& = \rho_{n\uparrow} + \rho_{n\downarrow} - \rho_{p\uparrow} - \rho_{p\downarrow} \, , \\
s_0 
& = \rho_{n\uparrow} - \rho_{n\downarrow} + \rho_{p\uparrow} - \rho_{p\downarrow} \, , \\
s_1 
& = \rho_{n\uparrow} - \rho_{n\downarrow} - \rho_{p\uparrow} + \rho_{p\downarrow} \, ,
\end{align}
\end{subequations}
The inverse relationships read
\begin{subequations}
\label{eq:INM:rhonpsnp}
\begin{align}
\rho_{n\uparrow} 
& = \tfrac{1}{4}\big( 1 + I_\tau + I_\sigma + I_{\sigma\tau} \big) \rho_0 \, , \\
\rho_{n\downarrow} 
& = \tfrac{1}{4}\big( 1 + I_\tau - I_\sigma - I_{\sigma\tau} \big) \rho_0 \, , \\
\rho_{p\uparrow} 
& = \tfrac{1}{4}\big( 1 - I_\tau + I_\sigma - I_{\sigma\tau} \big) \rho_0 \, , \\
\rho_{p\downarrow} 
& = \tfrac{1}{4}\big( 1 - I_\tau - I_\sigma + I_{\sigma\tau} \big) \rho_0 \, ,
\end{align}
\end{subequations}
where the relative isospin \mbox{$I_\tau \equiv \rho_1/\rho_0$}, 
spin \mbox{$I_\sigma \equiv s_0/\rho_0$} and spin-isospin 
\mbox{$I_{\sigma\tau} \equiv s_1/\rho_0$} excesses, taking values
\mbox{$-1 \leq I_i \leq 1$} have been introduced. Typical cases of 
interest are 
(i) symmetric nuclear matter ($I_\tau=I_\sigma=I_{\sigma\tau}=0$), 
(ii) isospin-asymmetric nuclear matter ($I_\tau \neq 0$), 
(iii) spin-polarized nuclear matter ($I_\sigma \neq 0$) and 
(iv) isospin-asymmetric spin-polarized nuclear matter 
($I_\tau \neq 0$, $I_\sigma \neq 0$ and $I_{\sigma\tau} \neq 0$), but the
definitions given above allow also for the coverage of all intermediate
cases.

In analogy to Eq.~(\ref{eq:INM:rho01s01}) one can also define isoscalar and 
isovector kinetic densities using Eqns.~(\ref{eq:INM:Intro:rho:npud})
and~(\ref{eq:INM:Intro:kin:npud})
\begin{subequations}
\label{eq:INM:Intro:kin}
\begin{align}
\tau_0 
& = \tau_{n\uparrow} + \tau_{n\downarrow} + \tau_{p\uparrow} + \tau_{p\downarrow} 
  = \frac{3}{5} c_s \rho_0^{\frac{5}{3}} \, F^{(0)}_{5/3}(I_\tau,I_\sigma,I_{\sigma\tau}) \, , 
    \\
\tau_1 
& = \tau_{n\uparrow} + \tau_{n\downarrow} - \tau_{p\uparrow} - \tau_{p\downarrow}
  =  \frac{3}{5} c_s \rho_0^{\frac{5}{3}} F^{(\tau)}_{5/3}(I_\tau,I_\sigma,I_{\sigma\tau}) \, , 
\\
T_0 
& = \tau_{n\uparrow} - \tau_{n\downarrow} + \tau_{p\uparrow} - \tau_{p\downarrow}
  = \frac{3}{5} c_s \rho_0^{\frac{5}{3}} F^{(\sigma)}_{5/3}(I_\tau,I_\sigma,I_{\sigma\tau}) \, , 
\\
T_1 
& = \tau_{n\uparrow} - \tau_{n\downarrow} - \tau_{p\uparrow} + \tau_{p\downarrow}
  = \frac{3}{5} c_s \rho_0^{\frac{5}{3}} F^{(\sigma\tau)}_{5/3}(I_\tau,I_\sigma,I_{\sigma\tau}) \, ,
\end{align}
\end{subequations}
where $\displaystyle c_s \equiv (3\pi^2/2)^{\frac{2}{3}}$ and  
$\displaystyle c_n \equiv (3\pi^2)^{\frac{2}{3}}$ and where functions $F$ of 
the relative isospin, spin, and spin-isospin excesses introduced in 
Ref.~\cite{bender02a} have been used. Their definitions are listed 
in Appendix~\ref{sec:F-functions} along with useful properties.

As mentioned above, the main quantity of interest is the EOS that can be 
easily calculated from Eqs.~(\ref{eq:Skyrme_int:1bodyEDF}), 
(\ref{eq:Skyrme_int:2bodyEDF:normal}), 
and~(\ref{eq:Skyrme_int:3bodyEDF:normal}). The fact that most of the local 
densities are zero in INM implies that quantities of interest will be 
expressed in terms of a limited number of couplings. 


\subsection{$F$-functions}
\label{sec:F-functions}

The kinetic densities in infinite nuclear matter can be 
expressed in a very compact manner in terms of functions 
$F^{(0)}_m(I_\tau,I_\sigma,I_{\sigma\tau})$, 
$F^{(\tau)}_m(I_\tau,I_\sigma,I_{\sigma\tau})$, 
$F^{(\sigma)}_m(I_\tau,I_\sigma,I_{\sigma\tau})$ and 
$F^{(\sigma\tau)}_m(I_\tau,I_\sigma,I_{\sigma\tau})$ that have 
been introduced in Ref.~\cite{bender02a}
\begin{subequations}
\begin{eqnarray}
F^{(0)}_{m}
&\equiv & \tfrac{1}{4} 
\Big[ (1+I_\tau+I_\sigma+I_{\sigma\tau})^m 
+ (1+I_\tau-I_\sigma-I_{\sigma\tau})^m
\nonumber \\ 
& &  
+(1-I_\tau+I_\sigma-I_{\sigma\tau})^m
+(1-I_\tau-I_\sigma+I_{\sigma\tau})^m \Big]
, \nonumber \\
& & \\
F^{(\tau)}_{m}
&\equiv& \tfrac{1}{4} 
\Big[ (1+I_\tau+I_\sigma+I_{\sigma\tau})^m 
+ (1+I_\tau-I_\sigma-I_{\sigma\tau})^m
\nonumber \\ 
& & 
-(1-I_\tau+I_\sigma-I_{\sigma\tau})^m
-(1-I_\tau-I_\sigma+I_{\sigma\tau})^m \Big]
, \nonumber \\
& & \\
F^{(\sigma)}_{m}
&\equiv& \tfrac{1}{4} 
\Big[ (1+I_\tau+I_\sigma+I_{\sigma\tau})^m 
- (1+I_\tau-I_\sigma-I_{\sigma\tau})^m
\nonumber \\ 
& & 
+(1-I_\tau+I_\sigma-I_{\sigma\tau})^m
-(1-I_\tau-I_\sigma+I_{\sigma\tau})^m \Big]
, \nonumber \\
& & \\ 
F^{(\sigma\tau)}_{m}
&\equiv& \tfrac{1}{4} 
\Big[ (1+I_\tau+I_\sigma+I_{\sigma\tau})^m 
- (1+I_\tau-I_\sigma-I_{\sigma\tau})^m
\nonumber \\ 
& & 
-(1-I_\tau+I_\sigma-I_{\sigma\tau})^m
+(1-I_\tau-I_\sigma+I_{\sigma\tau})^m \Big]
\, . \nonumber \\
& & 
\end{eqnarray}
\end{subequations}
Their first derivatives with respect to spin, isospin 
and spin-isospin excesses that are needed for the 
derivation of some nuclear matter properties are
\begin{subequations}
\begin{align}
\frac{\partial F^{(\tau)}_{m}}{\partial I_\tau} 
& = \frac{\partial F^{(\sigma)}_{m}}{\partial I_\sigma} 
  = \frac{\partial F^{(\sigma\tau)}_{m}}{\partial I_{\sigma\tau}} 
  = m \, F^{(0)}_{m-1} \, , 
    \\
\frac{\partial F^{(0)}_{m}}{\partial I_\tau} 
& = \frac{\partial F^{(\sigma)}_{m}}{\partial I_{\sigma\tau}} 
  = \frac{\partial F^{(\sigma\tau)}_{m}}{\partial I_\sigma} 
  = m \, F^{(\tau)}_{m-1} \, , \\
\frac{\partial F^{(0)}_{m}}{\partial I_\sigma} 
& = \frac{\partial F^{(\tau)}_{m}}{\partial I_{\sigma\tau}} 
  = \frac{\partial F^{(\sigma\tau)}_{m}}{\partial I_\tau} 
  = m \, F^{(\sigma)}_{m-1} \, , \\
\frac{\partial F^{(0)}_{m}}{\partial I_{\sigma\tau}} 
& = \frac{\partial F^{(\tau)}_{m}}{\partial I_\sigma} 
  = \frac{\partial F^{(\sigma)}_{m}}{\partial I_\tau} 
  = m \, F^{(\sigma\tau)}_{m-1} \, ,
\end{align}
\end{subequations}
whereas their second derivatives are given by
\begin{equation}
\frac{\partial^2 F^{(j)}_{m}}{\partial I_i^2} 
= m \, (m-1) \, F^{(j)}_{m-2} \, ,
\end{equation}
for any $i$, $j \in \{0,\tau,\sigma,\sigma\tau\}$. Special values that
are appear in the nuclear matter properties discussed below are
\begin{subequations}
\begin{align}
& F^{(0)}_{0}(I_\tau,I_\sigma,I_{\sigma\tau}) =1 \, , \quad F^{(i)}_{0}(I_\tau,I_\sigma,I_{\sigma\tau}) =0 \, ,  \\
& F^{(0)}_{1}(I_\tau,I_\sigma,I_{\sigma\tau}) =1 \, , \quad F^{(i)}_{1}(I_\tau,I_\sigma,I_{\sigma\tau}) =I_i \, ,
\end{align}
\end{subequations}
and
\begin{subequations}
\begin{align}
& F^{(0)}_{m} (0,0,0) = 1 \, ,   \\ 
& F^{(i)}_{m} (0,0,0) = 0 \, ,   \\
& F^{(\tau)}_{m} (0,1,0) = F^{(\tau)}_{m} (0,0,1) = 0 \, ,  \\
& F^{(\sigma)}_{m} (1,0,0) = F^{(\sigma)}_{m} (0,0,1) = 0 \, ,   \\
& F^{(\sigma\tau)}_{m} (1,0,0) = F^{(\sigma\tau)}_{m} (0,1,0) = 0 \, ,  \\
& F^{(0)}_{m} (1,0,0) = F^{(0)}_{m} (0,1,0) = F^{(0)}_{m} (0,0,1) = 2^{m-1} \, ,  \\
& F^{(\tau)}_{m} (1,0,0) = F^{(\sigma)}_{m} (0,1,0) = F^{(\sigma\tau)}_{m} (0,0,1) = 2^{m-1} \, ,  \\
& F^{(0)}_{m} (1,1,1) = F^{(i)}_{m} (1,1,1) = 4^{m-1} \, ,
\end{align}
\end{subequations}
where $i \in \{\tau,\sigma,\sigma\tau\}$.


\subsection{Symmetric nuclear matter (SNM)}

Symmetric nuclear matter is characterized by an equal number of protons and 
neutrons as well as of spin up and spin down particles in both nucleons 
species, $\rho_1=I_\tau=0$ and $I_\sigma = I_{\sigma\tau} = 0$. Only $\rho_0$ 
and $\tau_0$ are non-zero, i.e.\ $\rho_n = \rho_p = \frac{1}{2}\rho_0$ and 
$\tau_n=\tau_p=\frac{1}{2}\tau_0$. The resulting energy per particle reads
\begin{align}
\label{eq:INM:SNM:E}
\frac{E_H}{A} \equiv \frac{{\cal E}_H}{\rho_0}  
=
\frac{3}{5}\,\frac{\hbar^2}{2m}\,c_s\,\rho_0^{\frac{2}{3}} 
& + ( A^\rho_0  + B^\rho_0\, \rho_0 ) \, \rho_0  
\nonumber  \\ 
& +\frac{3}{5} \, c_s \, ( A^\tau_0 + B^\tau_0 \, \rho_0 ) \, 
  \rho_0^{\frac{5}{3}}    \, .
\end{align}
Symmetric nuclear matter presents a stable state such that a minimum energy 
is obtained for some finite value of the density $\rho_{\text{sat}}$. The 
pressure of the fluid relates to the first derivative of the EOS with respect 
to the isoscalar density, which in SNM reads
\begin{eqnarray}
\label{eq:INM:SNM:P}
P
& \equiv & \rho_0^2 \frac{\partial E_H/A }{\partial \rho_0} \Big|_A
\nonumber \\
& = & \frac{2}{5}\,\frac{\hbar^2}{2m}\,c_s\,\rho_0^{\frac{5}{3}}
      +( A^\rho_0\; + 2 B^\rho_0\; \rho_0 ) \, \rho_0^{2}  
      \nonumber \\ 
&   & \phantom{ \frac{2}{5}\,\frac{\hbar^2}{2m}\,c_s\,\rho_0^{\frac{5}{3}} } 
     + \,c_s  \big( A^\tau_0\; + \tfrac{8}{5} B^\tau_0\; \rho_0 \big) \, 
       \rho_0^{\frac{8}{3}}    
 \, .
\end{eqnarray}
The saturation density $\rho_{\text{sat}}$ is naturally obtained as the 
solution of $P(\rho_{\text{sat}}) = 0$.

The incompressibility of the nuclear fluid relates to the second derivative 
of the EOS with respect to the isoscalar density and expresses the energy 
cost to compress the nuclear fluid. It is defined as
\begin{equation}
\label{eq:INM:SNM:K}
K
\equiv \frac{18P}{\rho_0} 
       + 9 \rho_0^2 \frac{\partial^2 E_H/A}{\partial \rho_0^2} \, ,
\end{equation}
such that at equilibrium
\begin{eqnarray}
\label{eq:INM:SNM:Kinf}
K_\infty 
& \equiv & 9 \rho_0^2 \frac{\partial^2 E_H/A}{\partial \rho_0^2} \Big|_{\rho_0=\rho_{\text{sat}}}
\nonumber \\
& = & -\frac{6}{5}\,\frac{\hbar^2}{2m} \, c_s\, \rho_{\text{sat}}^{\frac{2}{3}}
      +18 B^\rho_0\; \rho_{\text{sat}}^{2}  
\nonumber \\ 
&   & 
      + 6 \,c_s  ( A^\tau_0\; + 4 B^\tau_0\; \rho_0 ) \, \rho_{\text{sat}}^{\frac{5}{3}}    
\, ,
\end{eqnarray}
which needs to be positive for the system to be stable against density fluctuations.

\subsection{Asymmetric nuclear matter (ANM)}

More general cases of homogeneous nuclear matter are characterized by 
(i) unequal proton and neutron matter densities, i.e.\ $I_\tau \neq 0$, 
(ii) a global spin polarization, i.e.\ $I_\sigma \neq 0$ and 
(iii) a spin polarization that differs for neutron and proton species, 
i.e.\ $I_{\sigma\tau} \neq 0$. Based on Eqs.~(\ref{eq:Skyrme_int:1bodyEDF}), 
(\ref{eq:Skyrme_int:2bodyEDF:normal}),
and~(\ref{eq:Skyrme_int:3bodyEDF:normal}), the EOS of such a nuclear fluid 
is given by
\begin{widetext}
\begin{eqnarray}
\frac{E_H}{A}
& = &
\frac{3}{5}\frac{\hbar^2}{2m}\,c_sF^{(0)}_{5/3}(I_{\tau},I_{\sigma},I_{\sigma \tau}) \, \rho_0^{\frac{2}{3}}
+( A^\rho_0 +B^\rho_0\; \rho_0 ) \, \rho_0  
+( A^\rho_1 +B^\rho_1\; \rho_0 ) \, \rho_0 \, I_{\tau}^{2}
+( A^s_0 +B^s_0\; \rho_0 ) \, \rho_0 \, I_{\sigma}^{2} 
+B^s_{10}\; \rho_0^{2}  \, I_{\sigma}I_{\tau}I_{\sigma \tau}
 \nonumber \\ 
& &
+( A^s_1 +B^s_1\; \rho_0 ) \, \rho_0 \, I_{\sigma\tau}^{2} 
+\frac{3}{5}  \Big[ ( A^\tau_0 + B^\tau_0\; \rho_0 + B^\tau_{10}\; \rho_0 \, I_{\tau}^{2} + B^{\tau s}_0\; \rho_0 \, I_{\sigma}^{2} + B^{\tau s}_{10}\; \rho_0 \, I_{\sigma \tau}^{2}) \, F^{(0)}_{5/3}(I_{\tau},I_{\sigma},I_{\sigma \tau}) 
 \nonumber \\ 
& &
+ ( A^\tau_1\; I_{\tau} + B^\tau_1\; \rho_0 \, I_{\tau} + B^{\tau s}_1\; \rho_0 \, I_{\sigma}I_{\sigma \tau} ) \, F^{(\tau)}_{5/3}(I_{\tau},I_{\sigma},I_{\sigma \tau}) 
+ ( A^T_0\; I_{\sigma} + B^T_0\; \rho_0 \, I_{\sigma} + B^T_{10}\; \rho_0 \, I_{\tau}I_{\sigma \tau} ) \, F^{(\sigma)}_{5/3}(I_{\tau},I_{\sigma},I_{\sigma \tau}) 
 \nonumber \\ 
& &
+ ( A^T_1\; I_{\sigma \tau} + B^T_1\; \rho_0 \, I_{\sigma \tau} + B^T_{01}\; \rho_0 \, I_{\sigma}I_{\tau} ) \, F^{(\sigma \tau)}_{5/3}(I_{\tau},I_{\sigma},I_{\sigma \tau}) \Big]  c_s \rho_0^{\frac{5}{3}} 
\, .
\end{eqnarray}
\end{widetext}
Spin, isospin and spin-isospin symmetry energies are analogues of 
$K_\infty$ with respect to spin, isospin and spin-isospin excesses, 
respectively, i.e.\ they characterize the stiffness of the EOS with 
respect to generating such non-zero excesses. At saturation of SNM, 
i.e.\ when $I_\sigma=I_\tau=I_{\sigma\tau}=0$ and $\rho_0=\rho_{\text{sat}}$, 
the three symmetry energies are given by 
\begin{subequations}
\begin{align}
a_\tau 
&\equiv \frac{1}{2}\frac{\partial^2 E_H/A}{\partial I_\tau^2} \Big|_{I_\sigma = I_\tau = I_{\sigma\tau} = 0} 
\label{eq:INM:ANM:a_i}  \\ 
&= 
\frac{1}{3}\,\frac{\hbar^2}{2m}\,c_s\rho_0^{\frac{2}{3}}
+ (A^\rho_1 +B^\rho_1\, \rho_0 )\; \rho_0  
\nonumber \\ &
+ \bigg[ \frac{1}{3}   ( A^\tau_0 + B^\tau_0\, \rho_0 ) 
+  A^\tau_1 
+ B^\tau_1\, \rho_0
+\frac{3}{5}   B^\tau_{10}\, \rho_0 \bigg] \,c_s \, \rho_0^{\frac{5}{3}}   
\nonumber 
\,,
\\
a_\sigma 
&\equiv \frac{1}{2}\frac{\partial^2 E_H/A}{\partial I_\sigma^2} \Big|_{I_\sigma = I_\tau = I_{\sigma\tau} = 0} 
\label{eq:INM:ANM:a_s}  \\ 
&=
\frac{1}{3}\,\frac{\hbar^2}{2m}\,c_s\rho_0^{\frac{2}{3}}
+ (A^s_0 + B^s_0\, \rho_0 )\; \rho_0  
\nonumber  \\ &
+ \bigg[ \frac{1}{3}   ( A^\tau_0 + B^\tau_0\, \rho_0 ) 
+ A^T_0 
+ B^T_0\, \rho_0 
+\frac{3}{5}   B^{\tau s}_0\, \rho_0 \bigg] \,c_s \, \rho_0^{\frac{5}{3}}   
\nonumber
\,,
\\
a_{\sigma\tau}
&\equiv \frac{1}{2}\frac{\partial^2 E_H/A}{\partial I_{\sigma\tau}^2} \Big|_{I_\sigma = I_\tau = I_{\sigma\tau} = 0} 
\label{eq:INM:ANM:a_si}  \\ 
&=
\frac{1}{3}\,\frac{\hbar^2}{2m}\,c_s\rho_0^{\frac{2}{3}}
+ (A^s_1 + B^s_1\, \rho_0 )\; \rho_0  
\nonumber  \\ &
+ \bigg[ \frac{1}{3}   ( A^\tau_0 + B^\tau_0\, \rho_0 ) 
+ A^T_1 
+ B^T_1\, \rho_0
+\frac{3}{5}   B^{\tau s}_{10}\, \rho_0 \bigg] \,c_s \, \rho_0^{\frac{5}{3}}   
\nonumber
\, ,
\end{align}
\end{subequations}
and must be positive for the minimum of the EOS to be stable. 

Two other quantities of interest are intimately connected to the skin 
thickness of heavy isospin-asymmetric nuclei, i.e.\ to the difference 
between their neutron and proton radii. These quantities are the 
density-symmetry coefficient $L$
\begin{align}
L & \equiv  
3 \rho 
\frac{\partial}{\partial \rho} \left(
\frac{1}{2}\frac{\partial^2 E_H/A}{\partial I_\tau^2} \right)
\Big|_{I_\sigma = I_\tau = I_{\sigma\tau} = 0} 
\nonumber \\
& =
 \frac{2}{3}\,\frac{\hbar^2}{2m}\,c_s\rho_0^{\frac{2}{3}}
+3  ( A^\rho_1 +2  B^\rho_1\, \rho_0 )\rho_0  
 \nonumber \\ &
+ \bigg(  \frac{5}{3} A^\tau_0 + \frac{8}{3} B^\tau_0\, \rho_0 + 5 A^\tau_1 + 8  B^\tau_1\, \rho_0 
+\frac{24}{5}  B^\tau_{10}\, \rho_0 
\bigg) \,c_s \, \rho_0^{\frac{5}{3}}   
\nonumber \,, \\
\end{align}
and the symmetry incompressibility coefficient
\begin{eqnarray}
K_{\text{sym}} 
& \equiv & 9 \rho^2 
           \frac{\partial^2}{\partial \rho^2} \left(
           \frac{1}{2}\frac{\partial^2 E_H/A}{\partial I_\tau^2} \right)
           \Big|_{I_\sigma = I_\tau = I_{\sigma\tau} = 0} 
 \nonumber  \\
&= &
-\frac{2}{3}\,\frac{\hbar^2}{2m}\,c_s\rho_0^{\frac{2}{3}}
+\frac{10}{3}  ( A^\tau_0 + 4 B^\tau_0\, \rho_0 ) \, c_s \, \rho_0^{\frac{5}{3}}  
 \nonumber \\ 
&  & + 18  B^\rho_1\; \rho_0^{2}  
     + \Big[  10 (A^\tau_1 + 4 B^\tau_1 \rho_0 )   
              +24 B^\tau_{10} \rho_0 \Big] c_s \, \rho_0^{\frac{5}{3}}  
\, . \nonumber \\
\end{eqnarray}

\subsection{Pure neutron matter (PNM)}

A particular case of isospin-asymmetric nuclear matter is (spin-saturated)
pure neutron matter (PNM) obtained for \mbox{$I_\tau=1$}, 
\mbox{$I_\sigma = I_{\sigma \tau} = 0$}. The EOS of PNM reads
\begin{eqnarray}
\label{eq:INM:PNM:E} 
\frac{E_H}{A} 
& = & \frac{3}{5} \frac{\hbar^2}{2m} \, c_n \, \rho_0^{\frac{2}{3}}
      +(A^\rho_0\; + B^\rho_0\; \rho_0 ) \, \rho_0  
      \nonumber \\ 
&   & +(A^\rho_1\; +B^\rho_1\; \rho_0 ) \, \rho_0  
      +\frac{3}{5}  \,c_n   (A^\tau_0\; + B^\tau_0\; \rho_0 ) \, \rho_0^{\frac{5}{3}}   
       \nonumber \\ 
&   & +\frac{3}{5}  \,c_n   (A^\tau_1\; + B^\tau_1\; \rho_0 + B^\tau_{10}\; \rho_0 ) \, \rho_0^{\frac{5}{3}}   
\,.
\end{eqnarray}

\subsection{Effective masses}

The average energy of a nucleon inside the nuclear medium is the sum of 
a kinetic term plus a momentum-dependent self-energy term\footnote{The 
standard notion of (physical) self-energy that appears in many-body 
theories is not to be confused with the notion of spurious self-interaction 
in energy functionals as invoked in the introduction.
}
coming from its interaction with all the other nucleons. This energy can 
be rewritten as a kinetic energy term involving an effective mass. One can
thus define four different effective masses for neutron or proton with spin 
up or down, i.e.\ $m^\ast_{n\uparrow}$, $m^\ast_{n\downarrow}$, $
m^\ast_{p\uparrow}$ and $m^\ast_{p\downarrow}$. The expressions for such 
effective masses at arbitrary values of the spin, isospin and spin-isospin 
excesses read as
\begin{eqnarray}
\label{eq:INM:SNM:m_qs}
\frac{m}{m^\ast_{q\sigma}} 
& \equiv & \frac{2m}{\hbar^2} 
           \frac{\partial {\cal E}_H}{\partial \tau_{q\sigma}} 
      \nonumber  \\
& = &
1
+
\frac{2m}{\hbar^2} 
\Big[ (A^\tau_0 + B^\tau_0 \, \rho_0) \, \rho_0
\nonumber \\ 
&   &
+(A^\tau_1 + B^\tau_1 \, \rho_0 ) \, \eta_q \, I_\tau \, \rho_0
\nonumber \\ 
&   &
+(A^T_0 + B^T_0 \, \rho_0 ) \, \eta_\sigma \, I_\sigma \, \rho_0
\nonumber \\ 
&   &
+(A^T_1 + B^T_1 \, \rho_0 ) \, \eta_\sigma \eta_q \, I_{\sigma\tau} \, \rho_0
\nonumber \\ 
&   &
+\Big(B^\tau_{10} \, I_\tau^2 
+ B^{\tau s}_{0} \, I_\sigma^2 
+ B^{\tau s}_{10} \, I_{\sigma\tau}^2 
+B^{\tau s}_{1} \, \eta_q \, I_\sigma \, I_{\sigma\tau} 
\nonumber \\ 
&   &
+B^{T}_{10} \,\eta_\sigma \, I_\tau \, I_{\sigma\tau} 
+B^{T}_{01} \, \eta_\sigma \, \eta_q \, I_\tau \, I_\sigma  
\Big) \, \rho_0^2 
\Big] \, ,
\end{eqnarray}
where 
\begin{eqnarray}
\eta_q
& = & \left\{ \begin{array}{ll} 
             +1 & \quad \text{for $q=n$} \\
             -1 & \quad \text{for $q=p$}
              \end{array} \right.
      \\
\eta_\sigma
& = & \left\{ \begin{array}{ll} 
             +1 & \quad \text{for $\sigma=\uparrow$} \\
             -1 & \quad \text{for $\sigma=\downarrow$} 
              \end{array} \right.
\end{eqnarray}
One can further define effective masses of the particle species $q$
\begin{subequations}
\begin{align}
\frac{m}{m^\ast_{q}} 
 \equiv &  \frac{1}{2} \left( \frac{m}{m^\ast_{q\uparrow}} + \frac{m}{m^\ast_{q\downarrow}}  \right) 
\nonumber\\ 
 = & 
1
+
\frac{2m}{\hbar^2} 
\Big[ (A^\tau_0 + B^\tau_0 \, \rho_0) \, \rho_0
+(A^\tau_1 + B^\tau_1 \, \rho_0 ) \, \eta_q \, I_\tau \, \rho_0
\nonumber \\ 
   & +(B^\tau_{10} \, I_\tau^2 + B^{\tau s}_{0} \, I_\sigma^2 + B^{\tau s}_{10} \, I_{\sigma\tau}^2 
      + B^{\tau s}_{1} \, \eta_q \, I_\sigma \, I_{\sigma\tau} ) \, \rho_0^2 
\Big]
, \nonumber \\
   & \\
\intertext{and of particles with spin orientation $\sigma$}
\frac{m}{m^\ast_{\sigma}} 
\equiv & \frac{1}{2} \left(  \frac{m}{m^\ast_{n\sigma}} + \frac{m}{m^\ast_{p\sigma}}   \right) 
\nonumber  \\ 
 = & 1
      + \frac{2m}{\hbar^2} 
       \Big[ (A^\tau_0 + B^\tau_0 \, \rho_0) \, \rho_0
       +(A^T_0 + B^T_0 \, \rho_0 ) \, \eta_\sigma \, I_\sigma \, \rho_0
\nonumber \\ 
  & + (B^\tau_{10} \, I_\tau^2 + B^{\tau s}_{0} \, I_\sigma^2 + B^{\tau s}_{10} \, I_{\sigma\tau}^2 
     + B^{T}_{10} \,\eta_\sigma \, I_\tau \, I_{\sigma\tau} ) \, \rho_0^2 
      \Big] \, .
\nonumber \\ 
\end{align}
\end{subequations}
Equivalently, one can define scalar-isoscalar, scalar-isovector, 
vector-isoscalar and vector-isovector effective masses $m^{\ast}_{s t}$
\begin{subequations}
\begin{align}
\frac{m}{m^\ast_{00}} 
& \equiv
\frac{1}{4} \left( 
\frac{m}{m^\ast_{n\uparrow}} + \frac{m}{m^\ast_{n\downarrow}} +\frac{m}{m^\ast_{p\uparrow}} + \frac{m}{m^\ast_{p\downarrow}}  
\right)
= \frac{2m}{\hbar^2} \frac{\partial {\cal E}_H}{\partial \tau_{0}} 
\,,
\\
\frac{m}{m^\ast_{01}} 
&\equiv 
\frac{1}{4} \left( 
\frac{m}{m^\ast_{n\uparrow}} + \frac{m}{m^\ast_{n\downarrow}} -\frac{m}{m^\ast_{p\uparrow}} - \frac{m}{m^\ast_{p\downarrow}}  
\right)
= \frac{2m}{\hbar^2} \frac{\partial {\cal E}_H}{\partial \tau_{1}} 
\,,
\\
\frac{m}{m^\ast_{10}} 
&\equiv 
\frac{1}{4} \left( 
\frac{m}{m^\ast_{n\uparrow}} - \frac{m}{m^\ast_{n\downarrow}} +\frac{m}{m^\ast_{p\uparrow}} - \frac{m}{m^\ast_{p\downarrow}}  
\right)
= \frac{2m}{\hbar^2} \frac{\partial {\cal E}_H}{\partial T_{0}} 
\,,
\\
\frac{m}{m^\ast_{11} }
&\equiv 
\frac{1}{4} \left( 
\frac{m}{m^\ast_{n\uparrow}} - \frac{m}{m^\ast_{n\downarrow}} -\frac{m}{m^\ast_{p\uparrow}} + \frac{m}{m^\ast_{p\downarrow}}  
\right)
= \frac{2m}{\hbar^2} \frac{\partial {\cal E}_H}{\partial T_{1}} 
\,.
\end{align}
\end{subequations}
which for our functional gives
\begin{subequations}
\begin{eqnarray}
\frac{m}{m^\ast_{00}} 
& = &  \, 
1
+
\frac{2m}{\hbar^2} 
\Big[ (A^\tau_0 + B^\tau_0 \, \rho_0) \, \rho_0
\nonumber \\ 
& &  +(B^\tau_{10} \, I_\tau^2 + B^{\tau s}_{0} \, I_\sigma^2 + B^{\tau s}_{10} \, I_{\sigma\tau}^2 ) \, \rho_0^2 
     \Big]
\,,
\\
\frac{m}{m^\ast_{01}} 
& = &  \, 
\frac{2m}{\hbar^2} 
\Big[ 
(A^\tau_1 + B^\tau_1 \, \rho_0 ) \, I_\tau \, \rho_0
+B^{\tau s}_{1} \, I_\sigma \, I_{\sigma\tau} \, \rho_0^2 
\Big]
\,,
\nonumber \\
\\
\frac{m}{m^\ast_{10}} 
& =&  \, 
\frac{2m}{\hbar^2} 
\Big[ 
(A^T_0 + B^T_0 \, \rho_0 ) \, I_\sigma \, \rho_0
+B^{T}_{10} \, I_\tau \, I_{\sigma\tau} \, \rho_0^2 
\Big]
\,,
\nonumber \\
\\
\frac{m}{m^\ast_{11} }
& =&  \, 
\frac{2m}{\hbar^2} 
\Big[ 
(A^T_1 + B^T_1 \, \rho_0 ) \, I_{\sigma\tau} \, \rho_0
+B^{T}_{01} \, I_\tau \, I_\sigma  \, \rho_0^2 
\Big]
\,.
\nonumber \\
\end{eqnarray}
\end{subequations}
The implication of these non-standard definitions of the effective masses 
will be expanded on in a forthcoming publication~\cite{meyerUnpublished}. 
Note that $m^\ast_{01}$ is different than the usual definition of the 
isovector effective mass. The various effective masses at the saturation 
point of SNM can be trivially obtained from the expressions given above by 
setting \mbox{$I_{\tau}=I_{\sigma}=I_{\sigma\tau}=0$}.

\subsection{Landau parameters}
\label{sec:landau}

\subsubsection{Introduction}

Landau parameters are interesting quantities~\cite{backman69,chang75a,cao10a} 
to compute for several reasons:

Two sum rules must be fulfilled by Landau parameters in order for the 
Pauli principle to be respected~\cite{migdal67a}. In the present case where 
the EDF kernel does derive from a pseudo-potential, the two sum rules 
are  fulfilled analytically by construction. 

There are also two other sum rules that derive from the 
antisymmetry of the scattering amplitude, which is determined by the
residual interaction. The antisymmetry of the residual interaction 
itself, to which Landau parameters are related, however, does not ensure 
the antisymmetry of the observable scattering amplitude~\cite{gogny77}. 

Landau parameters can also be used to detect and control infinite-wavelength 
instabilities. Such instabilities appear when Landau parameters do not 
respect the stability conditions~\cite{migdal67a,pomeranchuk59} 
\begin{equation}
\label{eq:INM:Landau:instabilities}
1 + \frac{X_l}{2l+1} > 0 \,, 
\end{equation}
where $X_l = \{F_l,F'_l,G_l,G'_l\}$ with $l=0,1$ denotes the Landau 
parameters. In particular, four of the Landau parameters, $F_0$, $F_0'$, 
$G_0$ and $G_0'$ are also related to the stiffness of the EOS, i.e.\ 
its second derivatives  with respect to density, isospin, spin and 
spin-isospin fluctuations. This leads to the following relationships 
at saturation
\begin{subequations}
\label{eq:INM:Landau:intro0}
\begin{align} 
K_\infty &= 6 \frac{\hbar^2 k_F^2}{2m^\ast_0} (1+F_0)
\,,\\
a_\tau &= \frac{1}{3} \frac{\hbar^2 k_F^2}{2m^\ast_0} (1+F_0')
\,,\\
a_\sigma &= \frac{1}{3} \frac{\hbar^2 k_F^2}{2m^\ast_0} (1+G_0)
\,,\\
a_{\sigma\tau} &= \frac{1}{3} \frac{\hbar^2 k_F^2}{2m^\ast_0} (1+G_0') \,.
\end{align}
\end{subequations}
For the EOS of SNM to have a stable minimum, all these second derivatives 
have to be larger than zero, such that $F_0$, $F_0'$, $G_0$ and $G_0'$ are
greater than $-1$, which is equivalent to the stability
conditions~(\ref{eq:INM:Landau:instabilities}). 

It has to be noted that parametrizations must not only be stable 
against infinite wavelength instabilities signaled by Landau parameters, 
but also against finite-size instabilities that probe gradient terms 
in the EDF~\cite{lesinski06a,pototzky10a,Schunck10a,hellemans12a,pastore12a}. 
The control of finite-size instabilities for the newly-proposed Skyrme-like 
parametrizations will be discussed in a forthcoming publication.

\subsubsection{Definition}

Landau parameters are calculated via the residual particle-hole interaction 
in INM, which in general is defined through
\begin{eqnarray}
v^{\text{res}}_{12} 
& \equiv & \langle \vec{r}^{\,\prime}_1 \sigma_1' q^{\,}_1 , \vec{r}^{\,\prime}_2 \sigma_2' q^{\,}_2  | \hat{v}^{\text{res}}_{12} | \vec{r}_1^{\,} \sigma_1^{\,} q_1^{\,} , \vec{r}_2^{\,} \sigma_2^{\,} q_2^{\,}  \rangle 
\nonumber \\
& = & \frac{\partial^2 {\cal E}}{\partial \rho(\vec{r}^{\,\prime}_2 \sigma_2^{\,\prime} q_2^{\,},\vec{r}_2^{\,} \sigma_2^{\,}q_2^{\,}) \partial \rho(\vec{r}^{\,\prime}_1 \sigma_1^{\,\prime}q_1^{\,},\vec{r}_1^{\,} \sigma_1^{\,}q_1^{\,})} \, ,
\end{eqnarray}
and can be written in infinite nuclear matter, for momenta lying on the 
Fermi surface, as
\begin{eqnarray}
\label{eq:INM:int_resToLandau}
v^{\text{res}}_{12} 
& = & N_0^{-1} \sum_l \Big[ F_l + F'_l \; {\bf \tau}_1 \circ {\bf \tau}_2  
      + G_l \; \vec{\sigma}_1 \cdot \vec{\sigma}_2 
    \nonumber \\ 
& & \hspace{1.2cm}
+ G'_l \; \vec{\sigma}_1 \cdot \vec{\sigma}_2 \; {\bf \tau}_1 \circ {\bf \tau}_2  \Big] P_l(\cos \theta)\, ,
\end{eqnarray}
where coefficients $F_l,F'_l,G_l$ and $G'_l$ are Landau parameters, 
$\displaystyle N_0 \equiv 2 m^{\ast}_0k_F /\pi^2 \hbar^2 $ is a normalization 
factor, $P_l(x)$ Legendre polynomials, and $\theta$ is the angle between the 
incoming momentum of nucleon~1 and the outgoing momentum of nucleon~2. In 
the present case, Landau parameters read explicitly as
\begin{subequations}
\label{eq:Skyrme_int:Landau:Final}
\begin{align}
f_0  
=& 
2 A^\rho_0\;  
+2 A^\tau_0\;  k_F^2 
+6 B^\rho_0\;  \rho_{0} 
+2 B^\tau_0\;  \tau_{0} 
+4 B^\tau_0\;  k_F^2 \rho_{0}
\,, \\
f'_0
=& 
2 A^\rho_1\;  
+2 A^\tau_1\;  k_F^2 
+2 B^\rho_1\;  \rho_{0} 
+2 B^\tau_{10}\;  \tau_{0} 
+2 B^\tau_1\;  k_F^2 \rho_{0}
\,, \\
g_0  
=& 
2 A^s_0\;  
+2 A^T_0\;  k_F^2 
+2 B^s_0\;  \rho_{0} 
+2 B^{\tau s}_0\;  \tau_{0} 
+2 B^T_0\;  k_F^2 \rho_{0} 
\,, \\
g'_0 
=& 
2 A^s_1\;  
+2 A^T_1\;  k_F^2 
+2 B^s_1\;  \rho_{0} 
+2 B^{\tau s}_{10}\;  \tau_{0} 
+2 B^T_1\;  k_F^2 \rho_{0} 
\,, \\
f_1
=& 
2 A^j_0\;  k_F^2 
+2 B^j_0\;  k_F^2 \rho_{0} 
\, , \\
f'_1
=&
2 A^j_1\;  k_F^2 
+2 B^j_1\;  k_F^2 \rho_{0} 
\, , \\
g_1
=& 
2 A^J_0\;  k_F^2 
+2 B^J_0\;  k_F^2 \rho_{0} 
\,, \\
g'_1
=&
2 A^J_1\;  k_F^2 
+2 B^J_1\;  k_F^2 \rho_{0}
\, .
\end{align}
\end{subequations}
where $f_l\equiv F_l/N_0$, $f_l'\equiv F_l'/N_0$, $g_l\equiv G_l/N_0$ 
and $g_l'\equiv G_l'/N_0$, and where we have used Eq.~(\ref{eq:INM:Intro:kin})
to express $\tau_0$ in terms of $k_F^2$ and $\rho_0$. 
The Landau parameters with \mbox{$l \geq 2$} are zero for a
Skyrme-type interaction with only up to two gradients. The expressions for 
the Landau parameters in terms of the pseudo-potential parameters are 
given in Tab.~\ref{tab:INM:Landau}.

\begin{table}[t!]
\begin{center}
\caption{\label{tab:INM:Landau}
Landau parameters expressed in terms of the pseudo-potential parameters.
Missing entries are zero.
}
\begin{tabular}{ l  r r r r r r r r }
\hline \hline \noalign{\smallskip}
& $f_0$ & $f_0'$ & $g_0$ & $g_0'$ & $f_1$ & $f_1'$ & $g_1$ & $g_1'$  \\
\noalign{\smallskip}  \hline \noalign{\smallskip}
$t_0$ & $\frac{3}{4}$ & $-\frac{1}{4}$ & $-\frac{1}{4}$ & $-\frac{1}{4}$ &  &  &  &  \\[0.5mm]
$t_0x_0$ &  & $-\frac{1}{2}$ & $\frac{1}{2}$ &  &  &  &  &  \\[0.5mm]
$t_1 k_F^2$ & $\frac{3}{8}$ & $-\frac{1}{8}$ & $-\frac{1}{8}$ & $-\frac{1}{8}$ & $-\frac{3}{8}$ & $\frac{1}{8}$ & $\frac{1}{8}$ & $\frac{1}{8}$ \\[0.5mm]
$t_1 x_1 k_F^2$ &    & $-\frac{1}{4}$ & $\frac{1}{4}$ &   &   & $\frac{1}{4}$ & $-\frac{1}{4}$ &   \\[0.5mm]
$t_2 k_F^2$ & $\frac{5}{8}$ & $\frac{1}{8}$ & $\frac{1}{8}$ & $\frac{1}{8}$ & $-\frac{5}{8}$ & $-\frac{1}{8}$ & $-\frac{1}{8}$ & $-\frac{1}{8}$\\[0.5mm]
$t_2 x_2 k_F^2$ & $\frac{1}{2}$ & $\frac{1}{4}$ & $\frac{1}{4}$ &   & $-\frac{1}{2}$ & $-\frac{1}{4}$ & $-\frac{1}{4}$ &     \\[0.5mm]
$u_0 \rho_0$ & $\frac{9}{8}$ & $-\frac{3}{8}$ & $-\frac{3}{8}$ & $-\frac{3}{8}$ &   &  &  &  \\[0.5mm]
$u_1 \rho_0 k_F^2$ & $\frac{39}{80}$ & $-\frac{13}{80}$ & $-\frac{13}{80}$ & $-\frac{13}{80}$ & $-\frac{3}{16}$ & $\frac{1}{16}$ & $\frac{1}{16}$ & $\frac{1}{16}$  \\[0.5mm]
$u_1y_1 \rho_0 k_F^2$ &  & $-\frac{1}{40}$ & $\frac{1}{40}$ &  &  & $\frac{1}{8}$ & $-\frac{1}{8}$ &  \\[0.5mm]
$u_2 \rho_0 k_F^2$ & $\frac{39}{32}$ & $-\frac{1}{32}$ & $-\frac{1}{32}$ & $-\frac{1}{32}$ & $-\frac{15}{32}$ & $-\frac{7}{32}$ & $-\frac{7}{32}$ & $-\frac{7}{32}$ \\ [0.5mm]
$u_2y_{21} \rho_0 k_F^2$ & $\frac{39}{40}$ & $\frac{1}{20}$ & $\frac{1}{20}$ & $-\frac{3}{40}$ & $-\frac{3}{8}$ & $-\frac{1}{4}$ & $-\frac{1}{4}$ & $-\frac{1}{8}$ \\[0.5mm]
$u_2y_{22} \rho_0 k_F^2$ & $\frac{39}{80}$ & $-\frac{31}{80}$ & $\frac{35}{80}$ & $-\frac{3}{80}$ & $-\frac{3}{16}$ & $-\frac{5}{16}$ & $\frac{1}{16}$ & $-\frac{1}{16}$ \\
\noalign{\smallskip} \hline \hline
\end{tabular}
\end{center}
\end{table}

\subsubsection{Sum rules from the residual interaction}
\label{sec:INM:Landau:sumrule1}

The EDF from which the residual interaction derives has been constructed from 
an antisymmetrized vertex such that the Pauli-principle is respected 
throughout. When the antisymmetrized vertex is a two-body pseudo-potential 
multiplied by a two-body antisymmetrizer, taking two derivatives of the EDF 
with respect to non-local densities gives back the original antisymmetrized 
vertex. When the antisymmetrized vertex is made of two- plus three-body 
pseudo-potentials multiplied by appropriate antisymmetrizers, the residual 
particle-hole interaction remains an antisymmetrized two-body vertex. 
Consequently, the exclusion principle demands that the residual 
interaction~Eq.~(\ref{eq:INM:int_resToLandau}) is antisymmetric under the 
exchange of incoming or outgoing particles. This is equivalent to requiring 
that incoming and outgoing two-body states carry odd values of \mbox{$L+S+T$},
where $L$ denotes the two-body orbital angular momentum of the relative 
motion, whereas $S$ and $T$ characterize the two-body spin and isospin,
respectively. Starting from Eq.~(\ref{eq:INM:int_resToLandau}) with 
$\vec{p}_1=\vec{p}_2^{\,_{'}}$, i.e.\ $\theta=0$, and requiring that the 
antisymmetry holds for each spin-isospin channel separately, provides two 
sum rules
\begin{subequations}
\label{eq:INM:Landau:param:sumrule}
\begin{eqnarray}
\label{eq:INM:Landau:param:sumrule1} 
\sum_l \big( F_l + F_l' + G_l + G_l' \big) 
& = & 0  \, ,
\\ 
\label{eq:INM:Landau:param:sumrule2} 
\sum_l \big( F_l -3 F_l' - 3 G_l + 9 G_l' \big)
& = & 0 \, ,
\end{eqnarray}
\end{subequations}
where we have used that $P_l(1)=1$ for all $l$. 
Equation~(\ref{eq:INM:Landau:param:sumrule1}) holds for spin and isospin 
triplet (\mbox{$S=T=1$}) two-body states, for which 
$\vec{\sigma}_1 \cdot \vec{\sigma}_2 = {\bf \tau}_1 \circ {\bf \tau}_2 = 1$. 
Equation~(\ref{eq:INM:Landau:param:sumrule2}) holds for spin and isospin 
singlet (\mbox{$S=T=0$}) two-body states for which 
$\vec{\sigma}_1 \cdot \vec{\sigma}_2 = {\bf \tau}_1 \circ {\bf \tau}_2 = -3$. 
In both cases the relative orbital angular-momentum of the two-body state 
is odd. 

Sum rules~(\ref{eq:INM:Landau:param:sumrule}) are fulfilled for Landau 
parameters derived from the presently-developed two- plus three-body 
pseudo-potential, see Tab.~\ref{tab:INM:Landau}. This property provides a 
stringent test that the derivation of the EDF and of the residual interaction 
are correct. 

Note that in the presence of tensor-type pseudo-potentials there are 
additional contributions to the sumrules \cite{cao10a,pastore12a,brown77a}.


\subsubsection{Sum rules from the scattering amplitude}
\label{sec:INM:Landau:sumrule2}

The residual particle-hole interaction is not a physically observable 
quantity in contrast to the scattering amplitude $\Gamma_{12}$ associated 
with the motion of a particle-hole pair~\cite{gogny77}. The latter is
related to the former through an integral equation, such that the 
particle-hole interaction can be seen as the irreducible vertex and the 
scattering amplitude as the total vertex. Analogues to 
Eq.~(\ref{eq:INM:Landau:param:sumrule}) can be derived from the 
antisymmetry of the scattering amplitude. Plugging the expansion  of 
the scattering amplitude on Legendre polynomials
\begin{eqnarray}
\label{eq:INM:Landau:scatt_amplitude}
\Gamma_{12} 
& \equiv & N_0^{-1} \sum_l \Big[ B_l + C_l \; {\bf \tau}_1 \circ {\bf \tau}_2  + D_l \; \vec{\sigma}_1 \cdot \vec{\sigma}_2 
 \\ 
&   & \hspace{1.9cm} \nonumber
      + E_l \; \vec{\sigma}_1 \cdot \vec{\sigma}_2 \; {\bf \tau}_1 \circ {\bf \tau}_2  \Big] P_l(\cos \theta)\,\,\, 
\end{eqnarray}
into the integral equation that relates it to the residual 
interaction~(\ref{eq:INM:int_resToLandau}), one obtains, in absence of 
tensor terms, the relationships~\cite{backman69,gogny77,friman79}
\begin{subequations}
\label{eq:INM:Landau:scatt_amplitude:Leg_exp_coeff}
\begin{eqnarray}
B_l &=& \frac{F_l}{1+ F_l/(2l+1)} \,, \\
C_l &=& \frac{F'_l}{1+F'_l/(2l+1)} \,, \\
D_l &=& \frac{G_l}{1+G_l/(2l+1)} \,, \\
E_l &=& \frac{G'_l}{1+G'_l/(2l+1)} \,.
\end{eqnarray}
\end{subequations}
The reasoning used in Sec.~\ref{sec:INM:Landau:sumrule1} now provides 
sum rules for the expansion coefficients of $\Gamma_{12}$
\begin{subequations}
\label{eq:INM:Landau:scatt_amplitude:sumrule}
\begin{eqnarray}
\sum_l \big( B_l + C_l + D_l + E_l \big)  & = & 0  \, ,
\label{eq:INM:Landau:scatt_amplitude:sumrule11} \\ 
\sum_l \big( B_l -3 C_l - 3 D_l + 9 E_l \big)  & = & 0 
\label{eq:INM:Landau:scatt_amplitude:sumrule12} \, ,
\end{eqnarray}
\end{subequations}
which can be rearranged as~\cite{friman79} 
\begin{subequations}
\label{eq:INM:Landau:scatt_amplitude:sumrule2}
\begin{eqnarray}
\sum_l \big( B_l + 3 E_l \big) &=& 0 
\label{eq:INM:Landau:scatt_amplitude:sumrule21} \, , \\ 
\sum_l \big( \tfrac{2}{3} B_l + C_l + D_l \big) &=& 0 
\label{eq:INM:Landau:scatt_amplitude:sumrule22} \, .
\end{eqnarray}
\end{subequations}
In Born approximation, i.e.\ when the magnitude of Landau parameters entering 
Eq.~(\ref{eq:INM:Landau:scatt_amplitude:Leg_exp_coeff}) are negligible 
compared to \mbox{$2l+1$}, Eq.~(\ref{eq:INM:Landau:scatt_amplitude:sumrule})
 reduces to Eq.~(\ref{eq:INM:Landau:param:sumrule}). However, Landau 
parameters are not small in nuclear matter, such that, physically speaking, 
sum rule~(\ref{eq:INM:Landau:param:sumrule}) cannot be justified starting 
from the scattering amplitude.

Interestingly, the antisymmetric character of the residual particle-hole 
interaction does not guarantee the antisymmetry of the scattering amplitude, 
which is frequently broken in practice. Through the iteration process of the 
integral equation, reducible diagrams might appear without their Pauli 
principle counterparts~\cite{gogny77,babu73}, which is a fingerprint of the 
lack of complexity of the irreducible residual interaction. Inserting density 
dependencies into the pseudo-potential has allowed in some cases to 
effectively compensate for such missing diagrams~\cite{gogny77} at the price 
of compromising the antisymmetry of the residual interaction itself and thus 
of violating Eq.~(\ref{eq:INM:Landau:param:sumrule}). In the end, fulfilling 
both the antisymmetry of the irreducible vertex and of the scattering 
amplitude is a difficult task. In the present case, the former is ensured 
analytically, whereas the latter is not. The extent to which it is violated 
will depend on the values of the parameters that result from a given fit.

\section{Steps to derive the EDF kernel}
\label{sect:howto}

This section lists the steps to derive the energy functional in 
proton-neutron representation from the pseudo-potentials defined in 
Eq.~(\ref{eq:Skyrme_int:Grad}). We limit the illustration to a few 
normal and anomalous terms resulting from the operator
$\hat{v}^{\text{ex}} = u_2 \, y_{21} \; \hat{P}^\sigma_{12} \; \hat{\vec{k}}^{\,\dagger}_{12} \, 
\hat{\delta}^r_{13}\hat{\delta}^r_{23} \, \cdot \hat{\vec{k}}^{\,}_{12}$,
one of the terms contained in $\hat{v}^2_{\overline{12}3}$, 
Eq.~(\ref{eq:Skyrme_int:3body:Grad_init:kp12k12}). Such an operator is 
used in Eqs.~(\ref{eq:Skyrme_int:Energy3:normal:3bodyA}) 
and~(\ref{eq:Skyrme_int:Energy3:anormal:3bodyB}) where one has to multiply 
it to the antisymmetrizers. For the normal part one must thus evaluate 
$\hat{P}^\sigma_{12} \hat{{\cal A}}_{123}$, which leads to
\begin{widetext}
\begin{eqnarray}
\hat{P}^\sigma_{12} \hat{{\cal A}}_{123} 
& = & 
\hat{P}^\sigma_{12} 
-\hat{P}^r_{12} \hat{P}^q_{12} 
-\hat{P}^r_{23} \hat{P}^\sigma_{12} \hat{P}^\sigma_{23} \hat{P}^q_{23} 
-\hat{P}^r_{13} \hat{P}^\sigma_{12} \hat{P}^\sigma_{13} \hat{P}^q_{13} 
+\hat{P}^r_{12} \hat{P}^r_{23} \hat{P}^\sigma_{23} \hat{P}^q_{12} \hat{P}^q_{23} 
+\hat{P}^r_{12} \hat{P}^r_{13} \hat{P}^\sigma_{13} \hat{P}^q_{12} \hat{P}^q_{13} 
\,.
\end{eqnarray}
Selecting only the third term as an example, i.e.\ $-\hat{v}^{ex} P_{23}$, 
which is also obtained in the pairing part when evaluating 
$\hat{{\cal A}}^{12}_{123} \hat{P}^\sigma_{12} \hat{{\cal A}}^{12}_{123}$, 
one computes its matrix elements by inserting closure relations on 
${\cal H}_3$ in the coordinate basis according to
\begin{align}
- \langle ijk | \hat{v}^{ex} P_{23} | lmn \rangle
=
-\int d\xi_1 d\xi_2 d\xi_3 d\xi_4 d\xi_5 d\xi_6 \;
\varphi^\dagger_i(\xi_1)
\varphi^\dagger_j(\xi_2)
\varphi^\dagger_k(\xi_3)
\langle \xi_1 \xi_2 \xi_3 | 
\hat{v}^{ex} P_{23} | \xi_4 \xi_5 \xi_6 \rangle 
\varphi_l(\xi_4)
\varphi_m(\xi_5)
\varphi_n(\xi_6) \, .
\end{align} 

\subsection{Spatial part of the matrix element}

Using Eqs.~(\ref{eq:Skyrme_int:gradients_opertors:2}) 
and~(\ref{eq:Skyrme_int:rsq_exchanges}), one obtains
\begin{align}
\langle \vec{r}_1 \vec{r}_2 \vec{r}_3 | 
\hat{\vec{k}}^{\,\dagger}_{12} \, \hat{\delta}^r_{13}\hat{\delta}^r_{23} \, \cdot \hat{\vec{k}}^{\,}_{12} \hat{P}^r_{23} | \vec{r}_4 \vec{r}_5 \vec{r}_6 \rangle
=
\cev{k}^{\,\ast}_{\vec{r}_1 \vec{r}_2}
\,
\langle \vec{r}_1 \vec{r}_2 \vec{r}_3 | 
\hat{\delta}^r_{13}\hat{\delta}^r_{23} | \vec{r}_4 \vec{r}_6 \vec{r}_5 \rangle
\,
\vec{k}^{\,}_{\vec{r}_4 \vec{r}_6}
\, .
\end{align}
Applying the gradients on the wave-functions to the right and to the left, 
one can write
\begin{subequations}
\begin{eqnarray}
-\langle ijk | \hat{v}^{ex} P_{23} | lmn \rangle 
&=& -\int \! d\xi_1 \, d\xi_2 \, d\xi_3 \, d\xi_4 \, d\xi_5 \, d\xi_6 \;
\langle \sigma_1 q_1 \sigma_2 q_2 \sigma_3 q_3 | 
 \hat{P}^\sigma_{12} \hat{P}^\sigma_{23} \hat{P}^q_{23} | \sigma_4 q_4 \sigma_5 q_5 \sigma_6 q_6 \rangle \;
\langle \vec{r}_1 \vec{r}_2 \vec{r}_3 | 
\hat{\delta}^r_{13}\hat{\delta}^r_{23} | \vec{r}_4 \vec{r}_6 \vec{r}_5 \rangle \nonumber \\
&&
\times \vec{k}^{\,\ast}_{\vec{r}_1 \vec{r}_2}
\vec{k}^{\,}_{\vec{r}_4 \vec{r}_6} \;
\varphi^\dagger_i(\xi_1) \,
\varphi^\dagger_j(\xi_2) \,
\varphi^\dagger_k(\xi_3) \,
\varphi_l(\xi_4) \,
\varphi_m(\xi_5) \,
\varphi_n(\xi_6) \\
&=& -\int \! d\xi_1 \, d\xi_2 \, d\xi_3 \, d\xi_4 \, d\xi_5 \, d\xi_6 \;
\langle \sigma_1 q_1 \sigma_2 q_2 \sigma_3 q_3 | 
 \hat{P}^\sigma_{12} \hat{P}^\sigma_{23} \hat{P}^q_{23}  | \sigma_4 q_4 \sigma_5 q_5 \sigma_6 q_6\rangle \;
\delta(\vec{r}_1-\vec{r}_4) \, 
\delta(\vec{r}_2-\vec{r}_6) \, 
\nonumber \\
&&
\times \delta(\vec{r}_3-\vec{r}_5) \, 
\delta(\vec{r}_4-\vec{r}_5) \,
\delta(\vec{r}_6-\vec{r}_5) \;
\vec{k}^{\,\ast}_{\vec{r}_1 \vec{r}_2}
\vec{k}^{\,}_{\vec{r}_4 \vec{r}_6} \;
\varphi^\dagger_i(\xi_1) \,
\varphi^\dagger_j(\xi_2) \,
\varphi^\dagger_k(\xi_3) \,
\varphi_l(\xi_4) \,
\varphi_m(\xi_5) \,
\varphi_n(\xi_6) \, .
\end{eqnarray} 
\end{subequations}
With that at hand, Eqs.~(\ref{eq:Skyrme_int:Energy3:normal:3bodyA}) 
and~(\ref{eq:Skyrme_int:Energy3:anormal:3bodyB}) become
\begin{subequations}
\begin{align}
E^{\rho\rho\rho}_{ex}
&= -\frac{1}{2} \sum_{ijklmn} 
\langle i j k | \hat{v}^{ex} P_{23} | l m n \rangle \, \rho _{li} \, \rho _{mj} \, \rho _{nk} \, ,
\\
&= -\frac{1}{2} \int \! d\xi_1 \, d\xi_2 \, d\xi_3 \, 
                        d\xi_4 \, d\xi_5 \, d\xi_6 \,
\langle  \hat{P}^\sigma_{12} \hat{P}^\sigma_{23} \hat{P}^q_{23} \rangle \;
\delta(\{\vec{r}\}=\vec{r}) \;
\vec{k}^{\,\ast}_{\vec{r}_1 \vec{r}_2}
\vec{k}^{\,}_{\vec{r}_4 \vec{r}_6} \;
\rho(\xi_4,\xi_1) \, \rho(\xi_5,\xi_2) \, \rho(\xi_6,\xi_3) \, ,
\\
E^{\kappa\kappa\rho}_{ex}
&= -\frac{1}{2} \sum_{ijklmn} 
\langle i j k | \hat{v}^{ex} P_{23} | l m n \rangle \, \kappa ^{*}_{ij} \, \kappa _{lm}  \, \rho _{nk} 
\\
\label{eq:example:Ekkr}
&= -\frac{1}{2} \int \! d\xi_1 \, d\xi_2 \, d\xi_3 \, 
                        d\xi_4 \, d\xi_5 \, d\xi_6 \,
\langle  \hat{P}^\sigma_{12} \hat{P}^\sigma_{23} \hat{P}^q_{23}  \rangle \;
\delta(\{\vec{r}\}=\vec{r}) \;
\vec{k}^{\,\ast}_{\vec{r}_1 \vec{r}_2}
\cdot 
\vec{k}^{\,}_{\vec{r}_4 \vec{r}_6} \;
\kappa ^{*}(\xi_1,\xi_2) \, \kappa(\xi_4 , \xi_5) \, \rho(\xi_6,\xi_3)
 \, ,
\end{align}
\end{subequations}
where we have introduced the shorthands
\begin{subequations}
\begin{eqnarray}
\langle  \hat{P}^\sigma_{12} \hat{P}^\sigma_{23} \hat{P}^q_{23}  \rangle 
&\equiv &
\langle \sigma_1 q_1 \sigma_2 q_2 \sigma_3 q_3 | 
 \hat{P}^\sigma_{12} \hat{P}^\sigma_{23} \hat{P}^q_{23}  | \sigma_4 q_4 \sigma_5 q_5 \sigma_6 q_6\rangle \, ,
 \\
\delta(\{\vec{r}\}=\vec{r})
&\equiv&
\delta(\vec{r}_1-\vec{r}_4) \,
\delta(\vec{r}_2-\vec{r}_6) \,
\delta(\vec{r}_3-\vec{r}_5) \,
\delta(\vec{r}_4-\vec{r}_5) \,
\delta(\vec{r}_6-\vec{r}_5)
\, .
\end{eqnarray}
\end{subequations}

\subsection{Isospin part of the matrix element}

The matrix element of the isospin-exchange operator is trivially evaluated using Eq.~(\ref{eq:Skyrme_int:rsq_exchanges})
\begin{equation}
\langle q_1 q_2 q_3 | \hat{P}^q_{23} | q_4 q_5 q_6 \rangle 
= \langle q_1 q_2 q_3 | q_4 q_6 q_5 \rangle 
= \delta_{q_1 q_4} \, \delta_{q_2 q_6} \, \delta_{q_3 q_5} \, ,
\end{equation}
Recalling that local densities are diagonal in isospin, the 
integrand is null if $q_4 \neq q_1$, $q_5 \neq q_2$, $q_6 \neq q_3$ for 
the normal part and if $q_1 \neq q_2$, $q_4 \neq q_5$, $q_6 \neq q_3$ for 
the anomalous part. One thus obtains
\begin{subequations}
\begin{align}
E^{\rho\rho\rho}_{ex}
&= -\frac{1}{2}  \int \! d\zeta_1 \, d\zeta_2 \, d\zeta_3 \, 
                         d\zeta_4 \, d\zeta_5 \, d\zeta_6 \sum_{q_1,q_2} \,
\langle  \hat{P}^\sigma_{12} \hat{P}^\sigma_{23} \rangle 
\delta(\{\vec{r}\}=\vec{r})
\vec{k}^{\,\ast}_{\vec{r}_1 \vec{r}_2}
\cdot 
\vec{k}^{\,}_{\vec{r}_4 \vec{r}_6}
\rho_{q_1}(\zeta_4,\zeta_1) \, \rho_{q_2}(\zeta_5,\zeta_2) \, \rho_{q_2}(\zeta_6,\zeta_3) \, ,
\\
E^{\kappa\kappa\rho}_{ex}
&= -\frac{1}{2} \int \! d\zeta_1 \, d\zeta_2 \, d\zeta_3 \, 
                        d\zeta_4 \, d\zeta_5 \, d\zeta_6 \sum_{q_1} \,
\langle  \hat{P}^\sigma_{12} \hat{P}^\sigma_{23}  \rangle \; 
\delta(\{\vec{r}\}=\vec{r}) \;
\vec{k}^{\,\ast}_{\vec{r}_1 \vec{r}_2}
\cdot 
\vec{k}^{\,}_{\vec{r}_4 \vec{r}_6} \;
\kappa ^{*}_{q_1}(\zeta_1,\zeta_2) \, \kappa_{q_1}(\zeta_4 , \zeta_5) \, 
\rho_{q_1}(\zeta_6,\zeta_3)
 \, ,
\end{align}
\end{subequations}
where $\zeta \equiv \vec{r} , \sigma$ and $\langle \hat{P}^\sigma_{12} \hat{P}^\sigma_{23} \rangle  \equiv \langle \sigma_1 \sigma_2 \sigma_3 | \hat{P}^\sigma_{12} \hat{P}^\sigma_{23} | \sigma_4 \sigma_5 \sigma_6 \rangle$.
More generally, matrix elements at play in the normal part of the EDF are
\begin{subequations}
\begin{align}
\langle q_1 q_2 q_3 | 1 | q_4 q_5 q_6 \rangle
\rightarrow & 
\sum_{q_1,q_2,q_3} 
\rho_{q_1}(\zeta_4,\zeta_1) \, \rho_{q_2}(\zeta_5,\zeta_2) \, \rho_{q_3}(\zeta_6,\zeta_3) 
\\
\langle q_1 q_2 q_3 | \hat{P}^q_{12} | q_4 q_5 q_6 \rangle
\rightarrow & 
\sum_{q_1,q_3} 
\rho_{q_1}(\zeta_4,\zeta_1) \, \rho_{q_1}(\zeta_5,\zeta_2) \, \rho_{q_3}(\zeta_6,\zeta_3)
\\
\langle q_1 q_2 q_3 | \hat{P}^q_{23} | q_4 q_5 q_6 \rangle
\rightarrow & 
\sum_{q_1,q_2} 
\rho_{q_1}(\zeta_4,\zeta_1) \, \rho_{q_2}(\zeta_5,\zeta_2) \, \rho_{q_2}(\zeta_6,\zeta_3)
\\
\langle q_1 q_2 q_3 | \hat{P}^q_{13} | q_4 q_5 q_6 \rangle
\rightarrow & 
\sum_{q_1,q_2} 
\rho_{q_1}(\zeta_4,\zeta_1) \, \rho_{q_2}(\zeta_5,\zeta_2) \, \rho_{q_1}(\zeta_6,\zeta_3)
\\
\langle q_1 q_2 q_3 | \hat{P}^q_{12} \hat{P}^q_{23} | q_4 q_5 q_6 \rangle
\rightarrow & 
\sum_{q_1} 
\rho_{q_1}(\zeta_4,\zeta_1) \, \rho_{q_1}(\zeta_5,\zeta_2) \, \rho_{q_1}(\zeta_6,\zeta_3)
\\
\langle q_1 q_2 q_3 | \hat{P}^q_{12} \hat{P}^q_{13} | q_4 q_5 q_6 \rangle
\rightarrow & 
\sum_{q_1} 
\rho_{q_1}(\zeta_4,\zeta_1) \, \rho_{q_1}(\zeta_5,\zeta_2) \, \rho_{q_1}(\zeta_6,\zeta_3)
\,,
\end{align}
\end{subequations}
whereas those at play for the pairing part are
\begin{subequations}
\begin{align}
\langle q_1 q_2 q_3 | 1 | q_4 q_5 q_6 \rangle
\rightarrow & 
\sum_{q_1,q_2} 
\kappa ^{*}_{q_1}(\zeta_1,\zeta_2) \, \kappa_{q_1}(\zeta_4 , \zeta_5) \, \rho_{q_2}(\zeta_6,\zeta_3)
\\
\langle q_1 q_2 q_3 | \hat{P}^q_{12} | q_4 q_5 q_6 \rangle
\rightarrow & 
\sum_{q_1,q_2} 
\kappa ^{*}_{q_1}(\zeta_1,\zeta_2) \, \kappa_{q_1}(\zeta_4 , \zeta_5) \, \rho_{q_2}(\zeta_6,\zeta_3)
\\
\langle q_1 q_2 q_3 | \hat{P}^q_{23} | q_4 q_5 q_6 \rangle
\rightarrow & 
\sum_{q_1} 
\kappa ^{*}_{q_1}(\zeta_1,\zeta_2) \, \kappa_{q_1}(\zeta_4 , \zeta_5) \, \rho_{q_1}(\zeta_6,\zeta_3)
\\
\langle q_1 q_2 q_3 | \hat{P}^q_{13} | q_4 q_5 q_6 \rangle
\rightarrow & 
\sum_{q_1} 
\kappa ^{*}_{q_1}(\zeta_1,\zeta_2) \, \kappa_{q_1}(\zeta_4 , \zeta_5) \, \rho_{q_1}(\zeta_6,\zeta_3)
\\
\langle q_1 q_2 q_3 | \hat{P}^q_{12} \hat{P}^q_{23} | q_4 q_5 q_6 \rangle
\rightarrow & 
\sum_{q_1} 
\kappa ^{*}_{q_1}(\zeta_1,\zeta_2) \, \kappa_{q_1}(\zeta_4 , \zeta_5) \, \rho_{q_1}(\zeta_6,\zeta_3)
\\
\langle q_1 q_2 q_3 | \hat{P}^q_{12} \hat{P}^q_{13} | q_4 q_5 q_6 \rangle
\rightarrow & 
\sum_{q_1} 
\kappa ^{*}_{q_1}(\zeta_1,\zeta_2) \, \kappa_{q_1}(\zeta_4 , \zeta_5) \, \rho_{q_1}(\zeta_6,\zeta_3)
\, .
\end{align}
\end{subequations}

\subsection{Spin part of the matrix element for the normal energy}

Using Eqs.~(\ref{eq:Skyrme_int:Double_ex_spin}), one arrives 
straightforwardly after one step of algebraic computation at
\begin{eqnarray}
E^{\rho\rho\rho}_{ex}
& = & -\frac{1}{8} \int \! d^3 r_1 \, d^3 r_2 \, d^3 r_3 \, d^3 r_4 \, 
                           d^3 r_5 \, d^3 r_6 \sum_{q_1,q_2} \,
\delta(\{\vec{r}\}=\vec{r}) \; 
\vec{k}^{\,\ast}_{\vec{r}_1 \vec{r}_2}
\cdot 
\vec{k}^{\,}_{\vec{r}_4 \vec{r}_6}
\; \nonumber \\
&  & \quad
\times \Big[
\rho_{q_1}(\vec{r}_4,\vec{r}_1) \, \rho_{q_2}(\vec{r}_5,\vec{r}_2) \, \rho_{q_2}(\vec{r}_6,\vec{r}_3)
+\vec{s}_{q_1}(\vec{r}_4,\vec{r}_1) \, \cdot \vec{s}_{q_2}(\vec{r}_5,\vec{r}_2) \, \rho_{q_2}(\vec{r}_6,\vec{r}_3) \nonumber
\\
& & \qquad
+\rho_{q_1}(\vec{r}_4,\vec{r}_1) \, \vec{s}_{q_2}(\vec{r}_5,\vec{r}_2) \, \cdot \vec{s}_{q_2}(\vec{r}_6,\vec{r}_3) \nonumber
+\vec{s}_{q_1}(\vec{r}_4,\vec{r}_1) \, \rho_{q_2}(\vec{r}_5,\vec{r}_2) \, \cdot \vec{s}_{q_2}(\vec{r}_6,\vec{r}_3) \nonumber
\\
& & \qquad 
+ i \sum_{\nu \kappa \lambda} \epsilon_{\nu \kappa \lambda} \, s_{q_1,\nu}(\vec{r}_4,\vec{r}_1) \, s_{q_2,\lambda}(\vec{r}_5,\vec{r}_2) \, s_{q_2,\kappa}(\vec{r}_6,\vec{r}_3)
\Big] \, ,
\end{eqnarray}
where Eqs.~(\ref{eq:Skyrme_int:nonlocdensities:rho}) and~(\ref{eq:Skyrme_int:nonlocdensities:s}) have been utilized under the form
\begin{subequations}
\begin{align}
\sum_{\sigma_1 \sigma_4} \langle \sigma_1 | 1 | \sigma_4 \rangle \rho_{q_1}(\zeta_4,\zeta_1)  &= \rho_{q_1} ( \vec{r}_4, \vec{r}_1 ) \, ,
\\
\sum_{\sigma_1 \sigma_4}  \langle \sigma_1 |  \hat{\sigma}_{\nu}  | \sigma_4 \rangle   \rho_{q_1}(\zeta_4,\zeta_1)    &= s_{q_1,\nu} ( \vec{r}_4, \vec{r}_1 ) \, .
\end{align}
\end{subequations}


\subsection{Spin part of the matrix element for the pairing energy}

Expressing the pairing part of the EDF kernel in terms of non-local pair-spin 
densities is trickier. Using Eq.~(\ref{eq:Skyrme_int:Double_ex_spin}) to 
express spin-exchange operators in terms of spin Pauli matrices, let us take 
one resulting term, i.e.\ the one proportional to 
$\hat{\vec{\sigma}}_1 \cdot \hat{\vec{\sigma}}_2$, to illustrate the 
procedure. One needs to compute
\begin{eqnarray}
\label{eq:spin:pair}
\lefteqn{
\sum_{\sigma_1 \sigma_2 \sigma_3 \sigma_4 \sigma_5 \sigma_6}
\langle  \hat{\vec{\sigma}}_1 \cdot \hat{\vec{\sigma}}_2  \rangle \;
\kappa ^{*}_{q_1}(\zeta_1,\zeta_2) \, \kappa_{q_1}(\zeta_4 , \zeta_5) \, \rho_{q_1}(\zeta_6,\zeta_3)
} \nonumber \\
 &= &
 \sum_{\sigma_1 \sigma_2 \sigma_3 \sigma_4 \sigma_5 \sigma_6}
  \langle \sigma_1 | \hat{\vec{\sigma}}  | \sigma_4 \rangle
  \cdot
   \langle \sigma_2 | \hat{\vec{\sigma}}  | \sigma_5 \rangle \;
 \delta_{\sigma_3\sigma_6} \,
 \kappa ^{*}_{q_1}(\zeta_1,\zeta_2) \, \kappa_{q_1}(\zeta_4 , \zeta_5) \, \rho_{q_1}(\zeta_6,\zeta_3) \nonumber
\\
&= &
 \sum_{\sigma_1 \sigma_2 \sigma_4 \sigma_5}
  \langle \sigma_1 | \hat{\vec{\sigma}}  | \sigma_4 \rangle
  \cdot
   \langle \sigma_2 | \hat{\vec{\sigma}}  | \sigma_5 \rangle \;
 \kappa ^{*}_{q_1}(\zeta_1,\zeta_2) \, \kappa_{q_1}(\zeta_4 , \zeta_5) \, 
 \rho_{q_1}(\vec{r}_6,\vec{r}_3)\, . 
\end{eqnarray}
To do so, one exploits the relations 
\begin{subequations}
\label{eq:kappa:rhotilde}
\begin{eqnarray}
\kappa_{q}(\vec{r} \sigma, \vec{r}\,' \sigma' )
& = & 2 \, \bar{\sigma}' \, \tilde{\rho}_{q}(\vec{r} \sigma, \vec{r}\,' \bar{\sigma}') \, ,
\\
\kappa^{\ast}_{q}(\vec{r} \sigma, \vec{r}\,' \sigma' )
& = & 2 \, \bar{\sigma}' \, 
       \tilde{\rho}^{\ast}_{q} (\vec{r} \sigma, \vec{r}\,' \bar{\sigma}') \, ,
\end{eqnarray}
\end{subequations}
Eqs.~(\ref{eq:density:matrices:su2:development}) 
and~(\ref{eq:Skyrme_int:nonlocdensities:symm}), as well as the following 
set of relations involving  matrix elements of spin Pauli matrices
\begin{subequations}
\begin{eqnarray}
\langle \sigma_1 | \hat{\sigma}_\mu | \sigma_2 \rangle 
&=& - 4 \, \sigma_1 \, \sigma_2 \, \langle \bar{\sigma}_2 | \hat{\sigma}_\mu | \bar{\sigma}_1 \rangle \, ,
\\
\hat{\sigma}_{\nu} \hat{\sigma}_{\nu} 
&=& \delta_{\nu \nu} \, ,
\\
\sum_{\sigma} \langle \sigma | \hat{\sigma}_{\nu}  | \sigma \rangle 
&=& 0 \, ,
\\
\sum_{\sigma} \langle \sigma | \sigma \rangle 
&=& 2 \, ,
\\
\sum_{\sigma} \langle \sigma | \hat{\sigma}_{\mu} \hat{\sigma}_{\nu}  \hat{\sigma}_{\lambda} \hat{\sigma}_{\kappa} | \sigma \rangle  
&=&  \delta_{\mu \nu} \, \delta_{\lambda \kappa} \,
    -\delta_{\mu \lambda} \, \delta_{\nu \kappa} \,
    +\delta_{\mu \kappa} \, \delta_{\nu \lambda}
\, ,
\end{eqnarray}
\end{subequations}
to perform the following algebraic manipulations
\begin{align}
& \sum_{\sigma_1\sigma_2\sigma_4\sigma_5} \langle \sigma_1 | \hat{\vec{\sigma}}  | \sigma_4 \rangle
  \cdot
   \langle \sigma_2 | \hat{\vec{\sigma}}  | \sigma_5 \rangle \;
 \kappa ^{*}_{q_1}(\zeta_1,\zeta_2) \, \kappa_{q_1}(\zeta_4 , \zeta_5) 
\nonumber \\
&=
 \sum_{\sigma_1\sigma_2\sigma_4\sigma_5}   \langle \sigma_1 | \hat{\vec{\sigma}}  | \sigma_4 \rangle
  \cdot
   \langle \sigma_2 | \hat{\vec{\sigma}}  | \sigma_5 \rangle \; 
 4 \, \sigma_2 \, \sigma_5 \,
\tilde{\rho}^{\ast}_{q_1} (\vec{r}_1 \sigma_1, \vec{r}_2 \bar{\sigma}_2) \,
\tilde{\rho}_{q_1} (\vec{r}_4 \sigma_4, \vec{r}_5 \bar{\sigma}_5) 
\nonumber  \\
&=
- \sum_{\sigma_1\sigma_2\sigma_4\sigma_5}   \langle \sigma_1 | \hat{\vec{\sigma}}  | \sigma_4 \rangle
  \cdot
   \langle \bar{\sigma}_5 | \hat{\vec{\sigma}}  | \bar{\sigma}_2 \rangle \;
\tilde{\rho}^{\ast}_{q_1} (\vec{r}_1 \sigma_1, \vec{r}_2 \bar{\sigma}_2) \,
\tilde{\rho}_{q_1} (\vec{r}_4 \sigma_4, \vec{r}_5 \bar{\sigma}_5) 
\nonumber \\
&=
- \sum_{\sigma_1\sigma_2\sigma_4\sigma_5}   \langle \sigma_1 | \hat{\vec{\sigma}}  | \sigma_4 \rangle
  \cdot
   \langle \sigma_5 | \hat{\vec{\sigma}}  | \sigma_2 \rangle \;
\tilde{\rho}^{\ast}_{q_1} (\vec{r}_1 \sigma_1, \vec{r}_2 \sigma_2) \,
\tilde{\rho}_{q_1} (\vec{r}_4 \sigma_4, \vec{r}_5 \sigma_5) 
\nonumber \\
&=
- \frac{1}{4}
\sum_{\sigma_1\sigma_2\sigma_4\sigma_5} \sum_{\nu}
   \langle \sigma_1 | \hat{\sigma}_{\nu} | \sigma_4 \rangle \;
   \langle \sigma_5 | \hat{\sigma}_{\nu}  | \sigma_2 \rangle \,
\Big(
 \tilde{\rho}^{\ast}_{q_1} ( \vec{r}_1, \vec{r}_2 ) \delta_{\sigma_1 \sigma_2}
+
\sum_{\kappa} \tilde{s}^{\ast}_{q_1,\kappa} ( \vec{r}_1, \vec{r}_2 ) \
\langle \sigma_2 | \sigma_\kappa | \sigma_1 \rangle
\Big)
\nonumber \\ & \hspace{1.0cm}
\times \Big(
 \tilde{\rho}_{q_1} ( \vec{r}_4, \vec{r}_5 ) \, \delta_{\sigma_4 \sigma_5}
+
\sum_{\lambda} \tilde{s}_{q_1,\lambda} ( \vec{r}_4, \vec{r}_5 ) \; \langle \sigma_4 | \sigma_\lambda | \sigma_5 \rangle
\Big)
\nonumber \\
&=
- \frac{1}{4}
\Big[
\sum_{\sigma} \sum_{\nu}
   \langle \sigma | \hat{\sigma}_{\nu} \hat{\sigma}_{\nu}  | \sigma \rangle \; 
   \tilde{\rho}^{\ast}_{q_1} ( \vec{r}_1, \vec{r}_2 ) \, 
   \tilde{\rho}_{q_1} ( \vec{r}_4, \vec{r}_5 )
   +
  \sum_{\sigma} \sum_{\nu\lambda}
   \langle \sigma | \hat{\sigma}_{\nu} \hat{\sigma}_{\nu} \hat{\sigma}_{\lambda}  | \sigma \rangle \;
   \tilde{\rho}^{\ast}_{q_1} ( \vec{r}_1, \vec{r}_2 ) \, 
   \tilde{s}_{q_1,\lambda} ( \vec{r}_4, \vec{r}_5 )
\nonumber \\
&   
   +
     \sum_{\sigma} \sum_{\nu\kappa}
   \langle \sigma | \hat{\sigma}_{\nu} \hat{\sigma}_{\nu} \hat{\sigma}_{\kappa}  | \sigma \rangle \; 
   \tilde{s}^{\ast}_{q_1,\kappa} ( \vec{r}_1, \vec{r}_2 ) \,
   \tilde{\rho}_{q_1} ( \vec{r}_4, \vec{r}_5 ) 
   +
     \sum_{\sigma} \sum_{\nu\lambda\kappa}
   \langle \sigma | \hat{\sigma}_{\nu} \hat{\sigma}_{\lambda} \hat{\sigma}_{\nu} \hat{\sigma}_{\kappa}  | \sigma \rangle  \;
  \tilde{s}^{\ast}_{q_1,\kappa} ( \vec{r}_1, \vec{r}_2 ) \,
  \tilde{s}_{q_1,\lambda} ( \vec{r}_4, \vec{r}_5 ) 
\Big]
\nonumber \\
&=
- \frac{1}{2}
\Big[
3 \, \tilde{\rho}^{\ast}_{q_1} ( \vec{r}_1, \vec{r}_2 ) \, 
     \tilde{\rho}_{q_1} ( \vec{r}_4, \vec{r}_5 )
-
   \sum_{\lambda}
     \tilde{s}^{\ast}_{q_1,\lambda} ( \vec{r}_1, \vec{r}_2 ) \, 
     \tilde{s}_{q_1,\lambda} ( \vec{r}_4, \vec{r}_5 ) 
\Big] \, .
\end{align}
The normal density matrix $\rho_{q_1}(\vec{r}_6,\vec{r}_3)$ in 
Eq.~(\ref{eq:spin:pair}) is not involved in these manipulations 
and has been omitted for brevity. Altogether, the evaluation of 
Eq.~(\ref{eq:example:Ekkr}) requires the identities
\begin{subequations}
\begin{align}
\sum_{\sigma_1 \sigma_2 \sigma_4 \sigma_5}
\langle \sigma_1 \sigma_2 | 1 | \sigma_4 \sigma_5 \rangle \;
\kappa ^{*}_{q_1}(\zeta_1,\zeta_2) \, \kappa_{q_1}(\zeta_4 , \zeta_5)
=&  \frac{1}{2}
\Big[
\tilde{\rho}^{\ast}_{q_1} ( \vec{r}_1, \vec{r}_2 ) \, 
\tilde{\rho}_{q_1} ( \vec{r}_4, \vec{r}_5 )
+
   \sum_{\nu}
     \tilde{s}^{\ast}_{q_1,\nu} ( \vec{r}_1, \vec{r}_2 ) \
     \tilde{s}_{q_1,\nu} ( \vec{r}_4, \vec{r}_5 ) 
\Big] \, ,
\\
\sum_{\sigma_1 \sigma_2 \sigma_4 \sigma_5}
\langle \sigma_1 \sigma_2 | \hat{\sigma}_{1,\nu} | \sigma_4 \sigma_5 \rangle \; 
\kappa ^{*}_{q_1}(\zeta_1,\zeta_2) \, \kappa_{q_1}(\zeta_4 , \zeta_5)
=& \frac{1}{2}
\Big[
\tilde{\rho}^{\ast}_{q_1} ( \vec{r}_1, \vec{r}_2 ) \, \tilde{s}_{q_1,\nu} ( \vec{r}_4, \vec{r}_5 )
+\tilde{s}^{\ast}_{q_1,\nu} ( \vec{r}_1, \vec{r}_2 ) \, \tilde{\rho}_{q_1} ( \vec{r}_4, \vec{r}_5 )
\nonumber \\
&
-i \sum_{\lambda \kappa} \epsilon_{\nu \lambda \kappa} \, \tilde{s}^{\ast}_{q_1,\lambda} ( \vec{r}_1, \vec{r}_2 ) \, \tilde{s}_{q_1,\kappa} ( \vec{r}_4, \vec{r}_5 )
\Big] \, ,
\\
\sum_{\sigma_1 \sigma_2 \sigma_4 \sigma_5}
\langle \sigma_1 \sigma_2 | \hat{\sigma}_{2,\nu} | \sigma_4 \sigma_5 \rangle \;
   \kappa ^{*}_{q_1}(\zeta_1,\zeta_2) \, \kappa_{q_1}(\zeta_4 , \zeta_5)
=&
  -\frac{1}{2}
\Big[
\tilde{\rho}^{\ast}_{q_1} ( \vec{r}_1, \vec{r}_2 ) \, \tilde{s}_{q_1,\nu} ( \vec{r}_4, \vec{r}_5 )
+\tilde{s}^{\ast}_{q_1,\nu} ( \vec{r}_1, \vec{r}_2 ) \, \tilde{\rho}_{q_1} ( \vec{r}_4, \vec{r}_5 )
\nonumber \\
&
+i \sum_{\lambda \kappa} \epsilon_{\nu \lambda \kappa} \, \tilde{s}^{\ast}_{q_1,\lambda} ( \vec{r}_1, \vec{r}_2 ) \,  \tilde{s}_{q_1,\kappa} ( \vec{r}_4, \vec{r}_5 )
\Big] \, ,
\\
\sum_{\sigma_1 \sigma_2 \sigma_4 \sigma_5}
\langle \sigma_1 \sigma_2 | \hat{\sigma}_{1,\nu} \hat{\sigma}_{2,\nu} | \sigma_4 \sigma_5 \rangle \; \kappa ^{*}_{q_1}(\zeta_1,\zeta_2) \, \kappa_{q_1}(\zeta_4 , \zeta_5)
=&
- \frac{1}{2}
\Big[
3\tilde{\rho}^{\ast}_{q_1} ( \vec{r}_1, \vec{r}_2 ) \, \tilde{\rho}_{q_1} ( \vec{r}_4, \vec{r}_5 )
-  \tilde{s}^{\ast}_{q_1,\nu} ( \vec{r}_1, \vec{r}_2 ) \, \tilde{s}_{q_1,\nu} ( \vec{r}_4, \vec{r}_5 ) 
\Big] \, .
\end{align}
\end{subequations}

\end{widetext}

\subsection{Applying gradient operators}
\label{sec:Skyrme_int:Rules}

Now that the matrix element has been evaluated, the integrand contains 
delta functions and differential operators acting on non-local densities. 
The latter must be evaluated prior to utilizing the former. Simple rules 
can be obtained that express the action of specific combinations of gradient 
operators on non-local densities  in terms of local 
densities~\cite{vautherin72a,engel75}. Those rules work identically for 
$\rho_q (\vec{r}  ,\vec{r}\,' )$, $s_{q, \mu} (\vec{r}  ,\vec{r}\,' )$, 
$\tilde{\rho}_q(\vec{r}\,',\vec{r}) $ or 
$\tilde{s}_{q,\nu}(\vec{r}\,',\vec{r})$. Defining
${\cal P}_{q,(\nu)}^{\vec{r}\vec{r}\,'}$, 
${\cal P}_{q,(\nu)}^{\vec{r}}$, 
${\cal T}_{q,(\nu)}^{\vec{r}}$ and
${\cal J}_{q,\mu(\nu)}^{\vec{r}}$ as generic notation for the densities,
for each column on the right-hand-side of the table
\begin{alignat*}{5}
{\cal P}_{q,(\nu)}^{\vec{r}^{ }\vec{r}'} 
& \equiv  \big\{ \rho_q(\vec{r}\,',\vec{r}) 
         \, & ; & \; s_{q,\nu}(\vec{r}\,',\vec{r}) 
         \, & ; & \; \tilde{\rho}_q(\vec{r}\,',\vec{r}) 
         \, & ; & \; \tilde{s}_{q,\nu}(\vec{r}\,',\vec{r}) & \big\} \, , \\
{\cal P}_{q,(\nu)}^{\vec{r}} 
&\equiv  \big\{ \rho_{q}(\vec{r})  
         \, & ; & \; s_{q, \nu}(\vec{r}) 
         \, & ; & \; \tilde{\rho}_{q}(\vec{r}) 
         \, & ; & \; \tilde{s}_{q, \nu}(\vec{r}) & \big\} \, ,
\\
{\cal T}_{q,(\nu)}^{\vec{r}} 
&\equiv  \big\{ \tau_{q}(\vec{r}) 
          \, & ; & \;  T_{q,\nu}(\vec{r})  
          \, & ; & \; \tilde{\tau}_{q}(\vec{r}) 
          \, & ; & \; \tilde{T}_{q,\nu}(\vec{r}) & \big\} \, ,
\\
{\cal J}_{q,\mu(\nu)}^{\vec{r}}  
&\equiv  \big\{  j_{q,\mu}(\vec{r}) 
          \, & ; & \; J_{q,\mu\nu}(\vec{r}) 
          \, & ; & \; \tilde{j}_{q,\mu}(\vec{r}) 
          \, & ; & \; \tilde{J}_{q,\mu\nu}(\vec{r}) & \big\} \, ,
\end{alignat*}
there is a set of four relations
\begin{subequations}
\label{eq:Skyrme_int:Rules}
\begin{align}
\nabla_{\vec{r},\mu} \; {\cal P}_{q,(\nu)}^{\vec{r}\vec{r}\,'} \; \Big|_{\vec{r}=\vec{r}'} 
=& \, \frac{1}{2} \nabla_{\mu} {\cal P}_{q,(\nu)}^{\vec{r}} + {\mathrm i} {\cal J}_{q,\mu(\nu)}^{\vec{r}}
\, , \\
\nabla_{\vec{r}\,',\mu} \; {\cal P}_{q,(\nu)}^{\vec{r}\vec{r}\,'} \; \Big|_{\vec{r}=\vec{r}'} 
=& \, \frac{1}{2} \nabla_{\mu} {\cal P}_{q,(\nu)}^{\vec{r}} - {\mathrm i} {\cal J}_{q,\mu(\nu)}^{\vec{r}}
\, , \\
\Delta_{\vec{r}} \; {\cal P}_{q,(\nu)}^{\vec{r}\vec{r}\,'} \; \Big|_{\vec{r}=\vec{r}'} 
=& \, \frac{1}{2} \Delta {\cal P}_{q,(\nu)}^{\vec{r}} - {\cal T}_{q,(\nu)}^{\vec{r}} + {\mathrm i} \vec{\nabla} \cdot \vec{{\cal J}}_{q,(\nu)}^{\vec{r}}
\, , \\
\Delta_{\vec{r}\,'} \; {\cal P}_{q,(\nu)}^{\vec{r}\vec{r}\,'} \; \Big|_{\vec{r}=\vec{r}'} 
=& \, \frac{1}{2} \Delta {\cal P}_{q,(\nu)}^{\vec{r}} - {\cal T}_{q,(\nu)}^{\vec{r}} - {\mathrm i} \vec{\nabla} \cdot \vec{{\cal J}}_{q,(\nu)}^{\vec{r}}
 \, ,
\end{align}
\end{subequations}
Applying those rules and exploiting the delta functions, one ends 
up with a local energy density expressed in terms of the local densities of 
interest.

\section{One-body fields} 
\label{Skyrme_int:fields}

Having the explicit expression of the EDF kernel at hand, its contributions 
to the one-body fields entering the HFB equations can be derived. Normal 
and anomalous fields are gathered into the HFB Hamiltonian 
matrix~\cite{ring80a,blaizotripka}
\begin{equation}
\mathcal{H} \equiv \left(
\begin{array} {cc}
h^{q} & \Delta^{q} \\
-\Delta^{q \ast} & -h^{q \ast}
\end{array}
\right) \, ,
\end{equation}
and are respectively defined as
\begin{alignat}{5}
\label{eq:Skyrme_int:field1}
h^{q}_{\beta\alpha} 
& \equiv & \frac{\delta E}{\delta \rho^q_{\alpha\beta}} \,  ,
  & \qquad & 
\Delta^{q}_{\alpha\beta} 
& \equiv &  \frac{\delta E}{\delta \kappa^{q \, \ast}_{\alpha\beta}} \,  , 
\end{alignat}
for \mbox{$\beta \leq \alpha$}. 
Field $h$ is hermitian,
\mbox{$h^q_{\beta\alpha}=h^{q *}_{\alpha\beta}$}, whereas $\Delta^q$ is 
skew symmetric \mbox{$\Delta^q_{\beta\alpha} = - \Delta^q_{\alpha\beta}$}. 
%
%
%
These fields can be specified either in a configuration basis 
\mbox{$\{ \alpha, \beta \} \in \{ i, j \}$} or in coordinate representation
\mbox{$\{ \alpha, \beta \} \in\{ \xi, \xi' \}$}.

Below, we explicitly provide contributions to the HFB Hamiltonian 
that derive from the energy functional defined by Eqs.~(\ref{ErhorhoNP}), 
(\ref{EkappakappaNP}), (\ref{ErhorhorhoNP})
and~(\ref{ErhokappakappaNP}), which constitutes just a part of the 
complete EDF kernel. In a realistic calculation, additional terms
contribute to the one-body fields in the HFB equation, such as   
the center-of-mass correction, the Coulomb interaction, as well as constraints,
in particular the obligatory one on neutron and proton numbers. 
None of these will be specified here.

The EDF being a functional of local densities, it 
is of advantage to calculate contributions to the matrix elements of 
the one-body fields in a configuration basis through the chain rule
\begin{subequations}
\label{eq:Skyrme_int:field}
\begin{eqnarray}
\label{eq:Skyrme_int:field:h} 
h^{q}_{ji}  
& = &  \int \! d^3r \, \Bigg[
      \frac{\delta \mathcal{E}}{\delta \rho_q (\vec{r})}
      \frac{\delta \rho_q (\vec{r})}{\delta \rho^q_{ij}}
      + \frac{\delta \mathcal{E}}{\delta \tau_q (\vec{r})}
      \frac{\delta \tau_q (\vec{r})}{\delta \rho^q_{ij}}
      \nonumber \\
&   & 
  + \sum_{\mu \nu}
      \frac{\delta \mathcal{E}}{\delta J_{q,\mu \nu}(\vec{r})}
      \frac{\delta J_{q, \mu \nu}(\vec{r})}
           {\delta \rho^q_{ij}}
  + \sum_{\mu}
      \frac{\delta \mathcal{E}}{\delta s_{q,\mu}(\vec{r})}
      \frac{\delta s_{q,\mu}(\vec{r})}
           {\delta \rho^q_{ij}}
      \nonumber \\
&  & + \sum_{\mu}
      \frac{\delta \mathcal{E}}{\delta T_{q,\mu}(\vec{r})}
      \frac{\delta T_{q,\mu}(\vec{r})}{\delta \rho^{q}_{ij}}
    + \sum_{\mu}
      \frac{\delta \mathcal{E}}{\delta j_{q,\mu} (\vec{r})}
      \frac{\delta j_{q,\mu}(\vec{r})}{\delta \rho^{q}_{ij}} \Bigg] \nonumber \, ,
 \nonumber  
 \\
& & \\
\Delta^{q}_{ij} 
& = &  \int \! d^3r \, \Bigg[
      \frac{\delta \mathcal{E}}{\delta \tilde{\rho}^{\ast}_q (\vec{r})}
      \frac{\delta \tilde{\rho}^{\ast}_q (\vec{r})}{\delta \kappa^{q \, \ast}_{ij}}
      + \frac{\delta \mathcal{E}}{\delta \tilde{\tau}^{\ast}_q (\vec{r})}
      \frac{\delta \tilde{\tau}^{\ast}_q (\vec{r})}{\delta \kappa^{q \, \ast}_{ij}}
\label{eq:Skyrme_int:field:D} \nonumber \\
&  & + \sum_{\mu \nu}
      \frac{\delta \mathcal{E}}{\delta \tilde{J}^{\ast}_{q,\mu \nu}(\vec{r})}
      \frac{\delta \tilde{J}^{\ast}_{q, \mu \nu}(\vec{r})}{\delta \kappa^{q \, \ast}_{ij}} 
      \Bigg]  \,  .
\end{eqnarray}
\end{subequations}
The functional derivatives of the local densities can be obtained for 
\mbox{$j \leq i$} as
\begin{subequations}
\begin{eqnarray}
\frac{\delta \rho_q (\vec{r})}{\delta \rho^q_{ij}}
& = & \sum_{\sigma}
      \varphi^{\ast}_{j} (\vec{r} \sigma q) \, \varphi_{i} (\vec{r} \sigma q) \, ,
      \\
\frac{\delta \tau_q (\vec{r})}{\delta \rho^q_{ij}} 
& = & \sum_{\sigma}
      \big[ \vec{\nabla} \varphi^{\ast}_{j} (\vec{r} \sigma q)\big] 
            \cdot \big[\vec{\nabla} \varphi_{i} (\vec{r} \sigma q)\big]   \, , \\
\frac{\delta J_{q, \mu \nu} (\vec{r})}{\delta \rho^q_{ij}}  
& = & - \sum_{\sigma \sigma'} \frac{{\mathrm i}}{2}
      \Big\{   \varphi^{\ast}_{j} (\vec{r} \sigma' q) \, 
       \langle \sigma' | \hat{\sigma}_\nu | \sigma \rangle \, \nabla_\mu \varphi_{i} (\vec{r} \sigma q)  
      \nonumber \\ 
&   & \hspace{0.3cm}
      - \nabla_\mu \varphi^{\ast}_{j} (\vec{r} \sigma' q) \, \langle \sigma' | \hat{\sigma}_\nu | \sigma \rangle \,  \varphi_{i} (\vec{r} \sigma q)  \Big\} \,,
      \\
\frac{\delta s_{q, \nu} (\vec{r})}{\delta \rho^q_{ij}}  
& = & \sum_{\sigma \sigma'}
      \varphi^{\ast}_{j} (\vec{r} \sigma' q) \, \langle \sigma' | \hat{\sigma}_\nu | \sigma \rangle \, \varphi_{i} (\vec{r} \sigma q)  \, ,
      \\
\frac{\delta T_{q, \nu} (\vec{r})}{\delta \rho^q_{ij}}  
& = & \sum_{\sigma \sigma'}
      \vec{\nabla} \varphi^{\ast}_{j} (\vec{r} \sigma' q) \, \langle \sigma' | \hat{\sigma}_\nu | \sigma \rangle \,
      \cdot \vec{\nabla} \varphi_{i} (\vec{r} \sigma q)  \, ,
      \\
\frac{\delta j_{q, \mu} (\vec{r})}{\delta \rho^q_{ij}}  
& = & - \sum_{\sigma} \frac{{\mathrm i}}{2}
     \Big\{ \varphi^{\ast}_{j} (\vec{r} \sigma q) \, \big[\nabla_{\mu} \varphi_{i} (\vec{r} \sigma q)\big]
      \nonumber \\ 
&  & \hspace{1cm}
      - \big[\nabla_{\mu} \varphi^{\ast}_{j} (\vec{r} \sigma q)\big] \, \varphi_{i} (\vec{r} \sigma q)
      \Big\}  \, ,    
      \\
\frac{\delta \tilde{\rho}^{\ast}_q (\vec{r})}{\delta \kappa^{q \, \ast}_{ij}}
& = & \sum_{\sigma} 4 \bar{\sigma} \,
      \varphi^{\ast}_{i} (\vec{r} \bar{\sigma} q) \, \varphi^{\ast}_{j} (\vec{r} \sigma q) \, ,
      \\      
\frac{\delta \tilde{\tau}^{\ast}_q (\vec{r})}{\delta \kappa^{q \, \ast}_{ij}} 
& = & \sum_{\sigma} 4 \bar{\sigma} \,
      \big[ \vec{\nabla} \varphi^{\ast}_{i} (\vec{r} \bar{\sigma} q)\big] 
            \cdot \big[\vec{\nabla} \varphi^{\ast}_{j} (\vec{r} \sigma q)\big]   \, , \\
\frac{\delta \tilde{J}^{\ast}_{q, \mu \nu} (\vec{r})}{\delta \kappa^{q \, \ast}_{ij}}  
& = & - \sum_{\sigma \sigma'} 4 \bar{\sigma}' \frac{{\mathrm i}}{2} 
      \Big\{   \varphi^{\ast}_{i} (\vec{r} \bar{\sigma}' q) \, 
      \langle \sigma' | \hat{\sigma}_\nu | \sigma \rangle \, \nabla_\mu \varphi^{\ast}_{j} (\vec{r} \sigma q)  
      \nonumber \\ 
&  & \hspace{0.3cm}
      - \nabla_\mu \varphi^{\ast}_{i} (\vec{r} \bar{\sigma}' q) \, \langle \sigma' | \hat{\sigma}_\nu | \sigma \rangle \,  \varphi^{\ast}_{j} (\vec{r} \sigma q)  \Big\} \,.
\end{eqnarray}
\end{subequations}
The functional derivatives of the local energy density ${\cal E}$ define
the local potentials
\begin{subequations}
\label{eq:Skyrme_int:fields:EDF}
\begin{eqnarray}
U_q (\vec{r}) & \equiv & \frac{\delta {\cal E}}{\delta \rho_q (\vec{r})} \, , \\
B_q (\vec{r}) & \equiv & \frac{\delta {\cal E}}{\delta \tau_q (\vec{r})} \, , \\
W_{q,\mu \nu} (\vec{r}) & \equiv & \frac{\delta {\cal E}}{\delta J_{q,\mu \nu} (\vec{r})} \, , \\
S_{q,\mu} (\vec{r}) & \equiv & \frac{\delta {\cal E}}{\delta s_{q,\mu} (\vec{r})} \, , \\
C_{q,\mu} (\vec{r}) & \equiv & \frac{\delta {\cal E}}{\delta T_{q,\mu} (\vec{r})} \, ,  \\
A_{q,\mu} (\vec{r}) & \equiv & \frac{\delta {\cal E}}{\delta j_{q,\mu} (\vec{r})} \, ,  \\
\tilde{U}_q (\vec{r}) & \equiv & \frac{\delta {\cal E}}{\delta \tilde{\rho}^{\ast}_q (\vec{r})} \, , \\
\tilde{B}_q (\vec{r}) & \equiv & \frac{\delta {\cal E}}{\delta \tilde{\tau}^{\ast}_q (\vec{r})}  \, ,\\
\tilde{W}_{q,\mu \nu} (\vec{r}) & \equiv & \frac{\delta {\cal E}}{\delta \tilde{J}^{\ast}_{q,\mu \nu} (\vec{r})} \, .
\end{eqnarray}
\end{subequations}
Matrix elements in the configuration basis can be related to those in the coordinate basis through
\begin{subequations}
\begin{align}
h^{q}_{ji} 
& \equiv \iint \! d^3r \,  d^3r' \sum_{\sigma \sigma'} 
         \varphi^{\ast}_{j} (\vec{r}\,' \sigma' q) \, 
         h^q (\vec{r} \sigma , \vec{r}\,' \sigma') \,
         \varphi_{i} (\vec{r} \sigma q) \, ,      
\\
\Delta^{q}_{ij} 
& \equiv \iint \! d^3r \, d^3r'  
         \sum_{\sigma \sigma'} \, 4 \bar{\sigma}' \, 
         \varphi^{\ast}_{j} (\vec{r}\,' \bar{\sigma}' q) \, 
         \tilde{h}^q (\vec{r} \sigma , \vec{r}\,' \sigma') \, 
         \varphi^{\ast}_{i} (\vec{r} \sigma q) \, .
\end{align}
\end{subequations}
In the present case,  fields are local in coordinate space representation, 
i.e.\
\begin{subequations}
\begin{eqnarray}
h^q (\vec{r} \sigma , \vec{r}\,' \sigma') 
&\equiv& \delta(\vec{r}-\vec{r}\,') \, h^q_{\sigma\sigma'} (\vec{r})  \, , \\
\tilde{h}^q (\vec{r} \sigma, \vec{r}\,' \sigma') 
&\equiv& \delta(\vec{r}-\vec{r}\,') \, \tilde{h}^q_{\sigma\sigma'} (\vec{r}) 
\, ,
\end{eqnarray}
\end{subequations}
with the generic structure
\begin{widetext}
\begin{subequations}
\begin{eqnarray}
\label{eq:Skyrme_int:fields:h}
h^q_{\sigma\sigma'} (\vec{r})
& = & U_q (\vec{r}) \delta_{\sigma \sigma'}
      - \sum_{\mu} \nabla_{\mu} \, B_q (\vec{r}) \nabla_{\mu} \delta_{\sigma \sigma'}
      - \frac{{\mathrm i}}{2} \sum_{\mu \nu}  \Big[ W_{q,\mu\nu} (\vec{r}) \, \nabla_\mu \, \langle \sigma' | \hat{\sigma}_\nu | \sigma \rangle
                     + \nabla_\mu \, \langle \sigma' | \hat{\sigma}_\nu | \sigma \rangle \, W_{q,\mu\nu} (\vec{r}) \Big]
 \nonumber \\ 
&  &
      + \sum_{\nu} S_{q,\nu} (\vec{r}) \, \langle \sigma' | \hat{\sigma}_\nu | \sigma \rangle
      - \sum_{\mu \nu} \nabla_\mu  \, C_{q,\nu} (\vec{r}) \, 
        \langle \sigma' | \hat{\sigma}_\nu | \sigma \rangle \, \nabla_\mu 
      - \sum_{\mu}  \frac{{\mathrm i}}{2} \Big[  A_{q,\mu} (\vec{r}) \, \nabla_{\mu}
                       + \nabla_{\mu} \, A_{q,\mu} (\vec{r}) \Big] \delta_{\sigma \sigma'} \, ,
\\
\tilde{h}^q_{\sigma\sigma'} (\vec{r})
& = &   
 \tilde{U}_{q} (\vec{r}) \delta_{\sigma \sigma'}
  - \sum_{\mu} \nabla_{\mu} \, \tilde{B}_q (\vec{r}) \nabla_{\mu} \delta_{\sigma \sigma'}
  - \frac{{\mathrm i}}{2} \sum_{\mu \nu}  \Big[ \tilde{W}_{q,\mu\nu} (\vec{r}) \, \nabla_\mu \, \langle \sigma' | \hat{\sigma}_\nu | \sigma \rangle
  + \nabla_\mu \, \langle \sigma' | \hat{\sigma}_\nu | \sigma \rangle \, \tilde{W}_{q,\mu\nu} (\vec{r}) \Big]
 \, .
\label{eq:Skyrme_int:fields:D}
\end{eqnarray}
\end{subequations}
In Eq.~(\ref{eq:Skyrme_int:fields:h}), gradient operators act to their 
right on both the local potentials and on the wave function the fields 
$h^q_{\sigma\sigma'} (\vec{r})$ and $\tilde{h}^q_{\sigma\sigma'} (\vec{r})$
are applied to. For the functional constructed here, the overall 
structure of the two fields is the same as for traditional Skyrme EDF 
parametrizations, the only difference being additional terms the local 
potentials.


\subsection{Local potentials}

Explicit expressions of the local potentials deriving from the EDF kernel 
defined through Eqs.~(\ref{ErhorhoNP}), (\ref{EkappakappaNP}), 
(\ref{ErhorhorhoNP}) and~(\ref{ErhokappakappaNP}) are given by
\begin{subequations}
\begin{eqnarray}
U_q 
&=&
 2 A^{\rho_1 \rho_1} \rho_{q}  
+2 A^{\rho_1 \rho_2} \rho_{\bar{q}}  
+A^{\tau_1 \rho_1} \tau_{q}  
+A^{\tau_1 \rho_2} \tau_{\bar{q}}  
-2 A^{\nabla \rho_1 \nabla \rho_1} \Delta \rho_{q}  
-2 A^{\nabla \rho_1 \nabla \rho_2} \Delta \rho_{\bar{q}} 
 \nonumber \\ 
& &
+2 B^{\rho_1 \rho_1 \rho_2} \rho_{q} \rho_{\bar{q}}  
+B^{\rho_1 \rho_1 \rho_2} \rho_{\bar{q}} \rho_{\bar{q}}  
+B^{s_1 s_1 \rho_2} \vec{s}_{\bar{q}} \cdot \vec{s}_{\bar{q}}  
+B^{\tilde{\rho}^*_1 \tilde{\rho}_1 \rho_2} \tilde{\rho}^*_{\bar{q}} \tilde{\rho}_{\bar{q}}  
+2 B^{\tau_1 \rho_1 \rho_1} \tau_{q} \rho_{q}  
 \nonumber \\ 
& &
+B^{\tau_1 \rho_1 \rho_2} \tau_{q} \rho_{\bar{q}}  
+B^{\tau_1 \rho_1 \rho_2} \tau_{\bar{q}} \rho_{\bar{q}}  
+2 B^{\tau_1 \rho_2 \rho_2} \tau_{\bar{q}} \rho_{q}  
+B^{T_1 s_1 \rho_2} \vec{T}_{\bar{q}} \cdot \vec{s}_{\bar{q}}  
+B^{T_1 s_2 \rho_1} \vec{T}_{q} \cdot \vec{s}_{\bar{q}}  
+B^{\tilde{\tau}^*_1 \tilde{\rho}_1 \rho_2} \tilde{\tau}^*_{\bar{q}} \tilde{\rho}_{\bar{q}}  
+B^{\tilde{\tau}_1 \tilde{\rho}^*_1 \rho_2} \tilde{\tau}_{\bar{q}} \tilde{\rho}^*_{\bar{q}}  
 \nonumber \\ 
& &
-2 B^{\nabla \rho_1 \nabla \rho_1 \rho_1} (\Delta \rho_{q}) \rho_{q}  
- B^{\nabla \rho_1 \nabla \rho_1 \rho_1} (\vec{\nabla} \rho_{q}) \cdot (\vec{\nabla} \rho_{q}) 
-2 B^{\nabla \rho_1 \nabla \rho_1 \rho_2} (\Delta \rho_{q}) \rho_{\bar{q}}  
-2 B^{\nabla \rho_1 \nabla \rho_1 \rho_2} (\vec{\nabla} \rho_{q}) \cdot (\vec{\nabla} \rho_{\bar{q}})
 \nonumber \\ 
& &
+B^{\nabla \rho_1 \nabla \rho_1 \rho_2} (\vec{\nabla} \rho_{\bar{q}}) \cdot (\vec{\nabla} \rho_{\bar{q}})
-B^{\nabla \rho_1 \nabla \rho_2 \rho_1} (\Delta \rho_{\bar{q}}) \rho_{q}  
-B^{\nabla \rho_1 \nabla \rho_2 \rho_1} (\Delta \rho_{\bar{q}}) \rho_{\bar{q}}  
-B^{\nabla \rho_1 \nabla \rho_2 \rho_1} (\vec{\nabla} \rho_{\bar{q}}) \cdot (\vec{\nabla} \rho_{\bar{q}})
 \nonumber \\ 
& &
+\sum_{\mu\nu} \Big[
  B^{\nabla s_1 \nabla s_1 \rho_1} (\nabla_\mu s_{q, \nu}) (\nabla_\mu s_{q, \nu}) 
+B^{\nabla s_1 \nabla s_1 \rho_2} (\nabla_\mu s_{\bar{q}, \nu}) (\nabla_\mu s_{\bar{q}, \nu})
+B^{\nabla s_1 \nabla s_2 \rho_1} (\nabla_\mu s_{q, \nu}) (\nabla_\mu s_{\bar{q}, \nu})  
\Big]
 \nonumber \\ 
& &
-B^{\nabla \rho_1 \nabla s_1 s_1} (\Delta \vec{s}_{q}) \cdot \vec{s}_{q}  
-B^{\nabla \rho_1 \nabla s_1 s_2} (\Delta \vec{s}_{q}) \cdot \vec{s}_{\bar{q}} 
-B^{\nabla \rho_1 \nabla s_2 s_1} (\Delta \vec{s}_{\bar{q}}) \cdot \vec{s}_{q}  
-B^{\nabla \rho_1 \nabla s_2 s_2} (\Delta \vec{s}_{\bar{q}}) \cdot \vec{s}_{\bar{q}}   
 \nonumber \\ 
& &
+\sum_{\mu\nu} \Big[
-B^{\nabla \rho_1 \nabla s_1 s_1} (\nabla_\mu s_{q, \nu}) (\nabla_\mu s_{q, \nu}) 
-B^{\nabla \rho_1 \nabla s_1 s_2} (\nabla_\mu s_{q, \nu}) (\nabla_\mu s_{\bar{q}, \nu})  
-B^{\nabla \rho_1 \nabla s_2 s_1} (\nabla_\mu s_{\bar{q}, \nu}) (\nabla_\mu s_{q, \nu})  
 \nonumber \\ 
& &
-B^{\nabla \rho_1 \nabla s_2 s_2} (\nabla_\mu s_{\bar{q}, \nu}) (\nabla_\mu s_{\bar{q}, \nu}) 
\Big]
+B^{\nabla \tilde{\rho}^*_1 \nabla \tilde{\rho}_1 \rho_1} (\vec{\nabla} \tilde{\rho}^*_{q}) \cdot (\vec{\nabla} \tilde{\rho}_{q}) 
+B^{\nabla \tilde{\rho}^*_1 \nabla \tilde{\rho}_1 \rho_2} (\vec{\nabla} \tilde{\rho}^*_{\bar{q}}) \cdot (\vec{\nabla} \tilde{\rho}_{\bar{q}})  
 \nonumber \\ 
& &
-B^{\nabla \tilde{\rho}^*_1 \nabla \rho_1 \tilde{\rho}_1} (\Delta \tilde{\rho}^*_{q}) \tilde{\rho}_{q}  
-B^{\nabla \tilde{\rho}^*_1 \nabla \rho_1 \tilde{\rho}_1} (\vec{\nabla} \tilde{\rho}^*_{q}) \cdot (\vec{\nabla} \tilde{\rho}_{q}) 
-B^{\nabla \tilde{\rho}^*_1 \nabla \rho_2 \tilde{\rho}_1} (\Delta \tilde{\rho}^*_{\bar{q}}) \tilde{\rho}_{\bar{q}}  
-B^{\nabla \tilde{\rho}^*_1 \nabla \rho_2 \tilde{\rho}_1} (\vec{\nabla} \tilde{\rho}^*_{\bar{q}}) \cdot (\vec{\nabla} \tilde{\rho}_{\bar{q}})
 \nonumber \\ 
& &
-B^{\nabla \tilde{\rho}_1 \nabla \rho_1 \tilde{\rho}^*_1} (\Delta \tilde{\rho}_{q}) \tilde{\rho}^*_{q}  
-B^{\nabla \tilde{\rho}_1 \nabla \rho_1 \tilde{\rho}^*_1} (\vec{\nabla} \tilde{\rho}_{q}) \cdot (\vec{\nabla} \tilde{\rho}^*_{q})
-B^{\nabla \tilde{\rho}_1 \nabla \rho_2 \tilde{\rho}^*_1} (\Delta \tilde{\rho}_{\bar{q}}) \tilde{\rho}^*_{\bar{q}}  
-B^{\nabla \tilde{\rho}_1 \nabla \rho_2 \tilde{\rho}^*_1} (\vec{\nabla} \tilde{\rho}_{\bar{q}}) \cdot (\vec{\nabla} \tilde{\rho}^*_{\bar{q}})
 \nonumber \\ 
& &
+B^{j_1 j_1 \rho_1} \vec{j}_{q} \cdot \vec{j}_{q}  
+B^{j_1 j_1 \rho_2} \vec{j}_{\bar{q}} \cdot \vec{j}_{\bar{q}}  
+B^{j_1 j_2 \rho_1} \vec{j}_{q} \cdot \vec{j}_{\bar{q}}  
+\sum_{\mu\nu} \Big[
  B^{J_1 J_1 \rho_1} J_{q, \mu \nu} J_{q, \mu \nu}  
+B^{J_1 J_1 \rho_2} J_{\bar{q}, \mu \nu} J_{\bar{q}, \mu \nu}  
 \nonumber \\ 
& &
+B^{J_1 J_2 \rho_1} J_{q, \mu \nu} J_{\bar{q}, \mu \nu}  
+B^{\tilde{J}^*_1 \tilde{J}_1 \rho_1} \tilde{J}^*_{q, \mu \nu} \tilde{J}_{q, \mu \nu}  
+B^{\tilde{J}^*_1 \tilde{J}_1 \rho_2} \tilde{J}^*_{\bar{q}, \mu \nu} \tilde{J}_{\bar{q}, \mu \nu}  
\Big]
\, ,
\\
 S_{q,\nu} 
& = &
 2 A^{s_1 s_1} s_{q, \nu}  
+2 A^{s_1 s_2} s_{\bar{q}, \nu}  
+A^{T_1 s_1} T_{q, \nu}  
+A^{T_1 s_2} T_{\bar{q}, \nu}  
-2 A^{\nabla s_1 \nabla s_1} \Delta s_{q, \nu}  
-2 A^{\nabla s_1 \nabla s_2} \Delta s_{\bar{q}, \nu}  
 \nonumber \\ 
& &
+2 B^{s_1 s_1 \rho_2} s_{q, \nu} \rho_{\bar{q}}  
+B^{T_1 s_1 \rho_2} T_{q, \nu} \rho_{\bar{q}}  
+B^{T_1 s_2 \rho_1} T_{\bar{q}, \nu} \rho_{\bar{q}}  
 \nonumber \\ 
& &
+2 B^{\tau_1 s_1 s_1} \tau_{q} s_{q, \nu}  
+B^{\tau_1 s_1 s_2} \tau_{q} s_{\bar{q}, \nu}  
+B^{\tau_1 s_1 s_2} \tau_{\bar{q}} s_{\bar{q}, \nu}  
+2 B^{\tau_1 s_2 s_2} \tau_{\bar{q}} s_{q, \nu}  
-2 B^{\nabla s_1 \nabla s_1 \rho_1} (\Delta s_{q, \nu}) \rho_{q}  
 \nonumber \\ 
& &
-2 B^{\nabla s_1 \nabla s_1 \rho_1} (\vec{\nabla} s_{q, \nu}) \cdot (\vec{\nabla} \rho_{q})
-2 B^{\nabla s_1 \nabla s_1 \rho_2} (\Delta s_{q, \nu}) \rho_{\bar{q}}  
-2 B^{\nabla s_1 \nabla s_1 \rho_2} (\vec{\nabla} s_{q, \nu}) \cdot (\vec{\nabla} \rho_{\bar{q}}) 
 \nonumber \\  
& &
-B^{\nabla s_1 \nabla s_2 \rho_1} (\Delta s_{\bar{q}, \nu}) \rho_{q}  
-B^{\nabla s_1 \nabla s_2 \rho_1} (\Delta s_{\bar{q}, \nu}) \rho_{\bar{q}}  
-B^{\nabla s_1 \nabla s_2 \rho_1} (\vec{\nabla} s_{\bar{q}, \nu}) \cdot (\vec{\nabla} \rho_{q})
-B^{\nabla s_1 \nabla s_2 \rho_1} (\vec{\nabla} s_{\bar{q}, \nu}) \cdot (\vec{\nabla} \rho_{\bar{q}})
 \nonumber \\  
& &
-B^{\nabla \rho_1 \nabla s_1 s_1} (\Delta \rho_{q}) s_{q, \nu}  
-B^{\nabla \rho_1 \nabla s_1 s_2} (\Delta \rho_{q}) s_{\bar{q}, \nu}  
-B^{\nabla \rho_1 \nabla s_1 s_2} (\vec{\nabla} \rho_{q}) \cdot (\vec{\nabla} s_{\bar{q}, \nu}) 
+B^{\nabla \rho_1 \nabla s_1 s_2} (\vec{\nabla} \rho_{\bar{q}}) \cdot (\vec{\nabla} s_{\bar{q}, \nu})
 \nonumber \\  
& &
-B^{\nabla \rho_1 \nabla s_2 s_1} (\Delta \rho_{\bar{q}}) s_{\bar{q}, \nu}  
-B^{\nabla \rho_1 \nabla s_2 s_1} (\vec{\nabla} \rho_{\bar{q}}) \cdot (\vec{\nabla} s_{\bar{q}, \nu})
+B^{\nabla \rho_1 \nabla s_2 s_1} (\vec{\nabla} \rho_{q}) \cdot (\vec{\nabla} s_{\bar{q}, \nu})  
-B^{\nabla \rho_1 \nabla s_2 s_2} (\Delta \rho_{\bar{q}}) s_{q, \nu}  
 \nonumber \\  
& &
+\sum_{\mu} \Big[
  B^{j_1 J_1 s_1} j_{q,\mu} J_{q, \mu \nu}  
+B^{j_1 J_1 s_2} j_{\bar{q},\mu} J_{\bar{q}, \mu \nu}  
+B^{j_1 J_2 s_1} j_{q,\mu} J_{\bar{q}, \mu \nu}  
+B^{j_1 J_2 s_2} j_{\bar{q},\mu} J_{q, \mu \nu} 
\Big]
 \nonumber \\  
& &
 +\sum_{\mu\lambda\kappa} \epsilon_{\nu \lambda \kappa}  \Big[
-B^{\nabla s_1 J_1 s_1} (\nabla_\mu J_{q, \mu \lambda}) s_{q, \kappa}  \, 
-B^{\nabla s_1 J_1 s_1} J_{q, \mu \lambda} (\nabla_\mu s_{q, \kappa})  \, 
+B^{\nabla s_1 J_1 s_1} (\nabla_\mu s_{q, \lambda}) J_{q, \mu \kappa}  \, 
 \nonumber \\  
& &
-B^{\nabla s_1 J_1 s_2} (\nabla_\mu J_{q, \mu \lambda}) s_{\bar{q}, \kappa}  \, 
-B^{\nabla s_1 J_1 s_2} J_{q, \mu \lambda} (\nabla_\mu s_{\bar{q}, \kappa})  \, 
+B^{\nabla s_1 J_1 s_2} (\nabla_\mu s_{\bar{q}, \lambda}) J_{\bar{q}, \mu \kappa}  \, 
-B^{\nabla s_1 J_2 s_1} (\nabla_\mu J_{\bar{q}, \mu \lambda}) s_{q, \kappa}  \,  
 \nonumber \\  
& &
+2B^{\nabla s_1 J_2 s_1} (\nabla_\mu s_{q, \lambda}) J_{\bar{q}, \mu \kappa}  \, 
-B^{\nabla s_1 J_2 s_2} (\nabla_\mu J_{\bar{q}, \mu \lambda}) s_{\bar{q}, \kappa}  \, 
-B^{\nabla s_1 J_2 s_2} J_{\bar{q}, \mu \lambda} (\nabla_\mu s_{\bar{q}, \kappa})  \, 
+B^{\nabla s_1 J_2 s_2} (\nabla_\mu s_{\bar{q}, \lambda}) J_{q, \mu \kappa}  \, 
 \nonumber \\  
& &
+{\mathrm i}B^{\tilde{J}^*_1 \tilde{J}_1 s_1} \tilde{J}^*_{q, \mu \lambda} \tilde{J}_{q, \mu \kappa}  \, 
+{\mathrm i}B^{\tilde{J}^*_1 \tilde{J}_1 s_2} \tilde{J}^*_{\bar{q}, \mu \lambda} \tilde{J}_{\bar{q}, \mu \kappa}  \, 
\Big]
+\sum_{\mu} \Big[
  {\mathrm i}B^{\nabla \tilde{\rho}^*_1 \tilde{J}_1 s_1} (\nabla_\mu \tilde{\rho}^*_{q}) \tilde{J}_{q, \mu \nu}  
+{\mathrm i}B^{\nabla \tilde{\rho}^*_1 \tilde{J}_1 s_2} (\nabla_\mu \tilde{\rho}^*_{\bar{q}}) \tilde{J}_{\bar{q}, \mu \nu}  
 \nonumber \\  
& &
+{\mathrm i}B^{\nabla \tilde{\rho}_1 \tilde{J}^*_1 s_1} (\nabla_\mu \tilde{\rho}_{q}) \tilde{J}^*_{q, \mu \nu}  
+{\mathrm i}B^{\nabla \tilde{\rho}_1 \tilde{J}^*_1 s_2} (\nabla_\mu \tilde{\rho}_{\bar{q}}) \tilde{J}^*_{\bar{q}, \mu \nu}  
-{\mathrm i}B^{\nabla s_1 \tilde{J}^*_1 \tilde{\rho}_1} (\nabla_\mu \tilde{J}^*_{q, \mu \nu}) \tilde{\rho}_{q}  
-{\mathrm i}B^{\nabla s_1 \tilde{J}^*_1 \tilde{\rho}_1} \tilde{J}^*_{q, \mu \nu} (\nabla_\mu \tilde{\rho}_{q})  
 \nonumber \\  
& &
-{\mathrm i}B^{\nabla s_2 \tilde{J}^*_1 \tilde{\rho}_1} (\nabla_\mu \tilde{J}^*_{\bar{q}, \mu \nu}) \tilde{\rho}_{\bar{q}}  
-{\mathrm i}B^{\nabla s_2 \tilde{J}^*_1 \tilde{\rho}_1} \tilde{J}^*_{\bar{q}, \mu \nu} (\nabla_\mu \tilde{\rho}_{\bar{q}}) 
-{\mathrm i}B^{\nabla s_1 \tilde{J}_1 \tilde{\rho}^*_1} (\nabla_\mu \tilde{J}_{q, \mu \nu}) \tilde{\rho}^*_{q}  
-{\mathrm i}B^{\nabla s_1 \tilde{J}_1 \tilde{\rho}^*_1} \tilde{J}_{q, \mu \nu} (\nabla_\mu \tilde{\rho}^*_{q})
 \nonumber \\  
& &
-{\mathrm i}B^{\nabla s_2 \tilde{J}_1 \tilde{\rho}^*_1} (\nabla_\mu \tilde{J}_{\bar{q}, \mu \nu}) \tilde{\rho}^*_{\bar{q}}  
-{\mathrm i}B^{\nabla s_2 \tilde{J}_1 \tilde{\rho}^*_1} \tilde{J}_{\bar{q}, \mu \nu} (\nabla_\mu \tilde{\rho}^*_{\bar{q}})  
\Big]
\, ,
\\
B_q 
&=&
 \frac{\hbar^2}{2m}
+A^{\tau_1 \rho_1} \rho_{q}  
+A^{\tau_1 \rho_2} \rho_{\bar{q}}  
+B^{\tau_1 \rho_1 \rho_1} \rho_{q} \rho_{q}  
+B^{\tau_1 \rho_1 \rho_2} \rho_{q} \rho_{\bar{q}}  
+B^{\tau_1 \rho_2 \rho_2} \rho_{\bar{q}} \rho_{\bar{q}}  
+B^{\tau_1 s_1 s_1} \vec{s}_{q} \cdot \vec{s}_{q}  
 \nonumber \\  
& &
+B^{\tau_1 s_1 s_2} \vec{s}_{q} \cdot \vec{s}_{\bar{q}}  
+B^{\tau_1 s_2 s_2} \vec{s}_{\bar{q}} \cdot \vec{s}_{\bar{q}}  
+B^{\tau_1 \tilde{\rho}^*_1 \tilde{\rho}_1} \tilde{\rho}^*_{q} \tilde{\rho}_{q}  
+B^{\tau_2 \tilde{\rho}^*_1 \tilde{\rho}_1} \tilde{\rho}^*_{\bar{q}} \tilde{\rho}_{\bar{q}}  
\, ,
\\
 C_{q,\nu} 
&=&
 A^{T_1 s_1} s_{q, \nu}  
+A^{T_1 s_2} s_{\bar{q}, \nu}  
+B^{T_1 s_1 \rho_2} s_{q, \nu} \rho_{\bar{q}}  
+B^{T_1 s_2 \rho_1} s_{\bar{q}, \nu} \rho_{q}  
\, ,
\\
 A_{q,\mu} 
&=&
 2 A^{j_1 j_1} j_{q,\mu}  
+2 A^{j_1 j_2} j_{\bar{q},\mu}  
+2 B^{j_1 j_1 \rho_1} j_{q,\mu} \rho_{q}  
+2 B^{j_1 j_1 \rho_2} j_{q,\mu} \rho_{\bar{q}}  
+B^{j_1 j_2 \rho_1} j_{\bar{q},\mu} \rho_{q}  
+B^{j_1 j_2 \rho_1} j_{\bar{q},\mu} \rho_{\bar{q}}  
 \nonumber \\  
& &
+\sum_{\nu} \Big( 
  B^{j_1 J_1 s_1} J_{q, \mu \nu} s_{q, \nu}  
+B^{j_1 J_1 s_2} J_{q, \mu \nu} s_{\bar{q}, \nu}  
+B^{j_1 J_2 s_1} J_{\bar{q}, \mu \nu} s_{q, \nu}  
+B^{j_1 J_2 s_2} J_{\bar{q}, \mu \nu} s_{\bar{q}, \nu}  
\Big)
 \nonumber  \\  
& &
+{\mathrm i}B^{\nabla \tilde{\rho}^*_1 j_1 \tilde{\rho}_1} (\nabla_\mu \tilde{\rho}^*_{q}) \tilde{\rho}_{q}  
+{\mathrm i}B^{\nabla \tilde{\rho}^*_1 j_2 \tilde{\rho}_1} (\nabla_\mu \tilde{\rho}^*_{\bar{q}}) \tilde{\rho}_{\bar{q}}  
+{\mathrm i}B^{\nabla \tilde{\rho}_1 j_1 \tilde{\rho}^*_1} (\nabla_\mu \tilde{\rho}_{q}) \tilde{\rho}^*_{q}  
+{\mathrm i}B^{\nabla \tilde{\rho}_1 j_2 \tilde{\rho}^*_1} (\nabla_\mu \tilde{\rho}_{\bar{q}}) \tilde{\rho}^*_{\bar{q}}  
\,,
\\
 W_{q,\mu\nu} 
&=&
 2 A^{J_1 J_1} J_{q, \mu \nu}  
+2 A^{J_1 J_2} J_{\bar{q}, \mu \nu}  
+2 B^{J_1 J_1 \rho_1} J_{q, \mu \nu} \rho_{q}  
+2 B^{J_1 J_1 \rho_2} J_{q, \mu \nu} \rho_{\bar{q}}  
 \nonumber \\  
& &
+B^{J_1 J_2 \rho_1} J_{\bar{q}, \mu \nu} \rho_{q}  
+B^{J_1 J_2 \rho_1} J_{\bar{q}, \mu \nu} \rho_{\bar{q}}  
+B^{j_1 J_1 s_1} j_{q,\mu} s_{q, \nu}  
+B^{j_1 J_1 s_2} j_{q,\mu} s_{\bar{q}, \nu}  
+B^{j_1 J_2 s_1} j_{\bar{q},\mu} s_{\bar{q}, \nu}  
 \nonumber \\  
& &
+B^{j_1 J_2 s_2} j_{\bar{q},\mu} s_{q, \nu}  
+B^{\tilde{J}^*_1 J_1 \tilde{\rho}_1} \tilde{J}^*_{q, \mu \nu} \tilde{\rho}_{q}  
+B^{\tilde{J}^*_1 J_2 \tilde{\rho}_1} \tilde{J}^*_{\bar{q}, \mu \nu} \tilde{\rho}_{\bar{q}}  
+B^{\tilde{J}_1 J_1 \tilde{\rho}^*_1} \tilde{J}_{q, \mu \nu} \tilde{\rho}^*_{q}  
+B^{\tilde{J}_1 J_2 \tilde{\rho}^*_1} \tilde{J}_{\bar{q}, \mu \nu} \tilde{\rho}^*_{\bar{q}}  
 \nonumber \\  
& &
+\sum_{\lambda \kappa} \epsilon_{\nu\lambda\kappa} \,  \Big[
-B^{\nabla s_1 J_1 s_1} (\nabla_\mu s_{q, \lambda}) s_{q, \kappa}  \, 
-B^{\nabla s_1 J_1 s_2} (\nabla_\mu s_{q, \lambda}) s_{\bar{q}, \kappa}  \, 
-B^{\nabla s_1 J_2 s_1} (\nabla_\mu s_{\bar{q}, \lambda}) s_{\bar{q}, \kappa}  \, 
 \nonumber \\  
& &
-B^{\nabla s_1 J_2 s_2} (\nabla_\mu s_{\bar{q}, \lambda}) s_{q, \kappa}  \, 
\Big]
\, ,\\
 \tilde{U}_q 
&=&
 A^{\tilde{\rho}^*_1 \tilde{\rho}_1} \tilde{\rho}_{q}  
+A^{\tilde{\tau}_1 \tilde{\rho}^*_1} \tilde{\tau}_{q}  
-A^{\nabla \tilde{\rho}^*_1 \nabla \tilde{\rho}_1} \Delta \tilde{\rho}_{q}  
+B^{\tilde{\rho}^*_1 \tilde{\rho}_1 \rho_2} \tilde{\rho}_{q} \rho_{\bar{q}}  
+B^{\tilde{\tau}_1 \tilde{\rho}^*_1 \rho_2} \tilde{\tau}_{q} \rho_{\bar{q}}  
+B^{\tau_1 \tilde{\rho}^*_1 \tilde{\rho}_1} \tau_{q} \tilde{\rho}_{q}  
+B^{\tau_2 \tilde{\rho}^*_1 \tilde{\rho}_1} \tau_{\bar{q}} \tilde{\rho}_{q}  
 \nonumber \\  
& &
-B^{\nabla \tilde{\rho}^*_1 \nabla \tilde{\rho}_1 \rho_1} (\Delta \tilde{\rho}_{q}) \rho_{q}  
-B^{\nabla \tilde{\rho}^*_1 \nabla \tilde{\rho}_1 \rho_1} (\vec{\nabla} \tilde{\rho}_{q}) \cdot (\vec{\nabla} \rho_{q})  
-B^{\nabla \tilde{\rho}^*_1 \nabla \tilde{\rho}_1 \rho_2} (\Delta \tilde{\rho}_{q}) \rho_{\bar{q}}  
-B^{\nabla \tilde{\rho}^*_1 \nabla \tilde{\rho}_1 \rho_2} (\vec{\nabla} \tilde{\rho}_{q}) \cdot (\vec{\nabla} \rho_{\bar{q}})  
 \nonumber \\  
& &
-B^{\nabla \tilde{\rho}^*_1 \nabla \rho_1 \tilde{\rho}_1} (\Delta \rho_{q}) \tilde{\rho}_{q}  
-B^{\nabla \tilde{\rho}^*_1 \nabla \rho_1 \tilde{\rho}_1} (\vec{\nabla} \rho_{q}) \cdot (\vec{\nabla} \tilde{\rho}_{q})  
-B^{\nabla \tilde{\rho}^*_1 \nabla \rho_2 \tilde{\rho}_1} (\Delta \rho_{\bar{q}}) \tilde{\rho}_{q}  
-B^{\nabla \tilde{\rho}^*_1 \nabla \rho_2 \tilde{\rho}_1} (\vec{\nabla} \rho_{\bar{q}}) \cdot (\vec{\nabla} \tilde{\rho}_{q})  
 \nonumber \\  
& &
+B^{\nabla \tilde{\rho}_1 \nabla \rho_1 \tilde{\rho}^*_1} (\vec{\nabla} \tilde{\rho}_{q}) \cdot (\vec{\nabla} \rho_{q})  
+B^{\nabla \tilde{\rho}_1 \nabla \rho_2 \tilde{\rho}^*_1} (\vec{\nabla} \tilde{\rho}_{q}) \cdot (\vec{\nabla} \rho_{\bar{q}})
+\sum_{\mu\nu} \Big[
  B^{\tilde{J}_1 J_1 \tilde{\rho}^*_1} \tilde{J}_{q, \mu \nu} J_{q, \mu \nu}  
+B^{\tilde{J}_1 J_2 \tilde{\rho}^*_1} \tilde{J}_{q, \mu \nu} J_{\bar{q}, \mu \nu}  
 \nonumber \\  
& &
-{\mathrm i}B^{\nabla \tilde{\rho}^*_1 \tilde{J}_1 s_1} \nabla_\mu \tilde{J}_{q, \mu \nu} s_{q, \nu}  
-{\mathrm i}B^{\nabla \tilde{\rho}^*_1 \tilde{J}_1 s_1} \tilde{J}_{q, \mu \nu} \nabla_\mu s_{q, \nu}  
-{\mathrm i}B^{\nabla \tilde{\rho}^*_1 \tilde{J}_1 s_2} \nabla_\mu \tilde{J}_{q, \mu \nu} s_{\bar{q}, \nu}  
-{\mathrm i}B^{\nabla \tilde{\rho}^*_1 \tilde{J}_1 s_2} \tilde{J}_{q, \mu \nu} \nabla_\mu s_{\bar{q}, \nu}  
\Big]
 \nonumber \\  
& &
-{\mathrm i}B^{\nabla \tilde{\rho}^*_1 j_1 \tilde{\rho}_1} (\vec{\nabla} \cdot \vec{j}_{q}) \tilde{\rho}_{q}  
-{\mathrm i}B^{\nabla \tilde{\rho}^*_1 j_1 \tilde{\rho}_1} \vec{j}_{q} \cdot (\vec{\nabla} \tilde{\rho}_{q})
-{\mathrm i}B^{\nabla \tilde{\rho}^*_1 j_2 \tilde{\rho}_1} (\vec{\nabla} \cdot \vec{j}_{\bar{q}}) \tilde{\rho}_{q}  
-{\mathrm i}B^{\nabla \tilde{\rho}^*_1 j_2 \tilde{\rho}_1} \vec{j}_{\bar{q}} \cdot (\vec{\nabla} \tilde{\rho}_{q})
  \\  
& & \nonumber
+{\mathrm i}B^{\nabla \tilde{\rho}_1 j_1 \tilde{\rho}^*_1} (\vec{\nabla} \tilde{\rho}_{q}) \cdot \vec{j}_{q}  
+{\mathrm i}B^{\nabla \tilde{\rho}_1 j_2 \tilde{\rho}^*_1} (\vec{\nabla} \tilde{\rho}_{q}) \cdot \vec{j}_{\bar{q}}  
+\sum_{\mu\nu} \Big[
   {\mathrm i}B^{\nabla s_1 \tilde{J}_1 \tilde{\rho}^*_1} (\nabla_\mu s_{q, \nu}) \tilde{J}_{q, \mu \nu}  
+{\mathrm i}B^{\nabla s_2 \tilde{J}_1 \tilde{\rho}^*_1} (\nabla_\mu s_{\bar{q}, \nu}) \tilde{J}_{q, \mu \nu}  
\Big]
\, ,
\\
 \tilde{B}_q 
&=&
 A^{\tilde{\tau}^*_1 \tilde{\rho}_1} \tilde{\rho}_{q}  
+B^{\tilde{\tau}^*_1 \tilde{\rho}_1 \rho_2} \tilde{\rho}_{q} \rho_{\bar{q}}  
\, ,
\\
 \tilde{W}_{q,\mu\nu} 
&=&
 A^{\tilde{J}^*_1 \tilde{J}_1}_1 \tilde{J}_{q, \mu \nu}  
+B^{\tilde{J}^*_1 \tilde{J}_1 \rho_1} \tilde{J}_{q, \mu \nu} \rho_{q}  
+B^{\tilde{J}^*_1 \tilde{J}_1 \rho_2} \tilde{J}_{q, \mu \nu} \rho_{\bar{q}}  
+B^{\tilde{J}^*_1 J_1 \tilde{\rho}_1} J_{q, \mu \nu} \tilde{\rho}_{q}  
 \nonumber \\  
& &
+B^{\tilde{J}^*_1 J_2 \tilde{\rho}_1} J_{\bar{q}, \mu \nu} \tilde{\rho}_{q}  
+{\mathrm i}B^{\nabla \tilde{\rho}_1 \tilde{J}^*_1 s_1} (\nabla_\mu \tilde{\rho}_{q}) s_{q, \nu}  
+{\mathrm i}B^{\nabla \tilde{\rho}_1 \tilde{J}^*_1 s_2} (\nabla_\mu \tilde{\rho}_{q}) s_{\bar{q}, \nu}  
+{\mathrm i}B^{\nabla s_1 \tilde{J}^*_1 \tilde{\rho}_1} (\nabla_\mu s_{q, \nu}) \tilde{\rho}_{q}  
 \nonumber \\  
& &
+{\mathrm i}B^{\nabla s_2 \tilde{J}^*_1 \tilde{\rho}_1} (\nabla_\mu s_{\bar{q}, \nu}) \tilde{\rho}_{q}  
+ \sum_{\lambda \kappa} \epsilon_{\nu \lambda \kappa} \,  \Big[
   {\mathrm i}B^{\tilde{J}^*_1 \tilde{J}_1 s_1} \tilde{J}_{q, \mu \lambda} s_{q, \kappa}  \, 
+{\mathrm i}B^{\tilde{J}^*_1 \tilde{J}_1 s_2} \tilde{J}_{q, \mu \lambda} s_{\bar{q}, \kappa}
\Big]
\, .
\end{eqnarray}
\end{subequations}
\end{widetext}


\section{Local gauge invariance}
\label{sec:Skyrme_int:Gauge}

\subsection{Gauge transformations}

The invariance of the energy under local gauge transformation traces 
back to the locality of the underlying interaction~\cite{blaizotripka}. 
Given that realistic nuclear interactions have no reason to be local, 
invariance of the diagonal EDF kernel under general local gauge 
transformations does not have to be required. On the other hand, invariance 
under Galilean transformations is mandatory. Given that Galilean 
transformations are nothing but a particular case of local gauge 
transformations, 
we now test the invariance of the nuclear EDF under the latter as a way 
to verify its invariance under the former. The the newly developed
EDF kernel happens to be invariant under general local gauge transformations 
indicates that the dependence of the pseudo potential up 
to second order in gradients represents an internally consistent 
approximation to a local finite-range three-body potential. 

Let us now characterize the behaviour of the EDF kernel under general gauge  
transformations~\cite{dobaczewski95a}. To do so, we first define the 
transformation law of the one-body density matrices, i.e.
\begin{subequations}
\label{eq:Skyrme_int:gauge:nonlocdensity}
\begin{align}
\rho^{\,\prime} (\vec{r}^{\,} \sigma q ,\vec{r}^{\,\prime} \sigma ' q') 
= e^{ {\mathrm i}\left( \phi(\vec{r}^{\,}) - \phi(\vec{r}^{\,\prime}) \right) } \,
  \rho^{\,} (\vec{r} \sigma q ,\vec{r}\,' \sigma ' q')  \, ,
\\
\kappa^{\,\prime} (\vec{r}^{\,} \sigma q ,\vec{r}^{\,\prime} \sigma ' q') 
= e^{ {\mathrm i}\left( \phi(\vec{r}^{\,}) + \phi(\vec{r}^{\,\prime}) \right) } \,
  \kappa^{\,} (\vec{r} \sigma q ,\vec{r}\,' \sigma ' q')  \, .
\end{align}
\end{subequations}
Galilean transformations are nothing but the particular gauge transformations obtained for $\phi(\vec{r}^{\,}) = \vec{p}\cdot\vec{r}/\hbar$, where $\vec{p}$ characterizes the Galilean boost. Based on Eq.~(\ref{eq:Skyrme_int:gauge:nonlocdensity}), the transformation 
law of the local densities from  which the EDF kernel is built is obtained as
\begin{subequations}
\label{eq:Skyrme_int:gauge:locdensities}
\begin{align}
\rho_{q}^{\, \prime}  
& = \rho_{q}  \, , \\
\tau_{q}^{\, \prime}  
& = \tau_{q}  + 2 \vec{j}_{q} \cdot (\vec{\nabla} \phi) + \rho_{q}  (\vec{\nabla} \phi)^2 \, ,  \\
\vec{j}_{q}^{\, \prime}  
& = \vec{j}_{q}  + \rho_{q}  (\vec{\nabla} \phi) 
 \, , \\
s_{q,\nu}^{\, \prime}  
& = s_{q,\nu}  \,, \\
T_{q,\nu}^{\, \prime}  
& = T_{q,\nu}  + \sum_{\mu} \big[ 2 J_{q,\mu \nu}  (\nabla_{\mu} \phi)  + s_{q,\nu} (\nabla_{\mu} \phi)^2 \big] \, , \\
J_{q,\mu \nu}^{\, \prime}  
& = J_{q,\mu \nu} + s_{q, \nu}  (\nabla_{\mu} \phi)
 \,  , \\
\tilde{\rho}_{q}^{\, \prime}  
& = e^{ 2 {\mathrm i} \phi} \, \tilde{\rho}_{q}  \, , \\
\tilde{\tau}_{q}^{\, \prime}  
& = e^{ 2 {\mathrm i} \phi} \,  \big[ \tilde{\tau}_{q} + {\mathrm i} (\vec{\nabla} \tilde{\rho}_{q}) \cdot (\vec{\nabla} \phi)
 - \tilde{\rho}_{q} (\vec{r}^{\,}) (\vec{\nabla} \phi)^2 \big] \, ,  \\
\tilde{J}_{q,\mu \nu}^{\, \prime}  
& = e^{ 2 {\mathrm i} \phi} \, \tilde{J}_{q,\mu \nu}  \,  . 
\end{align}
\end{subequations}
Although Eq.~(\ref{eq:Skyrme_int:gauge:locdensities}) makes use of neutron and proton densities, the same transformation laws hold for isoscalar and isovector densities. The latter are used in the following to characterize the gauge invariance of the EDF kernel.

\subsection{Normal part of the EDF kernel}

The gauge invariance of the normal part of the EDF kernel requires that
\begin{subequations}
\label{eq:Skyrme_int:gauge:condition_init}
\begin{align}
\label{eq:Skyrme_int:2body:gauge:condition_init}
{\cal E}^{\rho\rho \, \prime} - {\cal E}^{\rho\rho} 
& \equiv \Big[ {\cal E}^{\rho\rho} \Big]_{{\cal G}} = 0 
\, , \\
\label{eq:Skyrme_int:3body:gauge:condition_init}
{\cal E}^{\rho\rho\rho \, \prime} - {\cal E}^{\rho\rho\rho} 
& \equiv \Big[ {\cal E}^{\rho\rho\rho} \Big]_{{\cal G}} = 0
 \, ,
\end{align}
\end{subequations}
where ${\cal E}^{\rho\rho \, \prime}$ and  ${\cal E}^{\rho\rho\rho \, \prime}$
denote energy densities computed from the gauge-transformed densities defined 
in Eq.~(\ref{eq:Skyrme_int:gauge:locdensities}). In 
Eq.~(\ref{eq:Skyrme_int:gauge:condition_init}), square brackets with 
index $\mathcal{G}$ have to be zero for the EDF kernel to be gauge invariant. 
Such conditions can be fulfilled only if specific correlations between 
coupling constants are at play. Gauge transformation only affects normal 
densities $\tau_{t}$, $T_{t,\nu}$, $j_{t,\mu}$ and $J_{t,\mu\nu}$, following 
Eq.~(\ref{eq:Skyrme_int:gauge:locdensities}). The fact that 
$\tau_{t}^{\, \prime} -\tau_{t}$, $T_{t,\nu}^{\, \prime} -T_{t,\nu}$, 
$j_{t,\mu}^{\, \prime} -j_{t,\mu}$ and $J_{t,\mu\nu}^{\, \prime} -J_{t,\mu\nu}$
depend on densities $j_{t,\mu}$, $J_{t,\mu\nu}$, $\rho_t$ and $s_{t,\nu}$ in 
addition to the gauge function $\phi(\vec{r})$, implies that correlations only 
involve coefficients multiplying densities $\tau_{t}$, $T_{t,\nu}$, 
$j_{t,\mu}$ and $J_{t,\mu\nu}$ having the same spin and isospin character.

For the bilinear functional, the two densities involved in a given term 
are either both isoscalar or isovector and both scalar or vector, such 
that each gauge invariant combination involves only two terms of the 
functional. As a result condition 
Eq.~(\ref{eq:Skyrme_int:2body:gauge:condition_init}) is equivalent 
to requiring that
\begin{subequations}
\label{eq:Skyrme_int:2body:gauge:condition}
\begin{align}
\Big[ A^{\tau}_t \; \tau_t \rho_t + A^{j}_t \; \vec{j}_t \cdot \vec{j}_t \Big]_{{\cal G}}
&= 0  \,, \\
\Big[ A^{T}_t \; \vec{s}_t \cdot \vec{T}_t + \sum_{\mu\nu} A^{J}_t \; J_{t,\mu\nu} J_{t,\mu\nu} \Big]_{{\cal G}}
&= 0  \,, 
\end{align}
\end{subequations}
for $t\in \{0,1\}$ and is fulfilled whenever~\cite{dobaczewski95a}
\begin{align}
A_t^{\tau}
=-A_t^{j} \quad , \quad A_t^{T}=-A_t^{J} 
\, .
\end{align}
For the trilinear functional, such combinations can involve many more terms 
as two isovector or vector densities are always multiplied by an isoscalar 
or scalar density. Condition~(\ref{eq:Skyrme_int:3body:gauge:condition_init}) 
gives rise to seven independent relations that read
\begin{widetext}
\begin{subequations}
\label{eq:Skyrme_int:3body:gauge:condition}
\begin{align}
0 = & \Big[ B^\tau_0 \, \rho_0  \tau_{0} \rho_{0}  
+  B^j_0 \, \rho_0 \vec{j}_{0} \cdot \vec{j}_{0} \Big]_{{\cal G}} 
\,, \label{eq:Skyrme_int:3body:gauge:condition:isosc_normal} \\
0 = & \Big[ B^T_0 \, \rho_0 \vec{T}_{0} \cdot \vec{s}_{0}  
+ B^{\tau s}_0 \, \vec{s}_{0} \tau_{0} \cdot \vec{s}_{0} 
+ \sum_{\mu\nu} B^J_0 \, \rho_0 J_{0, \mu\nu} J_{0, \mu\nu}  
+ \sum_{\mu\nu} B^{Js}_0 \, s_{0, \nu} j_{0,\mu} J_{0, \mu\nu} \Big]_{{\cal G}} 
\,,\label{eq:Skyrme_int:3body:gauge:condition:isosc_spin}\\
0 = & \Big[ B^\tau_1 \, \rho_0 \tau_{1} \rho_{1}
+B^\tau_{10} \, \rho_1 \tau_{0} \rho_{1}   
+ B^j_1 \, \rho_0 \vec{j}_{1} \cdot \vec{j}_{1} 
+ B^j_{10} \, \rho_1 \vec{j}_{1} \cdot \vec{j}_{0}  \Big]_{{\cal G}}  
\,, \label{eq:Skyrme_int:3body:gauge:condition:isovec_normal} \\
0 = & \Big[  B^T_1 \, \rho_0 \vec{T}_{1} \cdot \vec{s}_{1}  
+ B^{\tau s}_{10} \, \vec{s}_{1} \tau_{0} \cdot \vec{s}_{1}    
+ \sum_{\mu\nu} B^J_1 \, \rho_0 J_{1, \mu\nu} J_{1, \mu\nu} 
+ \sum_{\mu\nu} B^{Js}_{01} \, s_{1, \nu} j_{0,\mu} J_{1, \mu\nu} \Big]_{{\cal G}} 
\,, \label{eq:Skyrme_int:3body:gauge:condition:isovec_spin1} \\
0 = & \Big[ B^T_{10} \, \rho_1 \vec{T}_{0} \cdot \vec{s}_{1} 
+ B^T_{01} \, \rho_1 \vec{T}_{1} \cdot \vec{s}_{0}        
+ B^{\tau s}_1 \, \vec{s}_{0} \tau_{1} \cdot \vec{s}_{1}  
+ \sum_{\mu\nu} B^J_{10} \, \rho_1 J_{1, \mu\nu} J_{0, \mu\nu}  
\nonumber \\ & \;
+ \sum_{\mu\nu} B^{Js}_1 \, s_{0, \nu} j_{1,\mu} J_{1, \mu\nu}   
+ \sum_{\mu\nu} B^{Js}_{10} \, s_{1, \nu} j_{1,\mu} J_{0, \mu\nu} \Big]_{{\cal G}}  
\,, \label{eq:Skyrme_int:3body:gauge:condition:isovec_spin2}\\
0 = & \Big[  \sum_{\mu \nu \lambda k} \epsilon_{\nu \lambda k} B^{\nabla s J}_0 \, s_{0,k}  (\nabla_{\mu}s_{0,\nu}) J_{0,\mu\lambda}
 \Big]_{{\cal G}}  
\,,\label{eq:Skyrme_int:3body:gauge:condition:isosc_3spin}
\\
0 = & \Big[  \sum_{\mu \nu \lambda k} \epsilon_{\nu \lambda k} \Big( B^{\nabla s J}_1 \, s_{0,k}  (\nabla_{\mu}s_{1,\nu}) J_{1,\mu\lambda}  
+ B^{\nabla s J}_{10} \, s_{1,k}  (\nabla_{\mu}s_{1,\nu}) J_{0,\mu\lambda}  
+ B^{\nabla s J}_{01} \, s_{1,k}  (\nabla_{\mu}s_{0,\nu}) J_{1,\mu\lambda} \Big) \Big]_{{\cal G}}  
\,.  \label{eq:Skyrme_int:3body:gauge:condition:isovec_3spin}
\end{align}
\end{subequations}
\end{widetext}
Condition~(\ref{eq:Skyrme_int:3body:gauge:condition:isosc_normal}) involves
 functional terms containing scalar-isoscalar densities. 
Condition~(\ref{eq:Skyrme_int:3body:gauge:condition:isosc_spin}) involves 
functional terms containing isoscalar densities among which two are vector 
densities. Condition~(\ref{eq:Skyrme_int:3body:gauge:condition:isovec_normal})
involves functional terms containing two isovector densities and no vector 
densities. Conditions~(\ref{eq:Skyrme_int:3body:gauge:condition:isovec_spin1}) 
and~(\ref{eq:Skyrme_int:3body:gauge:condition:isovec_spin2}) involve 
functional terms containing two isovector densities and two vector densities. 
Condition~(\ref{eq:Skyrme_int:3body:gauge:condition:isosc_3spin}) involves 
functional terms containing three vector-isoscalar densities. Finally, 
condition~(\ref{eq:Skyrme_int:3body:gauge:condition:isovec_3spin}) involves 
all the functional terms with three spin densities among which two are 
isovector. Correlations between coupling constants resulting from conditions 
Eq.~(\ref{eq:Skyrme_int:3body:gauge:condition}) read
\begin{subequations}
\label{eq:Skyrme_int:3body:gauge:combination}
\begin{align}
&\text{Eq.~(\ref{eq:Skyrme_int:3body:gauge:condition:isosc_normal})}
 \Rightarrow
B^\tau_0 + B^j_0 = 0   \, ,
\\
&\text{Eq.~(\ref{eq:Skyrme_int:3body:gauge:condition:isosc_spin})} 
\Rightarrow
\left\{
\begin{array}{l}
2 B^{\tau s}_0 + B^{Js}_0 = 0 \\
2 B^T_0 + 2 B^J_0 + B^{Js}_0 = 0 \\
\end{array}\right. \, ,
\\
&\text{Eq.~(\ref{eq:Skyrme_int:3body:gauge:condition:isovec_normal})}
\Rightarrow
\left\{
\begin{array}{l}
2 B^\tau_{10} + B^j_{10} = 0 \\
2 B^\tau_1 + 2 B^j_1 + B^j_{10} = 0 \\
\end{array}\right. \, ,
\\
&\text{Eq.~(\ref{eq:Skyrme_int:3body:gauge:condition:isovec_spin1})}
\Rightarrow
\left\{
\begin{array}{l}
2 B^{\tau s}_{10} + B^{Js}_{01} = 0 \\
2 B^T_1 + 2 B^J_1 + B^{Js}_{01} = 0 \\
\end{array}\right. \,\,,
\\
&\text{Eq.~(\ref{eq:Skyrme_int:3body:gauge:condition:isovec_spin2})}
\Rightarrow
\left\{
\begin{array}{l}
2 B^{\tau s}_1 + B^{Js}_1 + B^{Js}_{10} = 0 \\
2 B^T_{10} + B^J_{10} + B^{Js}_{10} = 0 \\
2 B^T_{01} + B^J_{10} + B^{Js}_1 = 0 \\
\end{array}\right. \, ,
\\
&\text{Eq.~(\ref{eq:Skyrme_int:3body:gauge:condition:isovec_3spin})}
\Rightarrow
B^{\nabla s J}_1 - B^{\nabla s J}_{10} = 0 
\, ,
\end{align}
\end{subequations}
while Eq.~(\ref{eq:Skyrme_int:3body:gauge:condition:isosc_3spin}) is respected for all $B^{\nabla s J}_0$. Conditions~(\ref{eq:Skyrme_int:3body:gauge:combination}) are fulfilled by our functional coefficients, see Tables~\ref{tab:Skyrme_int:2bodyEDF:coeff} and~\ref{tab:Skyrme_int:3bodyEDF:coeff}. 


\subsection{Anomalous part of the EDF kernel}

The same strategy is followed for the anomalous part of the EDF kernel. 
The analogue of condition~(\ref{eq:Skyrme_int:gauge:condition_init}) is 
\begin{subequations}
\label{eq:Skyrme_int:gauge:condition_init_p}
\begin{align}
{\cal E}^{\kappa\kappa \, \prime} - {\cal E}^{\kappa\kappa} \equiv \Big[ {\cal E}^{\kappa\kappa} \Big]_{{\cal G}} &= 0 
\label{eq:Skyrme_int:2body:gauge:condition_init_p}
\,\,\,, \\
{\cal E}^{\kappa\kappa\kappa \, \prime} - {\cal E}^{\kappa\kappa\kappa} \equiv \Big[ {\cal E}^{\kappa\kappa\kappa} \Big]_{{\cal G}} &= 0
\label{eq:Skyrme_int:3body:gauge:condition_init_p}
 \,\,\,.
\end{align}
\end{subequations}
As seen from Eq.~(\ref{eq:Skyrme_int:gauge:locdensities}), all anomalous 
local densities are affected by gauge transformations. However each pairing 
density enters the energy density with the complex conjugate of another 
pairing density, such that bilinear products of the form 
$\breve{\rho}^{*} \breve{\rho}$, $\breve{J}^{*} \breve{J}$, 
$\breve{\rho}^{*} \breve{J}$ or $\breve{J}^{*} \breve{\rho}$ are effectively 
gauge invariant. As a result only $\breve{\tau}$ or derivatives of 
$\breve{\rho}$ and $\breve{J}$ have to be explicitly dealt with. For 
trilinear terms a gauge dependence might also arise from the third, then 
normal, local density. Again, correlations will only involve coefficients 
multiplying densities of same spin and isospin character.

For the bilinear functional, condition~(\ref{eq:Skyrme_int:2body:gauge:condition_init_p}) is equivalent to requiring
\begin{align}
\label{eq:Skyrme_int:2body:gauge:condition_p}
& \Big[ 
A^{\breve{\tau}^*} \sum_{\frak{a}=1,2} \breve{\tau}^*_{1,\frak{a}} \breve{\rho}_{1,\frak{a}}   
+A^{\breve{\tau}} \sum_{\frak{a}=1,2} \breve{\tau}_{1,\frak{a}} \breve{\rho}^*_{1,\frak{a}}   
 \\
& \hspace{2.0cm} + A^{\nabla \breve{\rho}} \sum_{\frak{a}=1,2} (\vec{\nabla} \breve{\rho}^*_{1,\frak{a}} ) \cdot (\vec{\nabla} \breve{\rho}_{1,\frak{a}})   
\Big]_{{\cal G}}
= 0  \,, \nonumber
\end{align}
which is fulfilled for
\begin{align}
A^{\nabla \breve{\rho}} 
= \frac{1}{2} A^{\breve{\tau}^*} 
= \frac{1}{2} A^{\breve{\tau}} \, .
\end{align}
For the trilinear functional, condition~(\ref{eq:Skyrme_int:3body:gauge:condition_init_p}) gives rise to eight independent gauge invariant conditions that read
\begin{widetext}
\begin{subequations}
\label{eq:Skyrme_int:3body:gauge:condition_p}
\begin{align}
0 = & \Big[ \sum_{\frak{a}=1,2} \Big\{
   B^{\breve{\tau}^*}_0 \breve{\tau}^*_{1,\frak{a}} \breve{\rho}_{1,\frak{a}} \rho_{0}   
+ B^{\breve{\tau}}_0 \breve{\rho}^*_{1,\frak{a}} \breve{\tau}_{1,\frak{a}} \rho_{0}   
+ B^{\breve{\rho} \tau}_0 \breve{\rho}^*_{1,\frak{a}} \breve{\rho}_{1,\frak{a}} \tau_{0}  
+ B^{\nabla \breve{\rho}}_0 (\vec{\nabla} \breve{\rho}^*_{1,\frak{a}}) \cdot (\vec{\nabla} \breve{\rho}_{1,\frak{a}}) \rho_{0}   
\nonumber \\ & \;
+ {\mathrm i} B^{\nabla \breve{\rho}^* j}_0 (\vec{\nabla} \breve{\rho}^*_{1,\frak{a}}) \breve{\rho}_{1,\frak{a}}  \cdot \vec{j}_{0} 
+ {\mathrm i} B^{\nabla \breve{\rho} j}_0 \breve{\rho}^*_{1,\frak{a}} (\vec{\nabla} \breve{\rho}_{1,\frak{a}}) \cdot \vec{j}_{0} 
\Big\} \Big]_{{\cal G}} 
\,, \label{eq:Skyrme_int:3body:gauge:condition_p:isosc_normal} \\
0 = & \Big[ \sum_{\frak{a}=1,2} \Big\{
    B^{\nabla \breve{\rho}^* \breve{\rho}}_0 (\vec{\nabla} \breve{\rho}^*_{1,\frak{a}}) \breve{\rho}_{1,\frak{a}} \cdot (\vec{\nabla} \rho_{0}) 
+ B^{\breve{\rho}^*  \nabla \breve{\rho}}_0 \breve{\rho}^*_{1,\frak{a}} (\vec{\nabla} \breve{\rho}_{1,\frak{a}}) \cdot (\vec{\nabla} \rho_{0}) 
\Big\} \Big]_{{\cal G}} 
\,, \label{eq:Skyrme_int:3body:gauge:condition_p:isosc_normal2} \\
0 = & \Big[ \sum_{\frak{a}=1,2} \sum_{\mu \nu} \Big\{
    B^{\breve{\rho}^* \breve{J}}_0 \breve{\rho}^*_{1,\frak{a}}  \breve{J}_{1,\frak{a}, \mu \nu} J_{0, \mu \nu} 
+ {\mathrm i} B^{\nabla \breve{\rho}^* \breve{J}}_0 (\nabla_\mu \breve{\rho}^*_{1,\frak{a}}) \breve{J}_{1,\frak{a}, \mu \nu} s_{0, \nu}   
\Big\} \Big]_{{\cal G}} 
\,, \label{eq:Skyrme_int:3body:gauge:condition_p:isosc_spin} \\
0 = & \Big[ \sum_{\frak{a}=1,2} \sum_{\mu \nu} \Big\{
    B^{\breve{J}^* \breve{\rho}}_0 \breve{J}^*_{1,\frak{a}, \mu \nu} \breve{\rho}_{1,\frak{a}}  J_{0, \mu \nu} 
+ {\mathrm i} B^{\breve{J}^* \nabla \breve{\rho}}_0 \breve{J}^*_{1,\frak{a}, \mu \nu} (\nabla_\mu \breve{\rho}_{1,\frak{a}}) s_{0, \nu}   
\Big\} \Big]_{{\cal G}} 
\,, \label{eq:Skyrme_int:3body:gauge:condition_p:isosc_spin2} \\
0 = & \Big[ \sum_{\frak{a},\frak{b}=1,2} \sum_{\frak{c}=3} \epsilon_{\frak{a}\frak{b}\frak{c}}  \Big\{
    {\mathrm i} B^{\breve{\tau}^*}_1 \breve{\tau}^*_{1,\frak{a}} \breve{\rho}_{1,\frak{b}} \rho_{1,\frak{c}}   
+ {\mathrm i} B^{\breve{\tau}}_1 \breve{\rho}^*_{1,\frak{a}} \breve{\tau}_{1,\frak{b}} \rho_{1,\frak{c}}   
+ {\mathrm i} B^{\breve{\rho} \tau}_1 \breve{\rho}^*_{1,\frak{a}} \breve{\rho}_{1,\frak{b}}  \tau_{1,\frak{c}} 
+ {\mathrm i} B^{\nabla \breve{\rho}}_1 (\vec{\nabla} \breve{\rho}^*_{1,\frak{a}}) \cdot (\vec{\nabla} \breve{\rho}_{1,\frak{b}}) \rho_{1,\frak{c}}   
\nonumber \\ & \;
+ B^{\nabla \breve{\rho}^* j}_1 (\vec{\nabla} \breve{\rho}^*_{1,\frak{a}}) \breve{\rho}_{1,\frak{b}}  \cdot \vec{j}_{1,\frak{c}}   
+ B^{\nabla \breve{\rho} j}_1  \breve{\rho}^*_{1,\frak{a}}  (\vec{\nabla} \breve{\rho}_{1,\frak{b}}) \cdot \vec{j}_{1,\frak{c}}   
\Big\} \Big]_{{\cal G}} 
\,, \label{eq:Skyrme_int:3body:gauge:condition_p:isovec_normal} \\
0 = & \Big[ \sum_{\frak{a},\frak{b}=1,2} \sum_{\frak{c}=3} \epsilon_{\frak{a}\frak{b}\frak{c}}  \Big\{
   {\mathrm i} B^{\nabla \breve{\rho}^* \breve{\rho}}_1 (\vec{\nabla} \breve{\rho}^*_{1,\frak{a}}) \breve{\rho}_{1,\frak{b}} \cdot (\vec{\nabla} \rho_{1,\frak{c}})
+ {\mathrm i} B^{\breve{\rho}^* \nabla \breve{\rho}}_1 \breve{\rho}^*_{1,\frak{a}}  (\vec{\nabla} \breve{\rho}_{1,\frak{b}}) \cdot (\vec{\nabla} \rho_{1,\frak{c}} )
\Big\} \Big]_{{\cal G}} 
\,, \label{eq:Skyrme_int:3body:gauge:condition_p:isovec_normal2} \\
0 = & \Big[ \sum_{\frak{a},\frak{b}=1,2} \sum_{\frak{c}=3} \epsilon_{\frak{a}\frak{b}\frak{c}}  \sum_{\mu \nu} \Big\{
    {\mathrm i} B^{\breve{\rho}^* \breve{J}}_1 \breve{\rho}^*_{1,\frak{a}}  \breve{J}_{1,\frak{b}, \mu \nu} J_{1,\frak{c}, \mu \nu}  
+ B^{\nabla \breve{\rho}^* \breve{J}}_1  (\nabla_\mu \breve{\rho}^*_{1,\frak{a}}) \breve{J}_{1,\frak{b}, \mu \nu} s_{1,\frak{c}, \nu}   
\Big\} \Big]_{{\cal G}} 
\,, \label{eq:Skyrme_int:3body:gauge:condition_p:isovec_spin} \\
0 = & \Big[ \sum_{\frak{a},\frak{b}=1,2} \sum_{\frak{c}=3} \epsilon_{\frak{a}\frak{b}\frak{c}}  \sum_{\mu \nu} \Big\{
    {\mathrm i} B^{\breve{J}^* \breve{\rho}}_1 \breve{J}^*_{1,\frak{a}, \mu \nu} \breve{\rho}_{1,\frak{b}}  J_{1,\frak{c}, \mu \nu}  
+ B^{\breve{J}^* \nabla \breve{\rho}}_1 \breve{J}^*_{1,\frak{a}, \mu \nu} (\nabla_\mu \breve{\rho}_{1,\frak{b}}) s_{1,\frak{c}, \nu}   
\Big\} \Big]_{{\cal G}} 
\,, \label{eq:Skyrme_int:3body:gauge:condition_p:isovec_spin2}
\end{align}
\end{subequations}
\end{widetext}
These relations must be independently fulfilled, which requires 
that the coupling constants satisfy
\begin{subequations}
\label{eq:Skyrme_int:3body:gauge:combination_p}
\begin{align}
&\text{Eq.~(\ref{eq:Skyrme_int:3body:gauge:condition_p:isosc_normal})}
 \Rightarrow
\left\{
\begin{array}{l}
- B^{\breve{\tau}^*}_0 
+ 2 B^{\nabla \breve{\rho}}_0   
+ B^{\nabla \breve{\rho}^* j}_0  
= 0
\\
B^{\breve{\tau}}_0 
- 2 B^{\nabla \breve{\rho}}_0   
+ B^{\nabla \breve{\rho} j}_0 
= 0
\\
B^{\breve{\rho} \tau}_0 
+ B^{\nabla \breve{\rho}^* j}_0 
- B^{\nabla \breve{\rho} j}_0 
= 0
\end{array}\right. \,\,, 
\\
&\text{Eq.~(\ref{eq:Skyrme_int:3body:gauge:condition_p:isosc_normal2})} 
\Rightarrow
B^{\nabla \breve{\rho}^* \breve{\rho}}_0
=
B^{\breve{\rho}^*  \nabla \breve{\rho}}_0 \,\,,
\\
&\text{Eq.~(\ref{eq:Skyrme_int:3body:gauge:condition_p:isosc_spin})} 
\Rightarrow
B^{\breve{\rho}^* \breve{J}}_0 
=
- 2 B^{\nabla \breve{\rho}^* \breve{J}}_0 \,\,,
\\
&\text{Eq.~(\ref{eq:Skyrme_int:3body:gauge:condition_p:isosc_spin2})} 
\Rightarrow
B^{\breve{J}^* \breve{\rho}}_0 
=
2 B^{\breve{J}^* \nabla \breve{\rho}}_0  \,\,,
\\
&\text{Eq.~(\ref{eq:Skyrme_int:3body:gauge:condition_p:isovec_normal})}
 \Rightarrow
\left\{
\begin{array}{l}
- B^{\breve{\tau}^*}_1 
+ 2 B^{\nabla \breve{\rho}}_1   
- B^{\nabla \breve{\rho}^* j}_1  
= 0
\\
B^{\breve{\tau}}_1 
- 2 B^{\nabla \breve{\rho}}_1  
- B^{\nabla \breve{\rho} j}_1 
= 0
\\
B^{\breve{\rho} \tau}_1 
- B^{\nabla \breve{\rho}^* j}_1 
+ B^{\nabla \breve{\rho} j}_1 
= 0
\end{array}\right. \,,
\\
&\text{Eq.~(\ref{eq:Skyrme_int:3body:gauge:condition_p:isovec_normal2})} 
\Rightarrow
B^{\nabla \breve{\rho}^* \breve{\rho}}_1
=
B^{\breve{\rho}^*  \nabla \breve{\rho}}_1 \, ,
\\
&\text{Eq.~(\ref{eq:Skyrme_int:3body:gauge:condition_p:isovec_spin})} 
\Rightarrow
B^{\breve{\rho}^* \breve{J}}_1 
=
2 B^{\nabla \breve{\rho}^* \breve{J}}_1 \, ,
\\
&\text{Eq.~(\ref{eq:Skyrme_int:3body:gauge:condition_p:isovec_spin2})} 
\Rightarrow
B^{\breve{J}^* \breve{\rho}}_1 
=
- 2 B^{\breve{J}^* \nabla \breve{\rho}}_1 \, ,
\end{align}
\end{subequations}
which is indeed the case our EDF kernel, see 
Tables~\ref{tab:Skyrme_int:2bodyEDF:coeffP} 
and~\ref{tab:Skyrme_int:3bodyEDF:coeffP}.


\end{appendix}


%
%

\end{document}